\documentclass[12pt,a4paper]{article}
\pdfoutput=1
\usepackage{jheppub}

\usepackage{epsfig,multirow} 
\usepackage{amscd}
\usepackage[matrix,arrow,curve]{xy}
\usepackage{verbatim}
\usepackage{latexsym}
\usepackage{amsfonts,amsthm,amsmath}

\newcommand {\non}{\nonumber}

\newcommand{\beq}{\begin{equation}}
\newcommand{\eeq}{\end{equation}}
\newcommand{\bea}{\begin{eqnarray}}
\newcommand{\eea}{\end{eqnarray}}
\newcommand{\ena}{\end{eqnarray}}

\newcommand{\firr}[1]{{}^{{\rm Irr}}\!{\cal F}^{\flat}_{#1}}

\subheader{UCSD-PTH-12-04}
\title{\begin{center}
Integrability on the Master Space
\end{center}}
\author[a]{Antonio Amariti,}
\author[b]{Davide Forcella,}
\author[c]{Alberto Mariotti}

\affiliation[a]{Department of Physics, University of California, \\
San Diego La Jolla, CA 92093-0354, USA}
\affiliation[b]{
Physique Th\'eorique et Math\'ematique and International Solvay Institutes\\
Universit\'e Libre de Bruxelles, C.P. 231, 1050 Bruxelles, Belgium
}
\affiliation[c]{
Theoretische Natuurkunde, Vrije Universiteit Brussel \\
and International Solvay Institutes Pleinlaan 2, B-1050 Brussels, Belgium
}

\emailAdd{antonio.amariti@physics.ucsd.edu}
\emailAdd{dforcell@ulb.ac.be}
\emailAdd{amariott@vub.ac.be}

\abstract{
It has been recently shown that every SCFT living on D3 branes at a toric Calabi-Yau singularity surprisingly also describes a complete integrable system. In this paper we use the Master Space as a bridge between the integrable system and the underlying field theory and we reinterpret the Poisson manifold of the integrable system in term of the geometry of the field theory moduli space. 
}

\begin{document}
\maketitle

\section{Introduction}

Recently the AdS/CFT correspondence has been successfully applied to many fields of research both in physics and in mathematics.  
One of the most understood and investigated realization of the correspondence concerns the relation between type IIB supergravity on AdS$_5 \times Y$ and four dimensional SCFT describing the low energy dynamics of a stack of N D$3$ branes probing the tip  of the Calabi-Yau (CY) cone $\mathcal{X}$=C(Y) over the five dimensional Sasaki Einstein manifold $Y$. When the CY singularity is toric, namely it has at least $U(1)^3$ isometry, many tools have been developed to deeply study the correspondence (see \cite{Kennaway:2007tq}
for reivew).

Indeed in this specific case the SCFT living on a stack of N D$3$ branes is a specific  type of quiver gauge theory: it has $SU(N)^g$ gauge group and matter fields in the bifundamental representation of the gauge group, that appear just two times in the superpotential, once with positive sign and once with negative sign. All the information of these SCFT can be encoded in a dual structure, a bipartite graph on a torus, called the brane tiling and it gives a relation between the SCFT and the statistical mechanics of this graph 
\cite{Hanany:2005ve,Franco:2005rj,Hanany:2005ss,Feng:2005gw}.
The partition function of the brane tiling contains the informations regarding the mesonic moduli space of the toric quiver gauge theory, namely the $\mathcal{X}$ geometry itself, and the toric data can be encoded in a polyhedral cone called the 2d toric diagram.
Thanks to these results the study of the moduli space of $\mathcal{N}=1$ SCFT
has been related to the mathematical literature of graph theory, and the developments on one side could foster the developments on the other side.

Very recently a new exciting result has been reported in \cite{Goncharov:2011hp}.
It has indeed been observed that starting from the bipartite graph for the SCFT we have just described, one can construct a completely integrable system, in which the oriented loops of this graph are the dynamical variables.
The Poisson manifold \cite{2008InMat.175..223F} associated to this system is the collection of bipartite diagrams (seeds) glued via Poisson cluster transformations. These transformations, also known as mutations, have been studied on quiver diagrams in \cite{2007arXiv0704.0649D}.

It is interesting to investigate if the Poisson manifold can be analyzed with the field theory language of toric $\mathcal{N}=1$ SCFT. The study of the physical implication of the
integrability of the bipartite graph in field theory started in \cite{Franco:2011sz,Eager:2011dp,Eager:2011ns}.
In \cite{Franco:2011sz} the connection between the 
brane tiling, the quantum  Teichmuller  theory and the Poisson structure 
was investigated.
Then in \cite{Eager:2011dp} the Poisson manifold was studied in 
the language of quiver gauge theory and interesting connections with 
five dimensional $\mathcal{N}=1$ theories were made. Then in \cite{Eager:2011ns}
a connection with cluster algebra and flavored quiver models was proposed.

In this paper we want to show that the coherent component of the Master Space $\firr~$ provides also a geometrical description of the integrable system introduced in \cite{Goncharov:2011hp}. In \cite{Forcella:2008bb} $\firr~$ was introduced as the natural variety to describe the full moduli space of the $\mathcal{N}=1$ SCFT and the spectrum of supersymmetric gauge invariant operators for any number N of D3 branes.
The master space $\mathcal{F}^{\flat}$ is the full moduli space of one D$3$ brane 
probing the CY cone and it is the union of the baryonic and mesonic moduli
spaces. $\mathcal{F}^{\flat}$ is a reducible $g+2$ dimensional algebraic variety, and its 
largest irreducible component is $\firr~$ that it has been shown in 
\cite{Forcella:2008bb} to be a $g+2$ dimensional toric CY manifold.
$\firr~$ is a toric variety and it has a natural action of the $U(1)^{g+2}$ 
symmetries coming from the field theory global $U(1)$ symmetries: 
three mesonic and $g-1$ baryonic.
Since it is a conical toric variety it can be represented as an integer $g+2$ 
dimensional fan of vectors in the $\mathbb{Z}^{g+2}$ lattice. Moreover, because $\firr~$ is CY, 
the vectors describing the fan are coplanar.
The crucial connection between this variety and the Poisson manifold consists in 
relating the Poisson brackets among the loops of the bipartite graph \cite{Goncharov:2011hp} with an induced algebra among the $U(1)$ charges encoded in $\firr~$, and in identifying the chemical potentials of these charges as the natural canonical coordinates on a patch of the Poisson manifold. The adjacency matrix of the quiver theory will be the link between the Poisson algebra on the loops and the Poisson algebra on the $U(1)$ charges, while the vectors defining $\firr~$ will provide the map between the loop variables of the system and the local coordinates given by the chemical potentials. On every patch the Hamiltonians and the Casimir operators will be naturally expressed in terms of these local coordinates and they will be crucial to construct the full Poisson manifold. Indeed the different patches should be related by canonical transformations obtained mapping the Casimir operators, the Hamiltonians and their flows between two $\firr~$ of Seiberg dual phases associated to the same SCFT.
\\

In section \ref{sec2} we review some useful material for the study of
four dimensional toric quiver gauge theories, and the description 
of the moduli space in terms of dimer models. 
In \ref{sec2.1} we introduce the master space and its irreducible component $\firr~$. In section \ref{sec3} we review the cluster integrable dimer model derived in 
 \cite{Goncharov:2011hp}. We discuss the Poisson structure, the Casimir operators and the Hamiltonians in terms of loops on the dimer.  In section \ref{sec4} we illustrate the main result of the paper. We explain the 
 relation among the toric diagram describing the master space and the 
 Poisson structure on the dimer model.
 We give the detailed dictionary among the coordinates 
 used in \cite{Goncharov:2011hp} and the $U(1)$ global charges 
of $\firr~$. In \ref{sec4.1} we define the explicit 
algorithm to compute the Poisson structure among 
the charges and we define the Hamiltonians 
and the Casimir operators in our coordinates.
In section \ref{sec5} we study some detailed examples. We use the cone over the 
dP$_0$ surface as an illustrative example to fix the notation and the general procedure. 
Then we study the two Seiberg dual phases of the cone over $\mathbb{F}_0$. We then
 move to the cases with more Hamiltonians: $Y^{30}$ and $Y^{40}$. In section \ref{sec6} 
we analyze in more detail the relation between Seiberg dualities and the canonical map 
between the $\firr~$ of the two phases of $\mathbb{F}_0$.
In section \ref{sec7} we conclude and discuss some interesting 
future directions.

In appendix \ref{appA} we review the derivation of the antisymmetric structure of \cite{Goncharov:2011hp} as an intersection pairing among the cycles on the tiling.
Then in appendix \ref{appB} we give some details about  Seiberg duality 
as explained in \cite{Goncharov:2011hp}. In the last appendix \ref{appC}
we explicitly  show the perfect matchings of $Y^{30}$ and $Y^{40}$.

\section{Toric quiver gauge theories and dimer models}
\label{sec2}

In this section we briefly review the topic of superconformal 
$\mathcal{N}=1$ four dimensional 
quiver gauge theories 
describing a stack of $N$ D$3$ branes probing toric CY$_3$ singularities.
To simplify the discussion we 
make use of an example, which will be useful for the rest of the paper.
The example we are referring to is the $\mathbb{Z}_3$ chiral orbifold of 
$\mathcal{N}=4$ SYM, also known as the cone over the 
dP$_0$ singularity.

\begin{figure}
\begin{center}
\includegraphics[width=5cm]{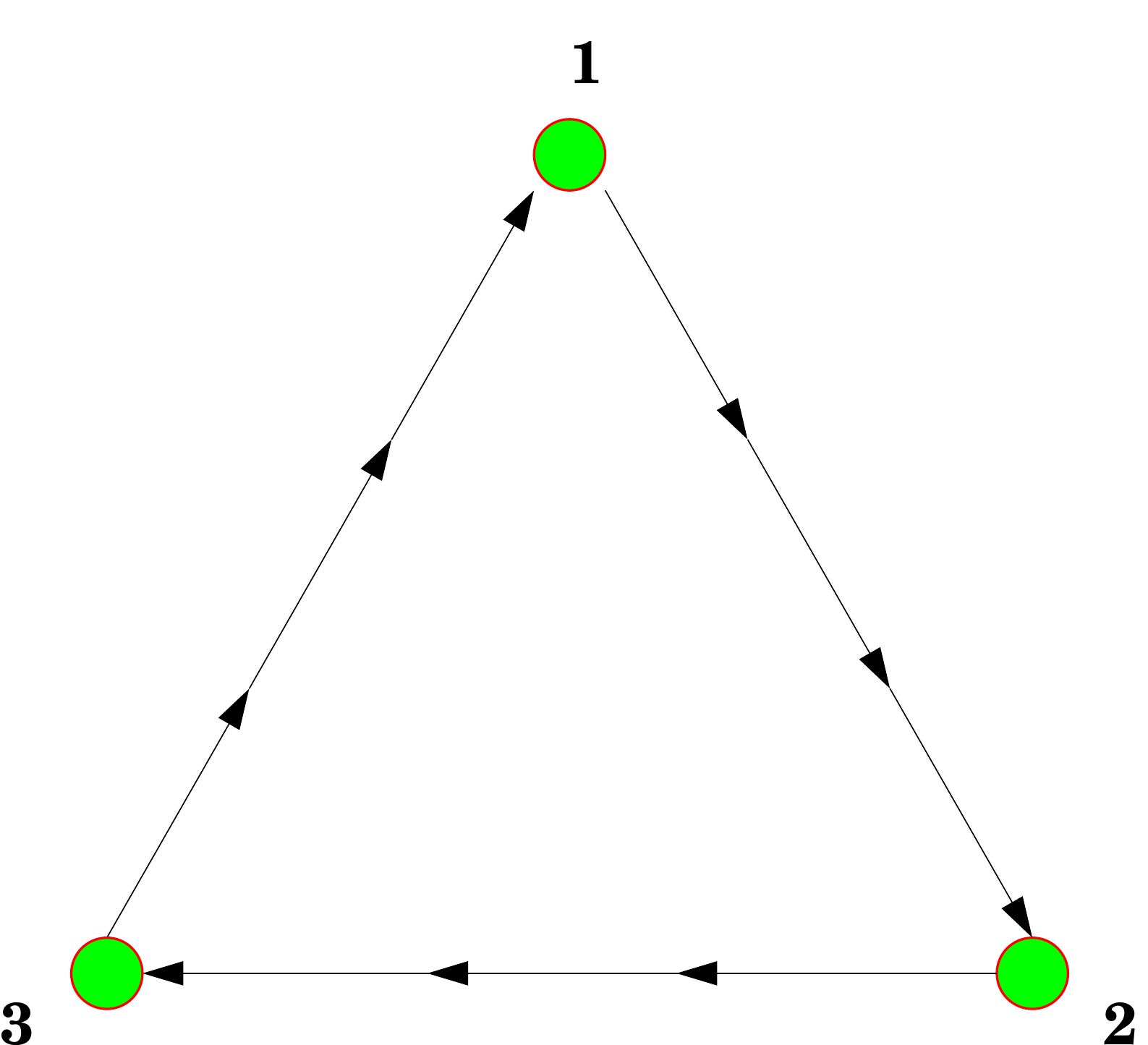}
\caption{Quiver diagram for dP$_0$}
\label{figquivdp0} 
\end{center}
\end{figure}

The quiver is a connected  two dimensional finite graph with nodes, 
which 
encodes all the information about the field content and the gauge symmetries of the 
theory.
The nodes
represent the $SU(N)$ gauge groups while the
oriented arrows connecting the nodes represent the $\mathcal{N}=1$
chiral multiplets.
The orientation of the arrows defines the representation of the field under the gauge groups.
The incoming arrows with respect of the $i-th$ node are associated with the 
fundamental representation under $SU(N)_i$
while the outcoming arrows are associated to the antifundamental representation. 
Every endpoint of each arrow ends on a node, and the only admissible representation are adjoint and bifundamental.
For instance in the case of dP$_0$ the quiver is given in the Figure \ref{figquivdp0} 
and it is associated to a product of three $SU(N)$  gauge groups with three chiral
bifundamental fields connecting each pair of nodes, $X_{12}^{(i)}$, $X_{23}^{(i)}$ and $X_{31}^{(i)}$,
with $i=1,2,3$.

Useful tools to describe the quiver are the oriented incidence and the adjacency matrices.
For a quiver with $g$ vertices (gauge groups)
and $\varepsilon $ oriented edges (bifundamental fields), the oriented
 incidence matrix $d$ is an $g \times \varepsilon$ matrix such
that the $(\bold{g}, e)$-th entry is $1$ (or $-1$)
 if the edge labelled $e$ is ingoing (or outgoing) to the  $\bold{g}$-th vertex 
 and zero otherwise. The adjoint fields always contribute as zero.
 Here for $dP_0$ the oriented incidence matrix is
 \begin{equation}
 d = \left(
 \begin{array}{c||ccccccccc}
  &X_{12}^{(1)}& X_{12}^{(2)}& X_{12}^{(3)}& X_{23}^{(1)}& X_{23}^{(2)}& X_{23}^{(3)}& X_{31}^{(1)}& X_{31}^{(2)}& X_{31}^{(3)}\\
 \hline
  1&1&1&1&0&0&0&-1&-1&-1\\
  2&-1&-1&-1&1&1&1&0&0&0\\
  3&0&0&0&-1&-1&-1&1&1&1
  \end{array}
 \right)
 \end{equation}
From the incidence matrix one can define the antisymmetric oriented adjacency matrix as 
$a=d |d^T|$.
This is a quadratic  $g\times g$ matrix  
such
that the $(\bold{g}_i, \bold{g}_j)$-th entry is the number of arrows
from $\bold{g}_i$ to $\bold{g}_j$, counted with their orientation where $\bold{g}_i$ and 
$\bold{g}_j$ are vertices.
For $dP_0$ it is
\begin{equation}
a=
\left(
\begin{array}{ccc}
0&3&-3\\
-3&0&3\\
3&-3&0\\
\end{array}
\right)
\end{equation}
Note that the row and columns of the adjacency matrix always sum to zero
for anomaly cancellation.

The quiver diagram, together with a superpotential $W$, completely defines the 
gauge theory describing the D$3$ branes probing the toric CY.
The superpotential is a function of the chiral fields associated to the edges and in the toric case it has a constrained structure. 
Every field appears precisely only twice in $W$ and with opposite sign.
The information about the superpotential can be directly added to the quiver diagram by defining a periodic 
graph, called the planar quiver.  
In this diagram the superpotential terms become the boundary of oriented polygons (plaquettes). 
The plaquettes are glued together along the fields that belong to both the 
superpotential terms. The orientation of the plaquettes determines the signs of the superpotential terms.

This geometrical structure is very useful because its dual graph is a polygonal tiling
of a torus, called the brane tiling \cite{Franco:2005rj}.
This graph is defined from the periodic quiver by replacing each faces with a vertex.
Then  edges separating two adjacent faces are
replaced by dual edges  and the vertices are replaced
by  faces, delimited by the dual edges.

This graph is bipartite (every vertex is  black or white) and this assignment is
defined by the orientation of the plaquettes. The vertices of the dimer represent the 
superpotential interactions, while the faces are related to the gauge groups.

\begin{figure}
\begin{center}
\includegraphics[width=15cm]{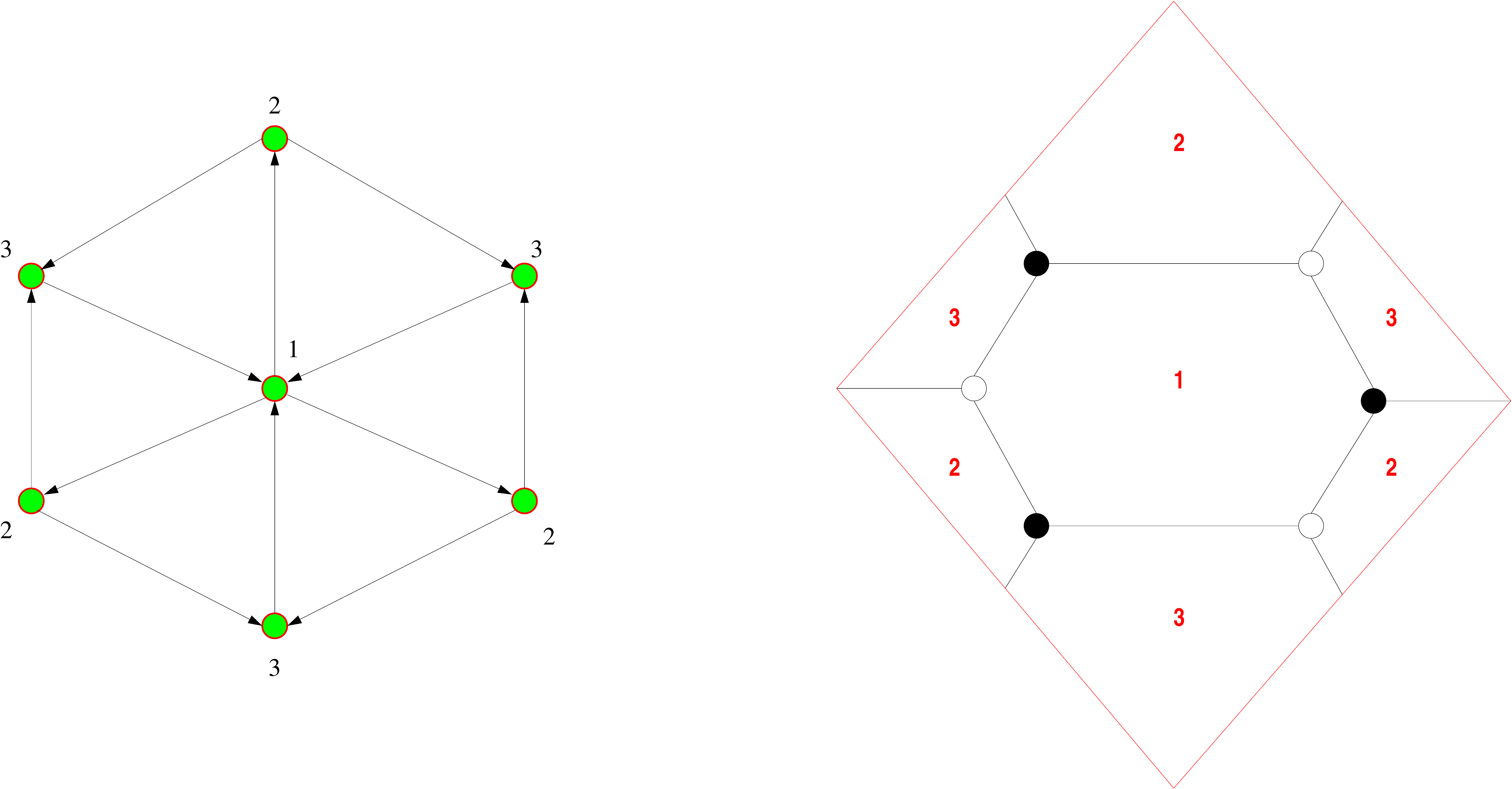}
\caption{Periodic quiver and bipartite diagram  for dP$_0$}
\label{dP0pertil} 
\end{center}
\end{figure}

\begin{figure}
\begin{center}
\begin{tabular}{ccc}
\includegraphics[width=3cm]{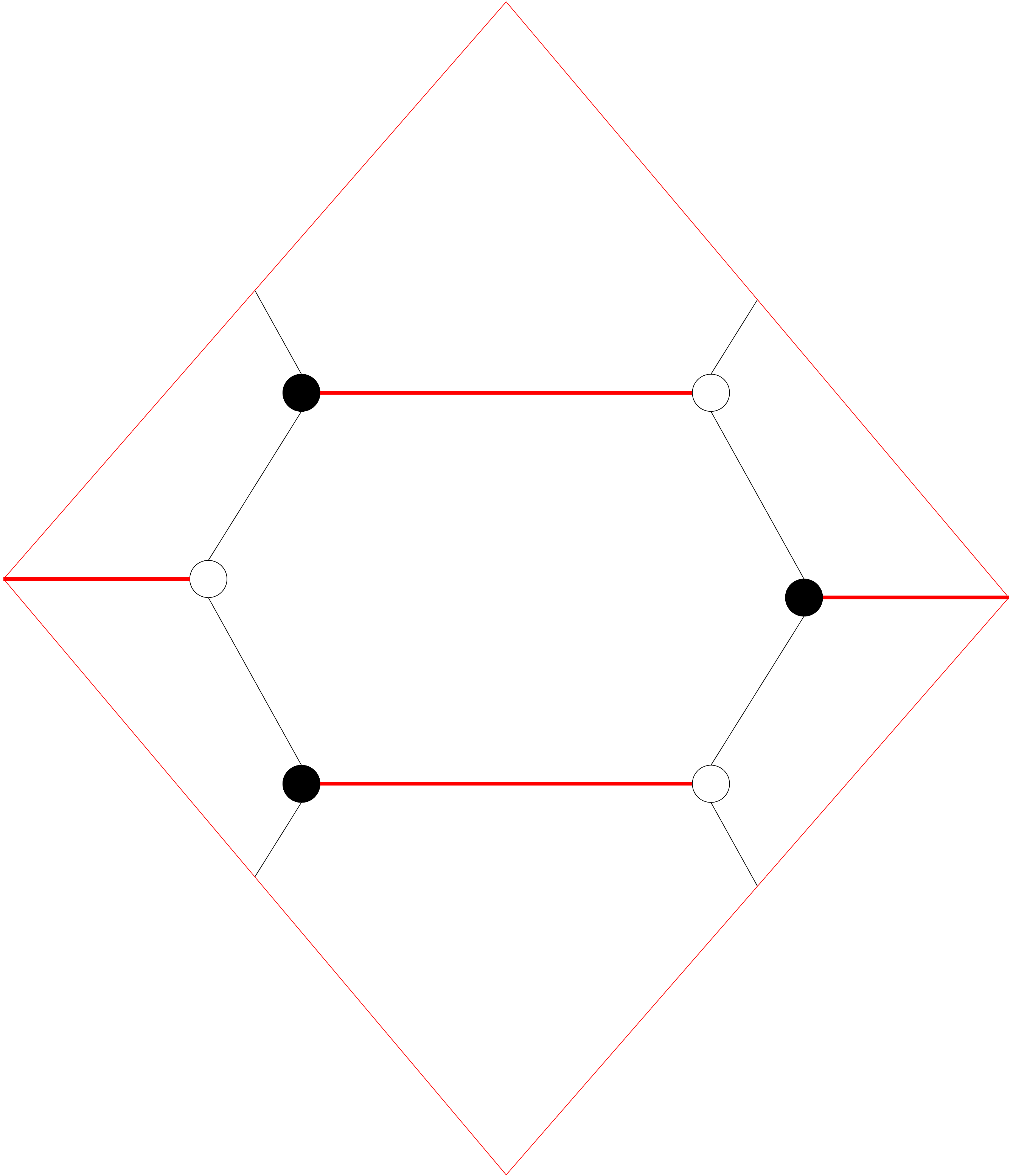}
&
\includegraphics[width=3cm]{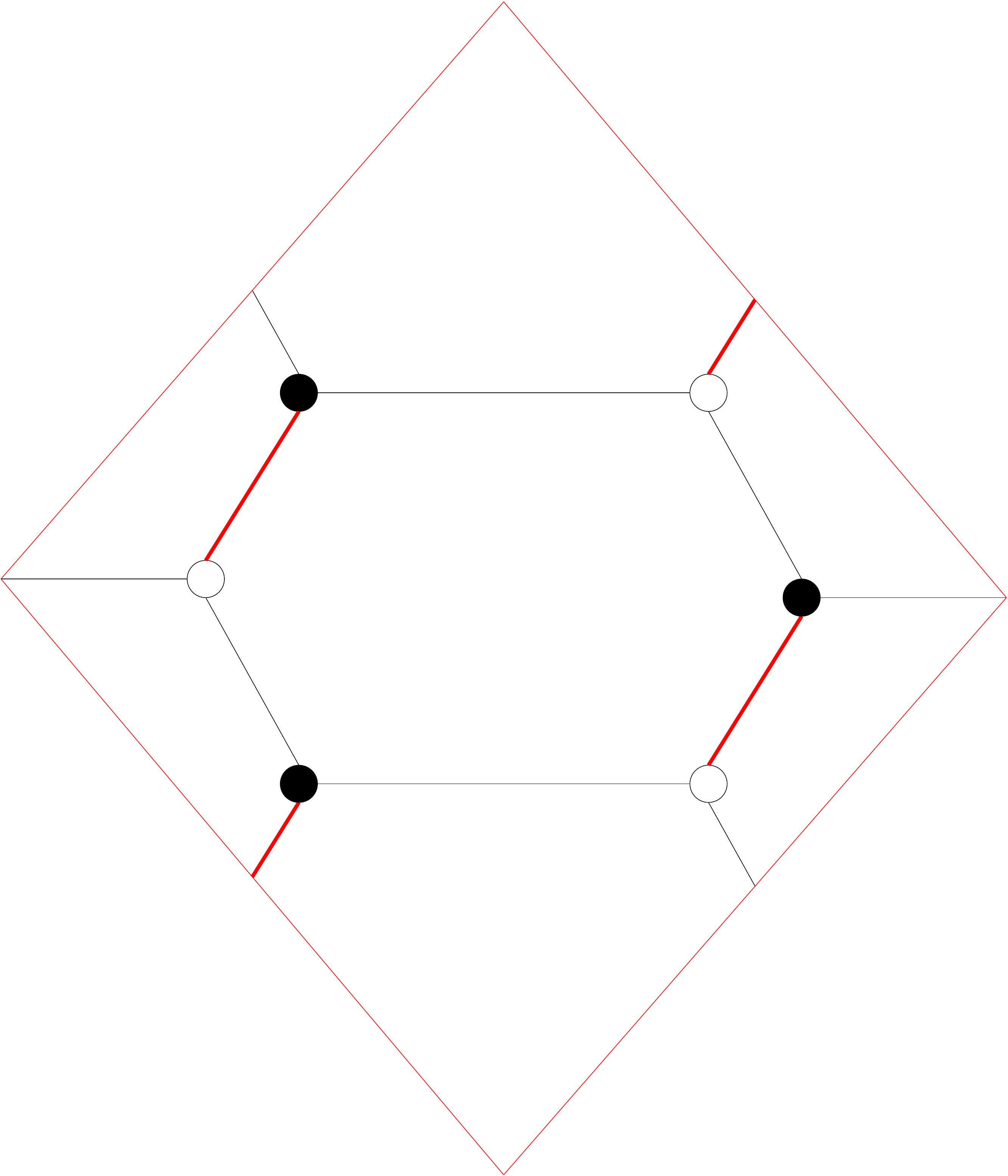}
&
\includegraphics[width=3cm]{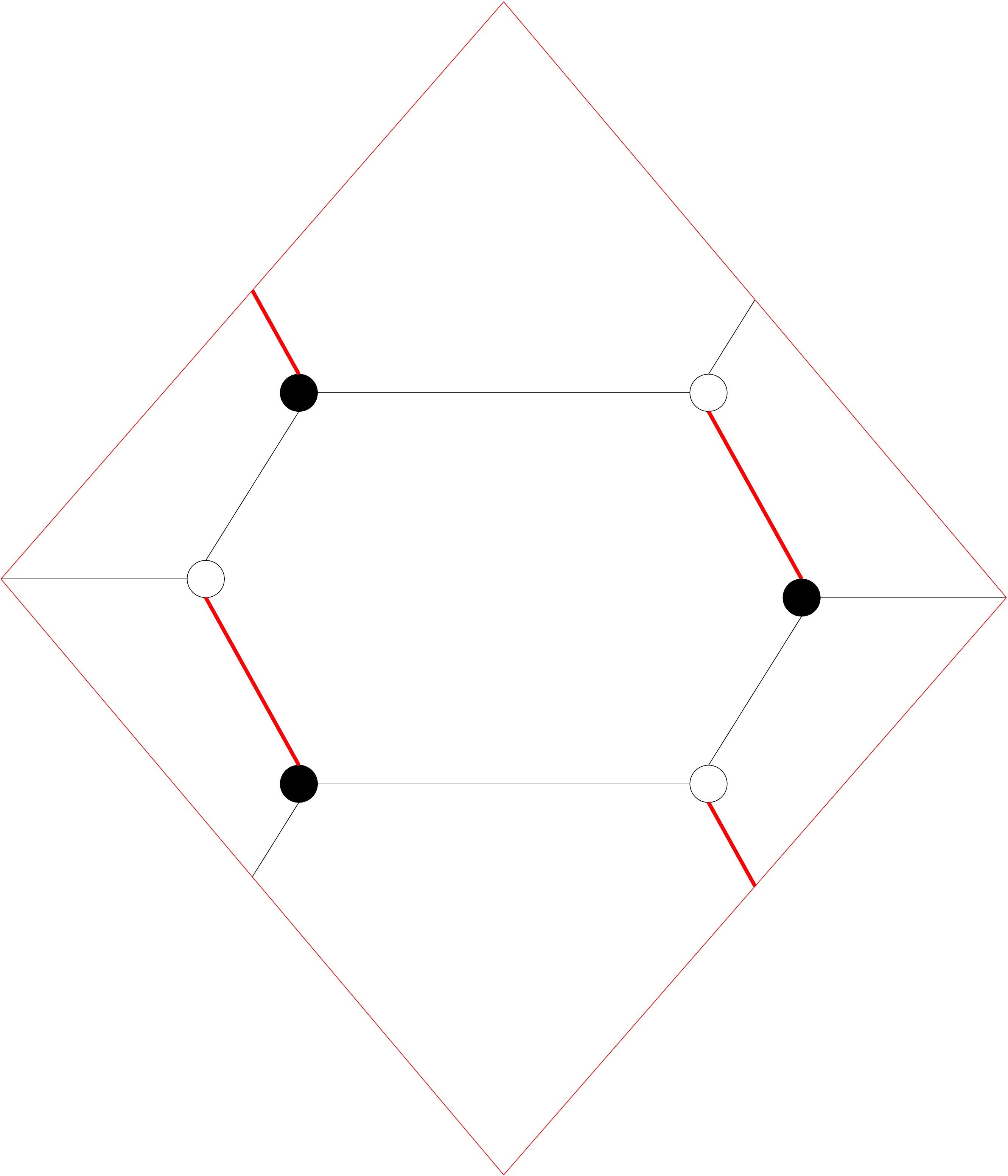}\\
&&\\
\includegraphics[width=3cm]{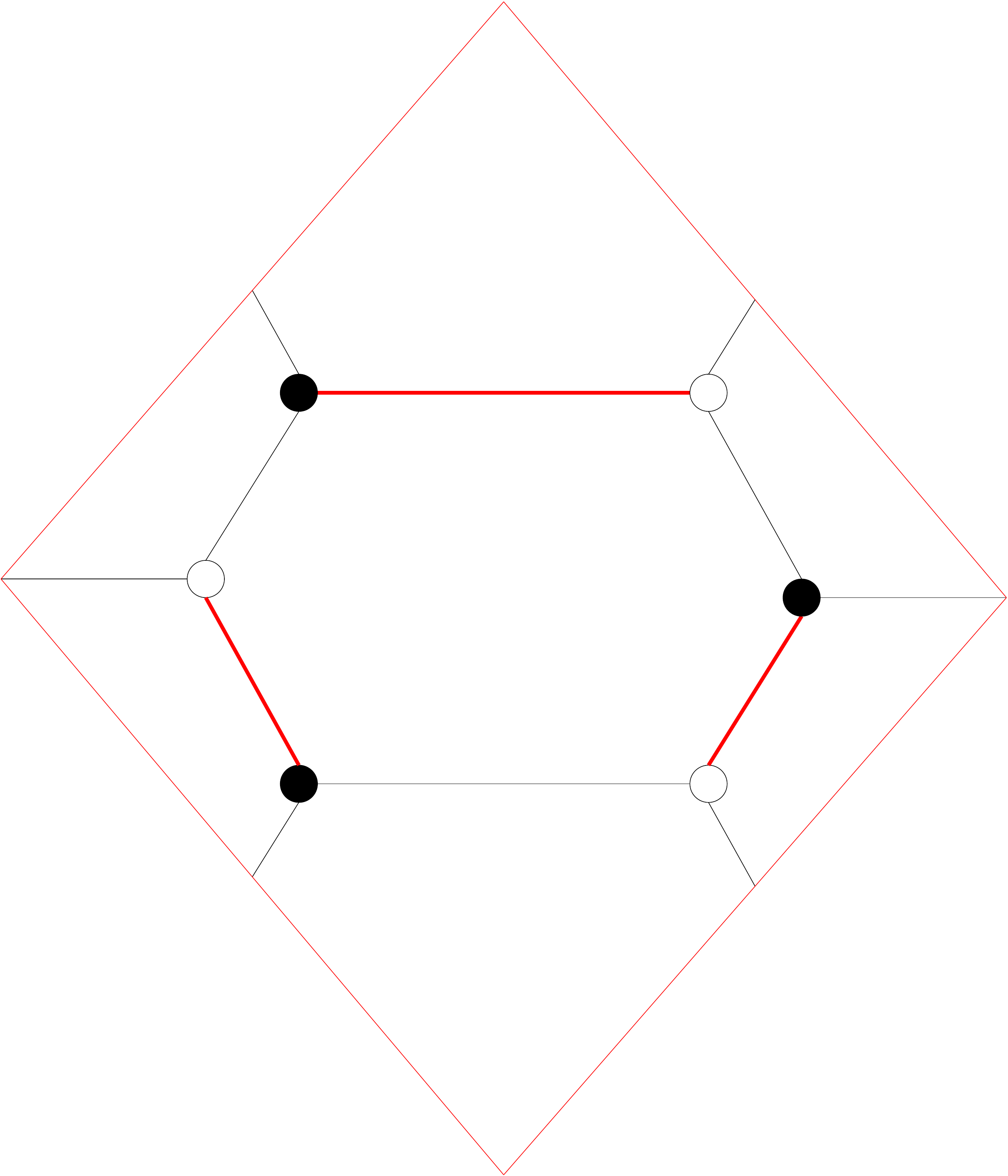}
&
\includegraphics[width=3cm]{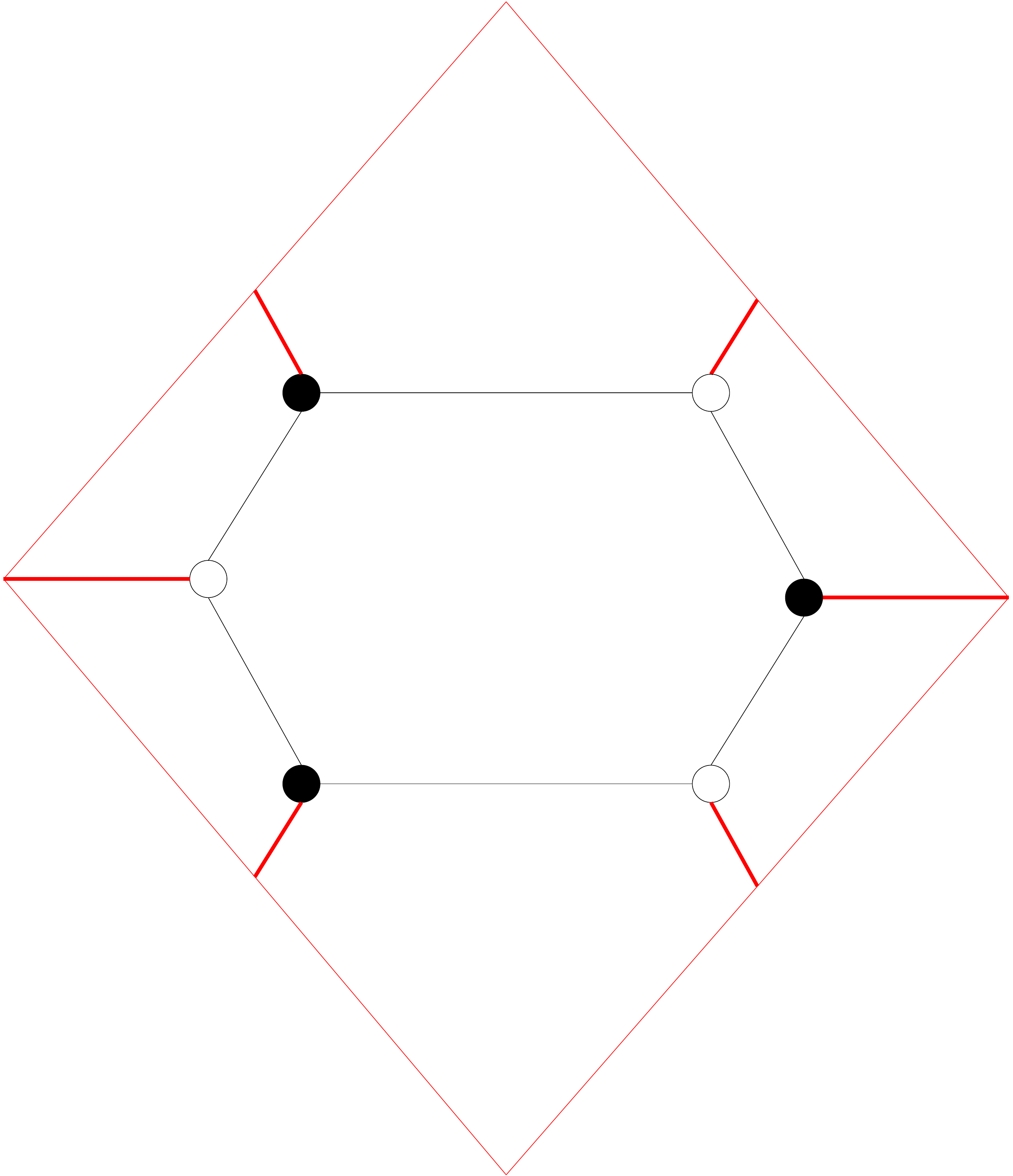}
&
\includegraphics[width=3cm]{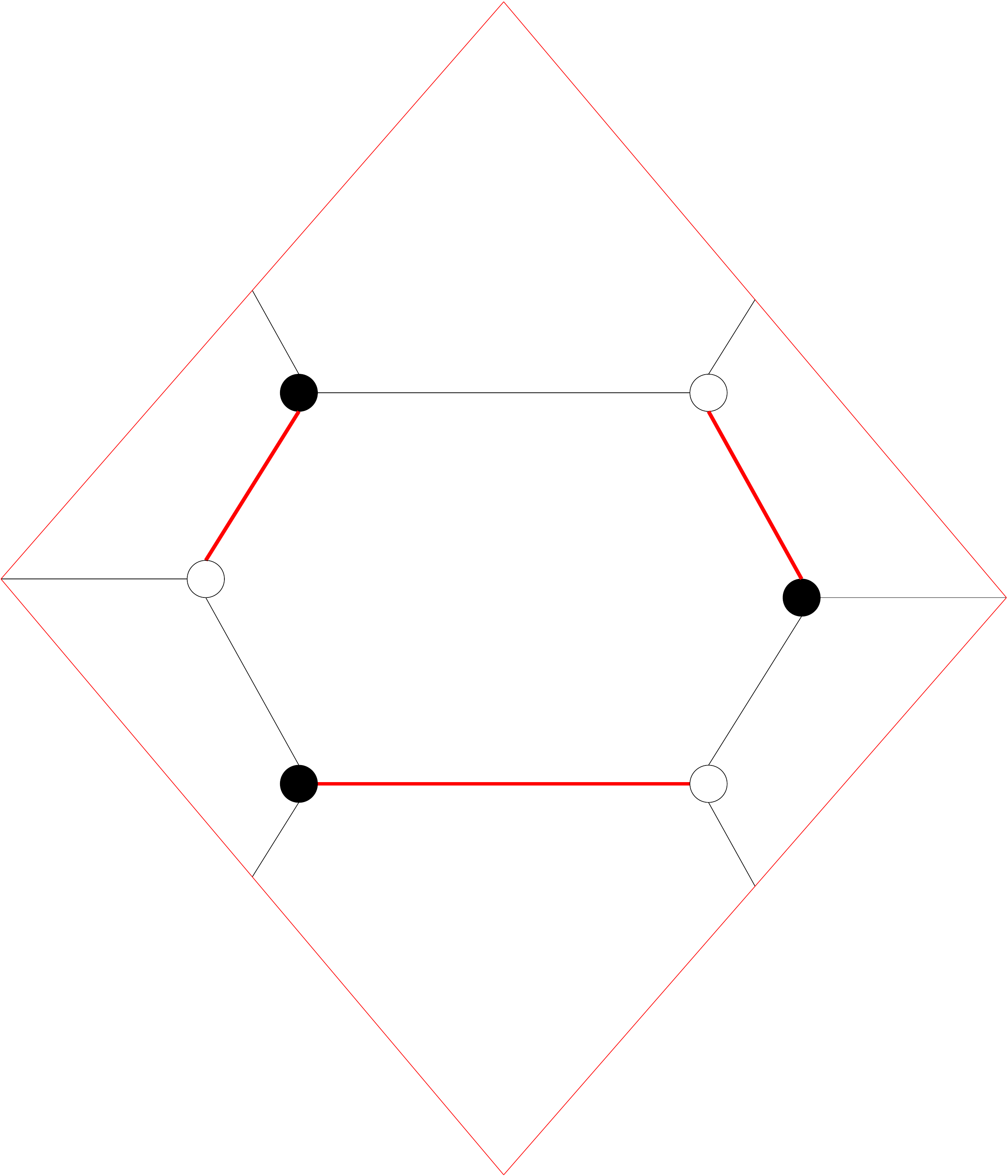}\\
\end{tabular}
\end{center}
\caption{Perfect matchings for dP$_0$: in the first line we represented the external perfect matchings, $\pi_1$, $\pi_2$ and $\pi_3$
while in the second line we represented the internal three perfect matchings, $\sigma_1$, $\sigma_2$ and $\sigma_3$ }
\label{PMdP0}
\end{figure}

In the case of dP$_0$ the periodic quiver and the bipartite diagram are shown in 
Figure \ref{dP0pertil}.
The superpotential can be easily read from these Figure and it is
\begin{equation}
W = \epsilon_{ijk} X_{12}^{(i)} X_{23}^{(j)} X_{31}^{(k)} 
\end{equation}
On this graph one can identify different sets of marked edges (dimers)
connecting the black and white nodes.
A perfect matching is a collection
of dimers chosen so that every vertex of the graph is covered by exactly one dimer.
The bipartite graph together with its perfect matchings defines the 
dimer model.
In this case we can identify the perfect matchings of dP$_0$ as in the Figure
\ref{PMdP0}.
The perfect matchings can be encoded in an $\varepsilon \times c$ matrix, 
where $\varepsilon$  represents the number of fields and 
$c$ is an index running on the perfect matchings. In this case
we have
\begin{equation} \label{pincopalla}
P=
\left(
\begin{array}{c|cccccc}
&\pi_1&\pi_2&\pi_3&\sigma_1&\sigma_2&\sigma_3\\
\hline
X_{12}^{(1)}& 1 & 0 & 0 & 1 & 0 & 0 \\
X_{12}^{(2)}&  0 & 1 & 0 & 1 & 0 & 0 \\
X_{12}^{(3)}&  0 & 0 & 1 & 1 & 0 & 0 \\
X_{23}^{(1)}&  1 & 0 & 0 & 0 & 1 & 0 \\
X_{23}^{(2)}&  0 & 1 & 0 & 0 & 1 & 0 \\
X_{23}^{(3)}&  0 & 0 & 1 & 0 & 1 & 0 \\
X_{31}^{(1)}&  1 & 0 & 0 & 0 & 0 & 1 \\
X_{31}^{(2)}&  0 & 1 & 0 & 0 & 0 & 1 \\
X_{31}^{(3)}&  0 & 0 & 1 & 0 & 0 & 1
\end{array}
\right)
\end{equation}
Finally,
on the dimer model one can define a partition function as the determinant of a matrix, called the Kasteleyn matrix \cite{Kasteleyn}.
This is a weighted signed adjacency matrix of the graph in which the rows and the columns represent the 
black and white nodes respectively.
The $ij$-th elements of the matrix are the fields connecting the pairs of black and 
white nodes associated to $i$ and $j$. Each element is then wighted by the intersection number of these fields 
with the homology classes $(1,0)$ and $(0,1)$ of $\gamma_w$ and $\gamma_z$ winding cycles of the torus.
In this case we have
\begin{equation}
\text{Kas} = 
\left(
\begin{array}{ccc}
X_{12}^{(1)}&X_{31}^{(3)}&X_{23}^{(2)} w z\\
X_{23}^{(3)} w^{-1}&X_{12}^{(2)}&X_{31}^{(3)}\\
X_{31}^{(2)}&X_{23}^{(1)}z^{-1}&X_{12}^{(3)}
\end{array}
\right)
\end{equation}
The permanent of this matrix counts the perfect matchings of the brane tiling and their homology.
In the case of dP$_0$  the permanent is
\begin{eqnarray} \label{perm}
\text{Perm}(\text{Kas})&=&
X_{12}^{(1)} X_{12}^{(2)} X_{12}^{(3)} 
+
X_{23}^{(1)} X_{23}^{(2)} X_{23}^{(3)} 
+
X_{31}^{(1)} X_{31}^{(2)} X_{31}^{(3)} 
\nonumber \\
&+&
\frac{1}{z}
X_{12}^{(1)}X_{23}^{(1)}X_{31}^{(1)}
+w z 
X_{12}^{(2)}X_{23}^{(2)}X_{31}^{(2)}
+\frac{1}{w}
X_{12}^{(3)}X_{23}^{(3)}X_{31}^{(3)}
\end{eqnarray}
This is a polynomial in the  $w$ and $z$ variables and 
one can associate a polyhedral on $\mathbb{Z}^2$ to this 
polynomial,  the toric diagram \cite{Fulton}.
This  rational polyhedral encodes the data of the conical 
toric Calabi-Yau and the information of the mesonic moduli space on the field theory
side. 

In this case the polynomial (\ref{perm}), once represented on $\mathbb{Z}^2$,
has three external points with coordinates
 $(1,1)$, $(-1,0)$ and $(0,-1)$.
There is also an internal point $(0,0)$ associated to the three perfect matchings
without 
$w$ and $z$ dependence in (\ref{perm}).
From now on we will refer to the perfect matching associated to the external points as 
external perfect matchings (denoted as $\pi$)
while the ones associated to the internal points are internal perfect 
matchings\footnote{Here and in the rest of the paper 
we are not considering models with points on the perimeter on the toric diagram. 
As observed in \cite{Eager:2011dp} they can be obtained from partial resolution 
and give origin to other integrable system. We comment on that in section \ref{sec7}}
(denoted as $\sigma$).
For example in Figure \ref{PMdP0} we 
distinguished the internal perfect matchings
 in the first line and the external ones in the second.

\subsection{The master space}
\label{sec2.1}

The moduli space of a supersymmetric field theory is the set of all the possible constant vacuum expectation values of the scalar gauge invariant operators of the theory that satisfy the zero energy condition. 
This variety contains a lot of information regarding the field theory and it is the solution of the zeros of the derivatives of the superpotential with respect to the elementary scalar fields (F-terms condition), modulo the 
action of the complexified gauge group.  
In \cite{Forcella:2008bb,Butti:2007jv} it was discovered that, in the particular case of N D3 branes at toric CY$_3$ singularities, the information of a peculiar branch of the moduli space for one brane, is enough to reconstruct the full moduli space of the theory for generic N. 
This branch is called the coherent component of the master space $\firr~$, it is a $g+2$ dimensional CY and it can be obtained as the symplectic quotient implementation of the linear relations 
among the $c$ perfect matchings of the dimer model associated to $\mathcal{X}$. Indeed the $c$ perfect matchings $p_i$ of a specific brane tiling are not free but they satisfy a set of $c-g-2$
 linear relations, and they can be thought as $c$ vectors $V_{p_i}$ in $\mathbb{Z}^{g+2}$ subjected to these relations
 \begin{equation}\label{linrel}
\sum_{i=1}^c Q_i^s V_{p_i}=0
\end{equation}
with $s=1,...,c-g-2$.  We can now assign a complex coordinate $x_{p_i}$ of $\mathbb{C}^c$ to every vector $V_{p_i}$ and obtain the coherent component of the master space $\firr~$ 
as the symplectic quotient
$$\firr~=\mathbb{C}^c // Q^s$$

In the example of dP$_0$ the relation among the perfect matching is simply
$\pi_1+\pi_2+\pi_3=\sigma_1+\sigma_2+\sigma_3$ as can be seen from the 
Figure \ref{PMdP0}.
The master space is then 
\beq
\mathbb{C}^6 // \{ -1,-1,-1,1,1,1\} \qquad 
\eeq
An explicit representation of this toric variety is a toric diagram in $g+2$ dimensions,
modulo a $SL(g+2,\mathbb{Z})$ transformation. Here  we have $g=3$ and the toric diagram is 
described by the matrix 
\begin{equation} \label{TDP0}
T=\left(
\begin{array}{cccccc}
\pi_1&\pi_2&\pi_3&\sigma_1&\sigma_2&\sigma_3\\
\hline
 0 & 0 & 1 & 1 & 0 & 0 \\
 0 & 0 & 1 & 0 & 1 & 0 \\
 1 & 0 & 0 & 0 & 0 & 1 \\
 0 & 1 & 0 & 0 & 0 & 1 \\
 0 & 0 & 1 & 0 & 0 & 1
\end{array}
\right)
\end{equation}
where we indicated the perfect matching associated to 
every column.
An useful way to think about $T$ is that every raw of the matrix determines the $U(1)$  charge of every perfect matching.
The $i$-th index of $T_{i,\rho}$ parameterizes one of the 
 the global $g+2$ $U(1)$ charge while the 
$\rho$-th index runs over the perfect matchings.

\section{Dimer models and integrable systems}\label{sec3}

In this section we review the results derived in \cite{Goncharov:2011hp},
relating the dimer models to integrable systems.

\subsection{Poisson manifold and Poisson structure}
An integrable system is usually described on a Poisson manifold, i.e. 
a manifold with an antisymmetric Poisson structure $\{,\}$.
A Poisson manifold is symplectic 
if the rank of the antisymmetric structure equals everywhere
the dimension of the Poisson manifold,
and the antisymmetric structure can be written locally in terms of canonical 
variables $\mathbf{q}$ and  $\mathbf{p}$.

However,
more generically, the Poisson structure can have lower rank over the Poisson manifold.
In this case there exist some operators which commute with everything:
the Casimir operators.
By fixing the value of these operators one usually restrict
to an even dimensional symplectic leaf.  This even dimensional subspace inherits the anticommuting structure of the 
Poisson manifold, and here the rank of the operator $\{,\}$ is maximal.
In these cases, given a $2I+d-1$ dimensional Poisson manifold,
 one can find local coordinates $(c_1,\dots,c_{d-1},\mathbf{q}_1,\dots,\mathbf{q}_I, \mathbf{p}_1,\dots,\mathbf{p}_I)$
such that $(\mathbf{q}_1,\dots,\mathbf{q}_I, \mathbf{p}_1,\dots,\mathbf{p}_I)$ satisfy canonical Poisson brackets and $(c_1,\dots,c_{d-1})$ commute with everything.

An integrable system on this manifold is defined by $I$ 
independent functions $H_a$ which are in involution.
Then on every symplectic leaf 
one can operate a change of variables. 
The new variables are usually named action and angle variables. 
The action variables commute among each other and generate the dynamics of the angle variables.

In \cite{Goncharov:2011hp} 
it was shown that a dimer model can be associated to a Poisson manifold.
Closed oriented loops $\alpha_i$ on the dimer corresponds to functions on the Poisson manifold,
and their Poisson brackets is given by 
\beq
\label{gonceinte}
\{ \alpha_i ,\alpha_j \}= \epsilon(\alpha_i,\alpha_j) \alpha_i \alpha_j
\eeq
where $\epsilon(\alpha_i,\alpha_j)$
is an antisymmetric intersection index specified by the construction 
that we review in the appendix \ref{appA}.
Observe that closed oriented loops on the dimer can be always obtained as difference of two perfect
matchings. 

A complete set of coordinates on a patch of the Poisson manifold is given by 
$w_A$ and $z_{1,2}$. The $w_A$ are the loops around the faces of the
tiling, while $z_{1}$ and $z_{2}$ represent two paths around the torus with homology  $(1,0)$ and $(0,1)$ 
respectively.
We stress that in this basis 
the anticommuting structure among the trivially homological loops $w_A$ is
\beq
\label{dabliu}
\{w_A,w_B\}=a_{AB} w_A w_B
\eeq
and it is simply specified by the adjacency matrix $a_{AB}$. Note that 
the product of the whole set of closed loops cover the torus and it becomes trivial, $w_1\cdot w_2 \dots
w_{g-1}\cdot w_g=1$.
The Poisson manifold has then dimension $g+1$, where $g$ is the number of gauge groups
in the corresponding gauge theory.

\subsection{Hamiltonian and Casimir operators}

The Casimir operators and the Hamiltonians which define the integrable system
consist of cycles on the bipartite graph.
These are specified  as differences of perfect matchings, depending on their homologies.
A straightforward way to describe these quantities appears after the toric diagram is considered.
Indeed different points of the diagram are related to different homologies of the perfect matchings in the tiling
(see the previous section for the procedure of finding the homology of each perfect matching).
We label as $\pi_k$  the $d$ perfect matchings associated to external points in the toric diagram, and with 
$\sigma_{k_a}^{a}$ the $r$ perfect matchings associated to internal points in the toric diagram.
The superscript $a$ defines the homology of each perfect matching (it is in 1-1 correspondence with the internal points of the toric diagram, counted without multiplicity) and it runs from $1$ to $I$.
Every internal point has a degeneracy $M_a$, which depends on the detail of the 
bipartite diagram.
The subscript $k_a$ runs from $1$ to $M_a$ such that the $\sum_{a=1}^{I} M_a=r$ and 
finally $d+r=c$.

We choose a reference perfect matching
among the external ones, that we label as $\pi_{ref}$.
The Casimir operators are the cycles obtained as differences of the other external
perfect matchings with respect to the reference one. There are thus $d-1$ independent Casimir operators \footnote{Note that we can redefine them in equivalent ways, for instance as differences of two consecutive external perfect matching on the toric diagram. The important property of the Casimir
operators  is that they are cycles defined as differences among external perfect matchings, 
and there are $d-1$ independent of them.}.

The Hamiltonians are composed by cycles built as differences of internal perfect matchings 
with the same
homology minus the reference perfect matching. Precisely,
there are $I$ Hamiltonians, one for each internal point with a defined homology $a$. 
Each Hamiltonian $H^{a}$ is made of
a sum of cycles, i.e. a sum of functions on the Poisson manifold. These cycles are built
as the difference of the perfect matchings
 associated to the internal point with homology $a$,
minus the reference perfect matching.
For instance,  the Hamiltonian corresponding to the $\sigma_{k_a}^a$ perfect matchings is
\beq
\label{hami1}
H^{a}=\sum_{k_a=1}^{M_a} \left(\sigma_{k_a}^a-\pi_{\text{ref}}   \right)
\eeq

It was proven in \cite{Goncharov:2011hp} 
 that, with respect to the Poisson structure defined by (\ref{gonceinte}),
the Casimir operators have vanishing Poisson brackets with everything, i.e. with every closed
loop on the dimer.
Moreover, the Hamiltonians built as in (\ref{hami1}) commute one each other, providing an
algebraic integrable system. 

In conclusion,
given a tiling with $g$ faces, the toric diagram has $I$ internal points and $d$ external points. The relation
among $g$, $d$ and $I$ is $g+1=d -1+2 I$.
This relation tell us that we are describing a  $g+1$-dimensional Poisson manifold (there are $g+2$ variables, but one constraint among 
the faces) with $d-1$ Casimir operators ($d-1$ are the independent relation among the differences of $d$ external perfect matchings)
and $2I$ phase space variables. 
There are $I$ commuting Hamiltonians and the system is classically (and quantum
\cite{,2008InMat.175..223F,Goncharov:2011hp}) integrable.

\section{Integrability on the master space}
\label{sec4}
In this section we illustrate the main result of the paper, the connection
between the integrable dimer models and the
moduli space of the 
associated $\mathcal{N}=1$ SCFT gauge theory, or more precisely the
coherent component of the master space $\firr~$.

By following \cite{Goncharov:2011hp}, we would like to point out that the same set of vectors $V_{p_i}$ defining the master space of the $\mathcal{N}=1$ SCFT describes also a patch of an integrable Poisson manifold. The full Poisson manifold is obtained by gluing together the patches associated to the master spaces of the different Seiberg dual phases of the $\mathcal{N}=1$ SCFT associated to the same CY cone $\mathcal{X}$.
We offer in such a way a "partial toric" description of the integrable cluster Poisson manifold.

Let us first analyze a patch. Because $\firr~$ is a CY$_{g+2}$ the $c$ vectors $V_{p_i}$ lie on a $g+1$ dimensional hyperplane \cite{Forcella:2008bb} and the set of differences of  perfect matchings 
is a convex $g+1$ dimensional polytope $\Omega$ in $\mathbb{Z}^{g+2}$. $\firr~$ has a natural projection to $\mathcal{X}$ obtained by disregarding the baryonic 
directions inside $\mathbb{Z}^{g+2}$: $\firr~ \rightarrow \mathcal{X}$.  This projection allows to divide the $p_i$ in 
the external perfect matchings $\pi_k$, $j=1,...,d$, the ones that are mapped to the boundary of the 2d toric diagram of $\mathcal{X}$, and the internal perfect matchings $\sigma^a_{k_a}$, $k_a=1,...,M_a$, the ones that are mapped inside the 2d toric diagram, where $a=1,...,I$ labels the internal points of the toric diagram and $\sum M_a=c-d$. 
Moreover, thanks to the algorithm \cite{Hanany:2005ss},
 it is possible to obtain the adjacency matrix $a_{AB}$, $A,B=1,...,g$, from the $\pi_k$ for the various Seiberg dual phases.

We can now consider the $g+1$ dimensional cone\footnote{This space is $\Omega$ once we add the origin.} defined by the $c-1$ vectors $V_{\delta p}$ obtained by subtracting to all the perfect matching vectors an arbitrary chosen external reference perfect matching vector $V_{\pi^{ref}}$. The $\delta p_\rho=p_\rho - \pi^{ext}$ are a set of $c-1$ cycles on the tiling. 
To every vector $V_{\delta p}$ we can now associate a local function: 
\begin{equation}\label{locfunc}
f[V_{\delta p_\rho}]= e^{\sum_{m=1}^{g+1} x_m V^m_{\delta p_\rho}}
\end{equation}
where $V^m_{\delta p_\rho}$ is the $m$-th coordinate associated to the $V_{\delta p_\rho}$ vector inside $\mathbb{Z}^{g+1}$, and $x_m$ is the associated chemical potential, one for every direction inside $\mathbb{Z}^{g+1}$. 
We would like to interpret the chemical potentials $x_m$ as a set of local coordinates inside a patch of an integrable Poisson manifold.

Given the matrix $a_{AB}$ there is a standard way to obtain the Poisson structure $J_{mn}$ on the local $x_m$ coordinates. Indeed the $a_{AB}$ is the natural Poisson structure \cite{Goncharov:2011hp} on a particular subset of vectors in $\Omega$ called $V_{w_A}$.
Actually there are $g$ different $w_A$ but they are constrained by $\prod w_A=1$,
so there are only $g-1$ independent vectors  $V_{w_A}$. 
The $w_A$ are a basis of zero homotopy cycles on the tiling: they are the closed loops around the $A$-th face of the tiling and they are obtained subtracting two internal perfect matchings belonging to the same internal point of the 2d toric diagram: $w_A=\sigma^a_{k_a} - \sigma^a_{k'_a}$. As vectors they are a basis in the space of anomalous and non anomalous baryonic charges and they are uncharged under the two mesonic directions. To complete the local basis of charges one has to add the two cycles $z_1$ and $z_2$ on the tiling that have respectively homotopy (1,0) and (0,1) along the two torus directions, and they are associated to the vectors $V_{z_1}$, $V_{z_2}$ inside $\Omega$ charged under the mesonic symmetries. The Poisson brackets among the $V_{w_A}$ is defined as:
\begin{equation}
\{f[V_{w_A}], f[V_{w_B}] \}= a_{AB} f[V_{w_A}] f[V_{w_B}]= \sum_{mn} \partial_{x_m}  f[V_{w_A}] \partial_{x_n} f[V_{w_B}] 
J_{mn}
\end{equation}
The matrix $a_{AB}$ has rank $2I$, equal to the number of anomalous charges in the SCFT. Indeed it distinguishes the anomalous from the non anomalous baryonic symmetries and it can be put in the natural block form: $J_{g+1}= 0_{d-1}  \bigoplus J_{2I}$. The $d-1$ null part of the matrix is related to the non anomalous charges, while the $J_{2I}$ acts also on the anomalous charges and can be brought in the canonical form $J_{2I}=( (0,\mathbb{I}_I),(- \mathbb{I}_I,0))$. The chemical potentials for the non anomalous charges are coordinates associated to the  Casimir
operators, $c_e$, with $e=1,\dots,d-1$ while the chemical potential for the anomalous charges are the generalized coordinates 
$\overrightarrow{\mathbf{q}}$, and conjugate momenta $\overrightarrow{\mathbf{p}}$, and together they are local coordinates of a Poisson manifold. 

Because the non anomalous charges are defined in toric geometry by the zero locus of the coordinates $x_{\pi_k}$ associated to the external perfect matchings, the $d-1$ functions that define the symplectic leaves of the manifolds can be identified with $f[V_{\delta \pi_e}]$.  We have right now two interesting sets of $g+1$ vectors: $\mathcal{B}^1=\{V_{w_A}, V_{z_1}, V_{z_2}\}$ and $\mathcal{B}^2=\{V_{\delta \pi_e}, V_{\delta \sigma_{\tilde m}} \}$, where the $V_{\delta \sigma_{\tilde m}}$, with 
$\tilde m=1,...,2I$, are a set of vectors perpendicular to $V_{\delta \pi_e}$ that form a basis of the anomalous charges inside $\mathbb{Z}^{g+1}$. 
Because every Poisson manifold admits a local set of coordinates given by the Casimir operators, the positions and conjugate momenta \cite{WEINSTEIN}, 
there should exist a change of coordinates from $\mathcal{B}^1$ to $\mathcal{B}^2$. The existence of this transformation is assured by the CY condition of the master space. Indeed, the $\mathcal{B}^1$ and $\mathcal{B}^2$ are subtraction of different perfect matchings in $\mathbb{Z}^{g+2}$ and they both live in a $g+1$ dimensional hyperplane. The CY condition assure that all the perfect matchings live on the same $g+1$ dimensional hyperplane implying that the two  hyperplanes of $\mathcal{B}^1$ and $\mathcal{B}^2$ are the same, and the existence of a linear transformation $\mathbb{M}$ between the two: $\mathcal{B}^2_i= \mathbb{M}_{ij} \mathcal{B}^1_j$.

The relation between  $\mathcal{B}^1$ and $\mathcal{B}^2$ corresponds to a linear map among the vectors $V_{\delta p}$ inside $\Omega$. The vectors $V_{\delta p}$ are a set that generates 
all the possible subtractions of perfect matchings and they look the natural variables to describe the completely integrable Hamiltonian dynamics of the closed cycles of the tilings. 
We characterize  the local patch of this completely integrable system as
follows\footnote{Observe the similarity of this procedure and the usual procedure used to obtain an affine algebraic variety starting from its dual toric diagram.}: to each vector $V_{\delta p_i}$ in $\Omega$ we associate a coordinate $y_{\delta p_i}$, 
and to the linear relations (\ref{linrel}) we associate the algebraic constraints: 
\begin{equation}\label{cony}
\prod_{i=1}^{c-1} y_{\delta p_i}^{Q_i^s}=1
\end{equation}
which can be locally solved by the functions $f[V_{\delta p_i}]$  defined above. The open dense subspace over which this algebraic intersection can be solved in term of the local coordinates $x_m$ is one patch of the Poisson manifold and (\ref{cony}) are rational transformations  among local coordinates of this patch. The dimension of the Poisson manifold is $g+1$.
$y_{\delta \pi_e}$ are the $d-1$ Casimir operators, they are the level functions of the symplectic foliation and they are associated to the non anomalous global symmetries. 
The symplectic leaf is $2I$ dimensional, where $2I$ 
corresponds to the number of anomalous symmetries of the field theory. There are $I$ Hamiltonians and $I$  associated flow vectors, given by linear functions of the $y_{\delta p_i}$ coordinates:
\begin{equation}
H^a=\sum_{k_a=1}^{M_a} y_{\delta \sigma^a_{k_a}} \hbox{  } \hbox{  ,  } \hbox{  } \{ H^a,\hbox{   }\} =\sum_{k_a=1}^{M_a}  \sum_{m,n} J_{mn} V^m_{\delta \sigma^a_{k_a}} y_{\delta \sigma^a_{k_a} } \partial_{x_n}
\end{equation}
for $a=1,...,I$ and $k_a$ running over the number of internal perfect matchings associated to the same internal point of the toric diagram of $\mathcal{X}$.  
In the paper, we will show, example by example, that the $I$ Hamiltonians $H^a$ are in involution: $\{H^a,H^b\}=0$, with respect to the $J_{2I}$ symplectic structure,
 and, as was proven in \cite{Goncharov:2011hp}, the system is integrable.

In \cite{Goncharov:2011hp} it is shown that the full Poisson manifold can be obtained by gluing together, with (cluster) Poisson transformations, the various patches associated to the closed loops in different Seiberg dual dimer model realizations of the SCFT living at $\mathcal{X}$. Indeed it is well known that to every $\mathcal{X}$ are associated different SCFT that are related by Seiberg duality (SD) transformations 
\cite{Feng:2000mi,Beasley:2001zp,Feng:2001bn}.
The $\firr~$ for every SD phase is in general a different algebraic variety \cite{Forcella:2008ng}. Following \cite{Goncharov:2011hp} we would like to argue that the various patches of the Poisson manifold, obtained from the master space with the procedure we have just explained, can be glued together, along open dense subsets, with transition functions that are non toric symplectic morphisms on the symplectic leaves and toric transformations on the Casimir operators.
Here we give some rules for the gluing procedure and in section \ref{sec6} we check them using the $\mathbb{F}_0$ example and the procedure discussed in
\cite{Goncharov:2011hp}.

The $\firr~$ of different SD phases are not isomorphic toric varieties \cite{Forcella:2008ng}, however the non anomalous charges of the SD phases can be mapped among themselves. Indeed there exist a linear transformation that maps the $d-1$ external perfect matchings vectors $V_{\delta \pi_e}$ of one phase to the $d-1$ external perfect matchings vectors $V_{\delta \pi^{S.d.}_e}$ of the dual Seiberg phase, and it can be translated into a rational map between local coordinates of two patches in the usual toric way\footnote{This map is defined on open dense subsets of the Poisson patches, and it is very similar to the usual gluing procedure of  the different patches of a toric variety.}:
\begin{equation}\label{casimirmap}
\sum_{e=1}^{d-1} D_e^k V_{\delta \pi_e} - \sum_{e=1}^{d-1}S^k_e V_{\delta \pi^{S.d.}_e}  = 0 \hbox{   } \rightarrow \hbox{  }  \prod_{e=1}^{d-1}y^{D_e^k}_{\delta \pi_e}= \prod_{e=1}^{d-1} (y_{\delta \pi_e^{S.d.}})^{S_e^k}
\end{equation}
We will see indeed that the map among the Casimir operators of the system is a linear map on the chemical potentials for the non anomalous charges. 
The map between the symplectic leaves is obtained by mapping among themselves the SD Hamiltonians associated to the same internal point in the 2d toric diagram and the associated Hamiltonian flows:
\begin{equation}\label{qpmap}
H^a=H^a_{S.d}, \hbox{      }  \hbox{      } \{H^a,...\}= \{H^a_{S.d},\hbox{   }\}_{S.d.}
\end{equation}
This second transformation on the leaves of the Poisson manifold  breaks the toricity of the map (it is not of the form monomial equal to monomial) and it translates, as we will see, in a non linear map among the canonical coordinates. More explicitly it can be written as:
\begin{equation}\label{qpmap2}
\sum_{k_a} y_{\delta \sigma^a_{k_a}} = \sum_k y^{S.d.}_{\delta \sigma^a_{k_a}} \hbox{      } \hbox{  ,  } \hbox{      } \sum_{m,k_a} J_{mn} V^m_{y_{\delta \sigma^a_{k_a}}} y_{\delta \sigma^a_{k_a}}= \sum_{m,s,k_a} J^{S.d}_{ms}\frac{\partial x_n}{\partial x^{S.d.}_s} V^m_{y^{S.d.}_{\delta \sigma^a_{k_a}}} y^{S.d.}_{\delta \sigma^a_{k_a}}
\end{equation}
These transformations are a set of algebraic and differential equations, they are defined on open dense subsets of the Poisson patches and are by definition Poisson isomorphisms.

\subsection{The computing algorithms}\label{sec4.1}

After we have stated the correspondence we have to provide a computing algorithm to obtain the explicit expression for the
Poisson structure and the Hamiltonians of the integrable systems in terms of the vectors defining the master space.

The only quantities necessary to study the Poisson structure and 
thus the integrability of the dimer models 
are the adjacency matrix, the matrix T defined in \ref{sec2.1} 
and the parametrization of the loops in terms of differences of perfect matchings.
Another necessary information  is the identification of the
perfect matchings corresponding to external and to internal points 
in  the matrix $T$.
Moreover, to build the Hamiltonians,
we should identify the set of perfect matchings which correspond to each internal point.
This is always possible by 
performing an $SL(g+2,\mathbb{Z})$ transformation on
this matrix  such  to isolate the $2d$ toric diagram 
on the first two rows.  
Then we can choose an order for the columns of $T$, i.e. the perfect matchings, such that
the first $d$ corresponds to the external points of the toric diagram and the others
are all internal points.

Note that, in this ordering of the perfect matching,
by applying another $SL(g+2,\mathbb{Z})$ transformation on  $T$ we can then 
have the first $d$ columns charged only under $d$ of the $g+2$ symmetries. 
This fact has a simple interpretation on the field theory side: the external points 
of the toric diagram are uncharged under the 
 $2I= g+2-d$ anomalous symmetries. 
 This is a signal that the Casimir operators, defined as differences of external perfect matching, 
 are associated with a 
 trivial (vanishing)  antisymmetric structure, 
 since the only non zero part of the antisymmetric structure comes from the anomalous $U(1)$'s.
 Anyway, in what follows we do not specify any particular basis for the $U(1)$ charges, i.e. for the
 rows of the matrix $ T$, and we proceed in complete generality.

We now define the loop matrix $\delta T$ as follows: 
first we fix one of the external perfect matchings 
as a reference perfect matching and then we subtract 
the associated columns of $ T$ to  all of the other columns of $ T$:
\begin{equation}
\delta T_{i,\rho} =  T_{i,\rho}- T_{i,\text{rif}}
\end{equation}
Without loose of generality from now on we fix "rif$=1$".
 To fix the notation we refer to the columns of the $ \delta T_{i,\rho}$ matrix as
$\delta T=(0_1,\mathcal{E}_{e},\mathcal{I}_{l})$, where $l=1,\dots,r$ runs over the number of internal perfect matchings
(counted with multiplicities),
$e=1,\dots,d-1$ runs over the external perfect matchings minus one,
and $\rho=1,\dots,c$ over all the perfect matchings.
The vectors of the matrix $ \delta T$ are associated to loops 
in the brane tiling.

The CY condition of the master space implies that the $g+2$ dimensional vectors 
$\mathcal{E}_{e}$ and $\mathcal{I}_{l}$ all lie on the same hyperplane of dimension $g+1$.
We define the perpendicular vector to this hyperplane
\beq
v=Ker[(\delta  T)^T]
\eeq
and we build a basis in the $g+2$ dimensional vector space
 \begin{equation}
 \label{base}
 B_{i,J}=(\mathcal{E},\text{Ker}[v,\mathcal{E}],v) \qquad J=1,\dots, g+2
 \end{equation}
 where Ker$[v,\mathcal{E}]$ is the orthogonal space to the  
 the subspace spanned by the vectors $v$ and $\mathcal{E}_{e}$  in the $g+2$ dimensional lattice.
This basis represents a mapping between the space of the $g+2$ $U(1)$ charges
and the $g+2$ dimensional vector space . We can invert this matrix in order
to obtain the inverse map $(B^{-1})_{J,i}$.

Every loop on the dimer is a difference of two perfect matchings which are determined
by their $U(1)$ charges. We can express the loop charges
in the basis of $g+2$ vectors via the map $B^{-1}$. Note that since the loop is a difference
of two perfect matchings this combination will never involve the vector $v$.
For instance, 
consider a cycle $\alpha$ corresponding to the difference of two perfect matchings 
$p_1-p_2$, which are themselves columns of the matrix $T$. This difference is a vector
in the $g+2$ dimensional vector space and it is orthogonal to the vector $v$. Hence it can be 
expressed uniquely on the basis (\ref{base}) as
\beq
(p_1-p_2)_i=B_{i,J} \alpha_J \qquad \alpha_J=B^{-1}_{J,i} (p_1-p_2)_i
\eeq
with coefficients $\alpha_J$. Note that because of the argument we have just explained,
the coefficient $\alpha_{g+2}$ will always be $0$.
The associated function on the $g+1$ dimensional Poisson manifold is then
\beq
\label{funzioneassociata}
f[\alpha]=e^{\sum_{m=1}^{g+1} x_m B^{-1}_{m,i} (p_1-p_2)_i}
\eeq
where now $m$ runs only up to $g+1$.
This gives the explicit realization of the map (\ref{locfunc}) on this special basis of charges.

The other information that we need to extract the Poisson structure
is the composition of the loops $w_A$ (with $A=1,\dots,g$)
 as differences of perfect matchings, i.e. of
 columns of  $ T$. 
 Every loop $w_{i,A}$ is then associated to a vector in the $g+2$ dimensional vector space, 
 which can be written in the basis of $g+1$ vectors (\ref{base}), as just explained. We can encode the parametrization of the cycles $w_A$ on the $g+2$ charges in a matrix
$A_{i,A}$. 
Then we can write the corresponding functions on the Poisson manifold as
 \begin{equation}
 \label{funzwA}
f[w_{A}]= e^{x_m B^{-1}_{m,i} A_{i,A}}
 \end{equation}
The $g\times g+2$ matrix A has rank $g-1$ since the $w_A$ satisfy
 the constraint $\prod_A w_A=1$, which translate in the functions as
 $\prod_A f[w_A]=1$. It encodes all the necessary data about the $w_A$ loops.
 
 The $x_m$ represents the local coordinates on the Poisson manifold.
 Given the ordering of the basis $B$, the local coordinates
 are already organized in Casimir and phase space variables as
 $x_m=(c_1,\dots,c_{d-1},q_1,\dots,q_{2I})$ .
These coordinates are in
correspondence with the $U(1)$ charges. 
The first $d-1$ elements are associated to global non anomalous $U(1)$ symmetries 
and the last $2I$ to the anomalous ones.
The Poisson structure for these local coordinates is given by
\begin{equation}
\{x_m,x_n\}=J_{mn}=
\left(
\begin{array}{cc}
0 & 0\\
0 & \tilde J_{\widetilde m \widetilde n}
\end{array}
\right)
\end{equation}
where the matrix $\tilde J$ has dimension $2I \times 2I$ and 
$\{q_{\widetilde m},q_{\widetilde n}\}=\tilde J_{\widetilde m \widetilde n}$ is the 
antisymmetric structure of the phase space variables, not necessarily canonical in this basis.
The antisymmetric structure $J_{\widetilde m \widetilde n}$ 
can be obtained 
by imposing the Poisson brackets (\ref{dabliu}) among the $w_A$ loops.
  Indeed 
  we can solve the following equations
 \begin{equation}
 \label{soluzione}
 a_{AB} = \frac{\{f[w_A],f[w_B]\}} {f[w_A] f[w_B]} = \{
 \sum_m x_m B^{-1}_{m,i} A_{i,A}
 ,
  \sum_n x_n B^{-1}_{n,i} A_{i,B}
 \}
=
 \sum_{m,n}  (B^{-1}_{m,i} A_{i,A})(  B^{-1}_{n,i} A_{i,B} ) J_{mn}  
 \end{equation}
 where $a_{AB}$ is the adjacency matrix,
 finding the antisymmetric structure $\tilde J_{\widetilde m \widetilde n}$.
 Note that the matrix $B^1_{m,i}$ has rank $g+1$, the matrix $A$ has rank $g-1$, and the matrix $a_{AB}$
 has rank $2I$. 
 For every toric diagram the inequality $d-3=g-1-2I\geq 0 $ is satisfied.
 Hence the matrix $\tilde J_{\widetilde m \widetilde n}$ has also rank $2I$.
  
  With this procedure we have obtained an explicit mapping between cycles, i.e. differences of two perfect
 matchings, and functions of local coordinates on the Poisson manifold. Moreover we have 
 extracted the Poisson structure for this coordinate system. 
 One can check that, taken two arbitrary closed cycles on the dimer, our mapping
 and Poisson structure reproduce the antisymmetric intersection algebra 
 based on (\ref{gonceinte}).

\subsection{Hamiltonians}\label{hamiltoniane}
Once the Poisson structure is known, we can define the Hamiltonians and
study the integrability of the system.
Consider an internal point with homology $a$,  related to a set
of perfect matchings $\sigma_{k_a}^a$, where $k_a=1,\dots,M_a$
and  $M_a$ is the multiplicity of the $a$-th internal point.
This set of perfect matchings correspond to a set of $M_a$ different cycles
with the same homology. They  correspond to 
columns in $\delta  T$, and we label them as $\delta T_{i,k_a}$.
Each of these elements is a vector in the basis of the $g+2$ charges 
which can be expressed in the basis (\ref{base})
via $B^{-1}$.
Each Hamiltonian function is computed by summing the cycles with the same homology
 $a$
\begin{equation}
H^a = \sum_{k_a=1}^{M_a}  e^{x_m B^{-1}_{m,i} \delta  T_{i,k_a} }   
\end{equation}
The commutation relation of two Hamiltonians is then given by 
\begin{eqnarray}
\{ H^a,H^b\} = 
&=&
\{  \sum_{k_a=1}^{M_a} e^{x_m B^{-1}_{m,i} \delta  T_{i,k_a} } 
,
 \sum_{k_b=1}^{M_b}  e^{x_m B^{-1}_{m,i} \delta  T_{i,k_b} } \}
 \\
&=&
\sum_{k_a,k_b} 
e^{x_m  B^{-1}_{m,i} \delta  T_{i,k_a} +x_n B^{-1}_{n,i} \delta \tilde T_{i,k_b}}
\sum_{m,n=1}^{g+1} (B^{-1}_{m,i} \delta  T_{i,k_a} )(B^{-1}_{n,i} \delta  T_{i,k_b}) J_{mn}   \nonumber
\end{eqnarray}

\section{Examples}
\label{sec5}
In this section we exemplify the general procedure explained in the previous sections in a 
series of examples of increasing complexity. We perform the case of the chiral gauge theory
associated to the $dP_0$ singularity  introduced before.
Then we describe the two phases of $\mathbb{F}_0$, which will be later used to 
discuss Seiberg duality. 
We then show two examples with multiple internal points $Y^{(3,0)}$ and $Y^{(4,0)}$ to
explicitly check that the Hamiltonians are in involution.

\subsection{Theories with one Hamiltonian}

\subsubsection{$dP_0$}
We here study the quiver gauge theory and the dimer associated to $N$
D$3$-branes probing the cone over the zeroth del Pezzo surface $dP_0$.
In the previous sections we used this theory to review the 
computation of the moduli space for $\mathcal{N}=1$ toric SCFT.
Here we apply the algorithm explained in section \ref{sec4.1} to obtain
the Poisson structure and the Hamiltonian.

The $T$ matrix  (\ref{TDP0})
encodes the charges of the $6$ perfect matchings under the $g+2=5$ $U(1)$ global symmetries of the theory.
Note that we already order it such that the first three columns correspond to the three
external perfect matchings.
Our construction is invariant under $SL(g+2,\mathbb{Z})$ transformation. Thus,
to render the final expressions simpler, we can
transform the matrix $T$ to a more easy form 
by acting with an $SL(5,\mathbb{Z})$ transformation, we obtain
\begin{equation}
T=
\left(
\begin{array}{cccccc}
\pi_1&\pi_2&\pi_3&\sigma_1&\sigma_2&\sigma_3\\
 1 & 0 & 0 & 0 & 0 & 1 \\
 0 & 1 & 0 & 0 & 0 & 1 \\
 0 & 0 & 1 & 0 & 0 & 1 \\
 0 & 0 & 0 & 1 & 0 & -1 \\
 0 & 0 & 0 & 0 & 1 & -1
\end{array}
\right)
\end{equation}
The external perfect matchings are uncharged under two of the 
$U(1)$ symmetries, which are identified with the anomalous $U(1)$'s of the theory.

From now one we fix the first external perfect matching as the reference perfect matching,
and the $\delta T$ matrix becomes
\begin{equation}
\label{deltaTdp0}
\delta T= \left(
\begin{array}{cccccc}
 0 & -1 & -1 & -1 & -1 & 0 \\
 0 & 1 & 0 & 0 & 0 & 1 \\
 0 & 0 & 1 & 0 & 0 & 1 \\
 0 & 0 & 0 & 1 & 0 & -1 \\
 0 & 0 & 0 & 0 & 1 & -1
\end{array}
\right)
\end{equation}
The last information we need to extract the Poisson structure is the structure of the $w_A$ loops
associated to the faces in the dimer.
We can identify the loops  
as the differences 
ratio of the $y$ variables defined in section \ref{sec4}. We have
\begin{equation}
w_1 = \frac{y_{\delta \sigma_3}}{y_{\delta \sigma_1}}, \quad
w_2 = \frac{y_{\delta \sigma_1}}{y_{\delta \sigma_2}}, \quad
w_3=\frac{y_{\delta \sigma_2}}{y_{\delta \sigma_3}}
\end{equation}
Thus, the matrix $B$ (\ref{base}) , its inverse and the matrix $A$ (\ref{funzwA})
are
\small
\begin{equation}
B=
\left(
\begin{array}{ccccc}
 -1 & -1 & -1 & -1 & 1 \\
 1 & 0 & -1 & -1 & 1 \\
 0 & 1 & -1 & -1 & 1 \\
 0 & 0 & 0 & 3 & 1 \\
 0 & 0 & 3 & 0 & 1
\end{array}
\right),
\quad
B^{-1}=-\frac{1}{15}
\left(
\begin{array}{ccccc}
 ~~5 & -10 &~ ~5 &~~ 0 & ~~0 \\
 ~~5 & ~~5 & -10 & ~~0 &~~ 0 \\
 ~~1 & ~~1 & ~~1 & ~~1 & -4 \\
 ~~1 &~ ~1 & ~~1 & -4 &~~ 1 \\
 -3 & -3 & -3 & -3 & -3
\end{array}
\right),
\quad
A=
\left(
\begin{array}{ccc}
 1 & 0 & -1 \\
 1 & 0 & -1 \\
 1 & 0 & -1 \\
 -2 & 1 & 1 \\
 -1 & -1 & 2
\end{array}
\right)
\end{equation}
\normalsize
The last column of $B$, i.e. the vector $v=(1,1,1,1,1)$, 
identifies the orthogonal direction in the master space, and it is 
perpendicular to all the vectors in (\ref{deltaTdp0}).

The CY condition of the master space
guarantees that the charges of 
every closed loop
in the dimer can be expressed as a 
linear combination of the first four columns of  $B$ only.
We then project out the orthogonal direction and associate 
to the four independent $U(1)$ charges
the four local coordinates of the Poisson manifold,  parametrized as
\begin{equation}
\{x_m\} = (c_1,c_2,q_1,q_2)
\end{equation}
The ordering we have chosen for the matrix $B$ (see eq. (\ref{base}))
implies that the first two coordinates $c_i$ are associated to Casimir operators, 
while $q_1$ and $q_2$ are the phase space variables.
By solving the relations (\ref{soluzione}),
we find the antisymmetric structure among these local coordinates as
\begin{equation}
\label{PSdp0}
\{c_e,c_f\} = 0,\quad  \{c_e,q_{\widetilde n}\}= \{q_{\widetilde m},c_f\}=0, \quad 
\{q_{\widetilde m},q_{\widetilde n}\}= -9 \, \epsilon_{\widetilde m \widetilde n}
\end{equation}
The $y$ variables are parametrized in terms of the $\{x_m\}=(c_e,q_{\widetilde m})$ variables as
\begin{equation}
y_{s} =  e^{x_{m} B^{-1}_{mi} \delta T_{is}}
\end{equation}
where $s$ runs over the external and internal perfect matchings.
Precisely $s=(1,\dots,6) \rightarrow (\pi_1,\pi_2,\pi_3,\sigma_1,\sigma_2,\sigma_3)$, and we have
\begin{equation}
B^{-1} \delta T=\left(
\begin{array}{c|cccccc}
&\pi_1&\pi_2&\pi_3&\sigma_1&\sigma_2&\sigma_3\\
\hline
c_1&0&1&0&  \frac{1}{3}& \frac{1}{3}  & \frac{1}{3}\\
c_2&0&0&1&  \frac{1}{3} &  \frac{1}{3} & \frac{1}{3}\\
q_1&0&0&0& 0 &   \frac{1}{3}&- \frac{1}{3}\\
q_2&0&0&0&  \frac{1}{3} & 0&- \frac{1}{3}\\
\end{array}
\right)
\end{equation}
With this matrix we  parameterize the $y_i$
as
\begin{eqnarray}
&&
y_{\delta \pi_1}=1,\quad \quad \quad \quad
y_{\delta \pi_2}=e^{c_1},\quad \quad \quad \,\,\,
y_{\delta \pi_3}=e^{c_2}\quad \quad \quad \nonumber \\
&&
y_{\delta \sigma_1}=e^{\frac{c_1+c_2+q_2}{3}},\quad
y_{\delta \sigma_2}= e^{\frac{c_1+c_2+q_1}{3}},\quad 
y_{\delta \sigma_3}= e^{\frac{c_1+c_2-q_1-q_2}{3}}\quad 
\end{eqnarray}
The only Hamiltonian associated with the single internal point is obtained with the procedure explained in section
\ref{hamiltoniane}
and it is
\begin{equation}
H=
y_{\delta \sigma_1}+y_{\delta \sigma_2}+y_{\delta \sigma_3}
\end{equation}
\\
Finally,
by introducing the paths $z_1$ and $z_2$ as 
\begin{equation}
z_1 =\frac{y_{\delta \sigma_{3}}}{y_{\delta \pi_{3}}}
,\quad
z_2 =\frac{y_{\delta \pi_{2}}}{y_{\delta \sigma_{3}}}
\end{equation}
we reconstruct the algebra of the $w_A$ and $z_1$ and $z_2$ defined in 
\cite{Goncharov:2011hp} from the algebra of the $c_i$ and $q_{\tilde m}$.
Indeed the cycles $w_A,z_1, z_2$ are associated to exponential functions of the local coordinates
via the mapping (\ref{funzioneassociata}) and, by knowing the antisymmetric structure (\ref{PSdp0}),
 their Poisson bracket
can be  obtained.
We have
\begin{equation}
\left(
\begin{array}{c|c}
\frac{\{w_A,w_B\}}{w_A w_B} & \frac{\{w_A,z_t\}}{w_A z_t} \\
\hline
\frac{\{z_u,w_B\}}{z_u w_B} & \frac{\{z_u,z_t\}}{z_u z_t} 
\end{array}
\right)
=
\left(
\begin{array}{ccccc}
 0 & 3 & -3 & 1 & -1 \\
 -3 & 0 & 3 & -2 & 2 \\
 3 & -3 & 0 & 1 & -1 \\
 -1 & 2 & -1 & 0 & 0 \\
 1 & -2 & 1 & 0 & 0
\end{array}
\right)\end{equation}
as in \cite{Franco:2011sz},
where here  $u,t=1,2$.

\subsubsection{$\mathbb{F}_0^{(I)}$}

\begin{figure}[htpb]
\begin{minipage}[b]{0.6\linewidth}
\begin{center}
\includegraphics[width=5cm]{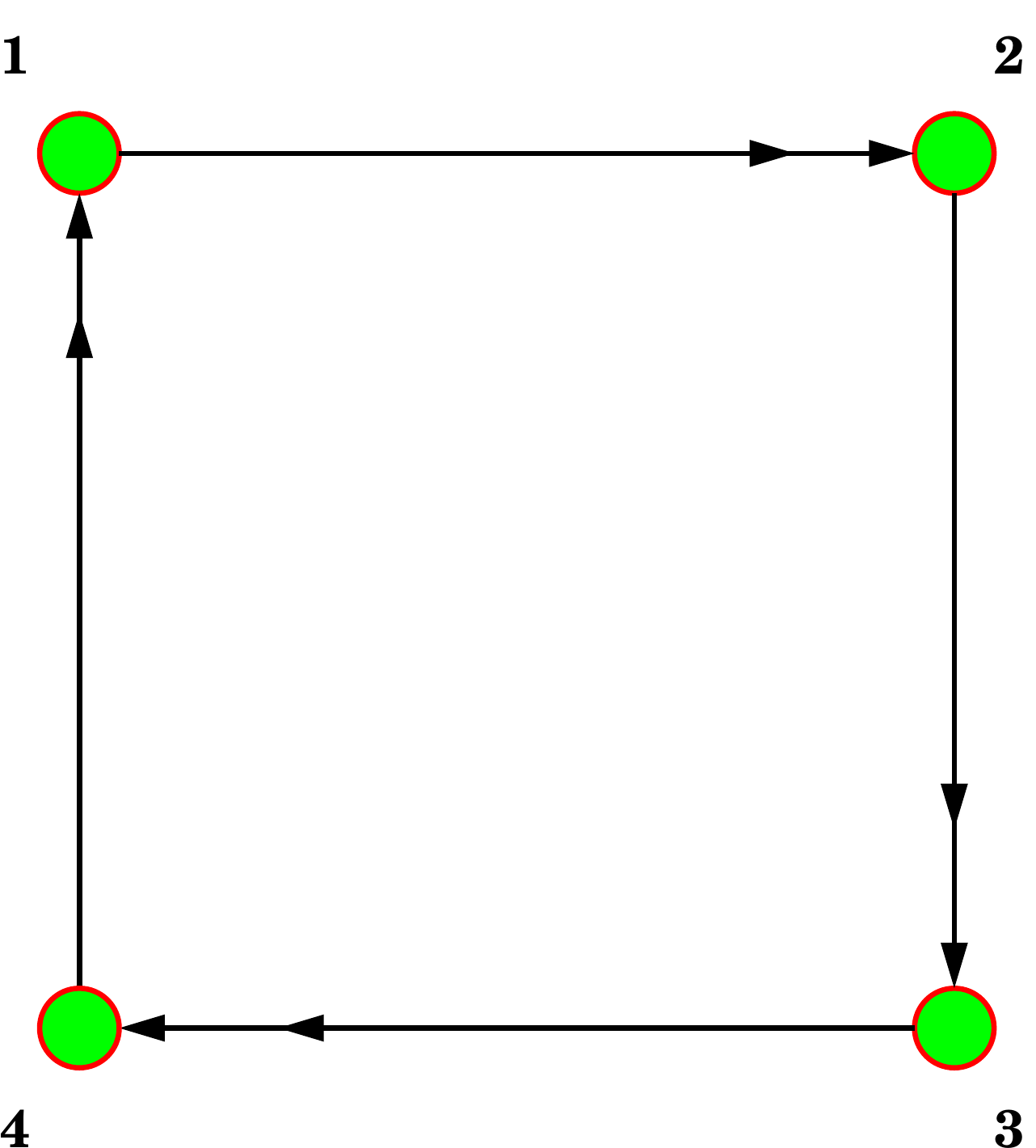}
\end{center}
\end{minipage}
\hspace{-0.5cm}
\begin{minipage}[b]{0.6\linewidth}
\includegraphics[width=5cm]{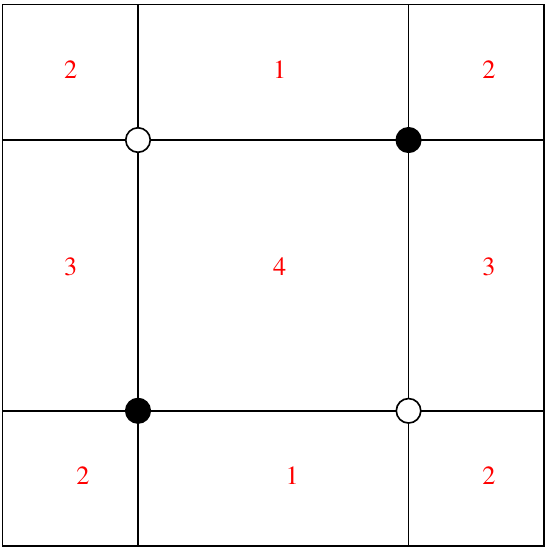}
\end{minipage}
\\
\\
\\
\begin{minipage}[b]{0.6\linewidth}
\begin{center}
\includegraphics[width=3cm]{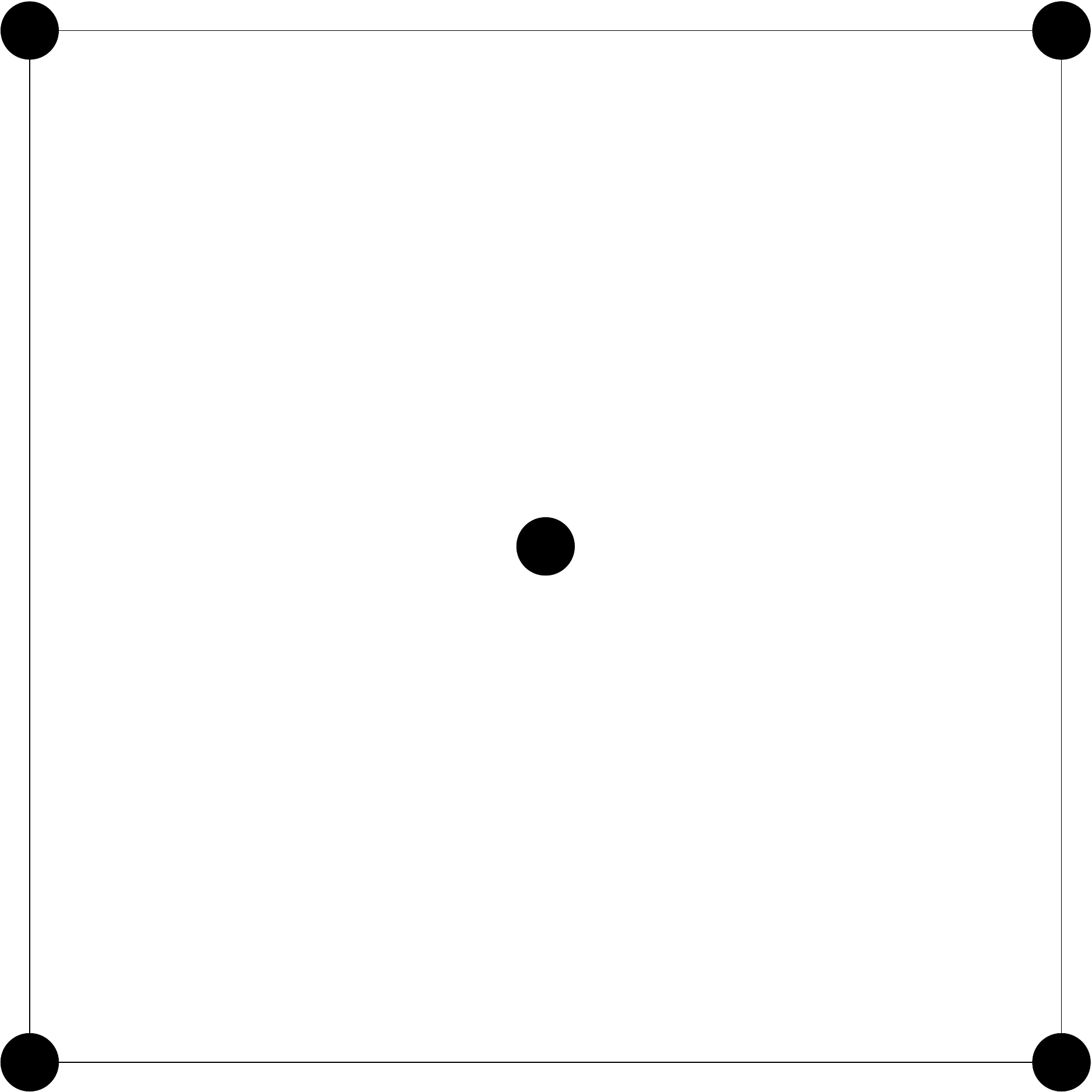}
\vspace{.5cm}
\end{center}
\end{minipage}
\hspace{-.5cm}
\begin{minipage}[b]{0.6\linewidth}
\large
$a=
\left(
\begin{array}{cccc}
0&2&0&-2\\
-2&0&2&0\\
0&-2&0&2\\
2&0&-2&0
\end{array}
\right)
$
\normalsize
\vspace{.5cm}
\end{minipage}
\caption{Quiver, brane tiling, toric diagram and adjacency matrix for $\mathbb{F}_0^{(I)}$}
\label{F0fig}
\end{figure}

\begin{figure}[htpb]
\begin{center}
\includegraphics[width=10cm]{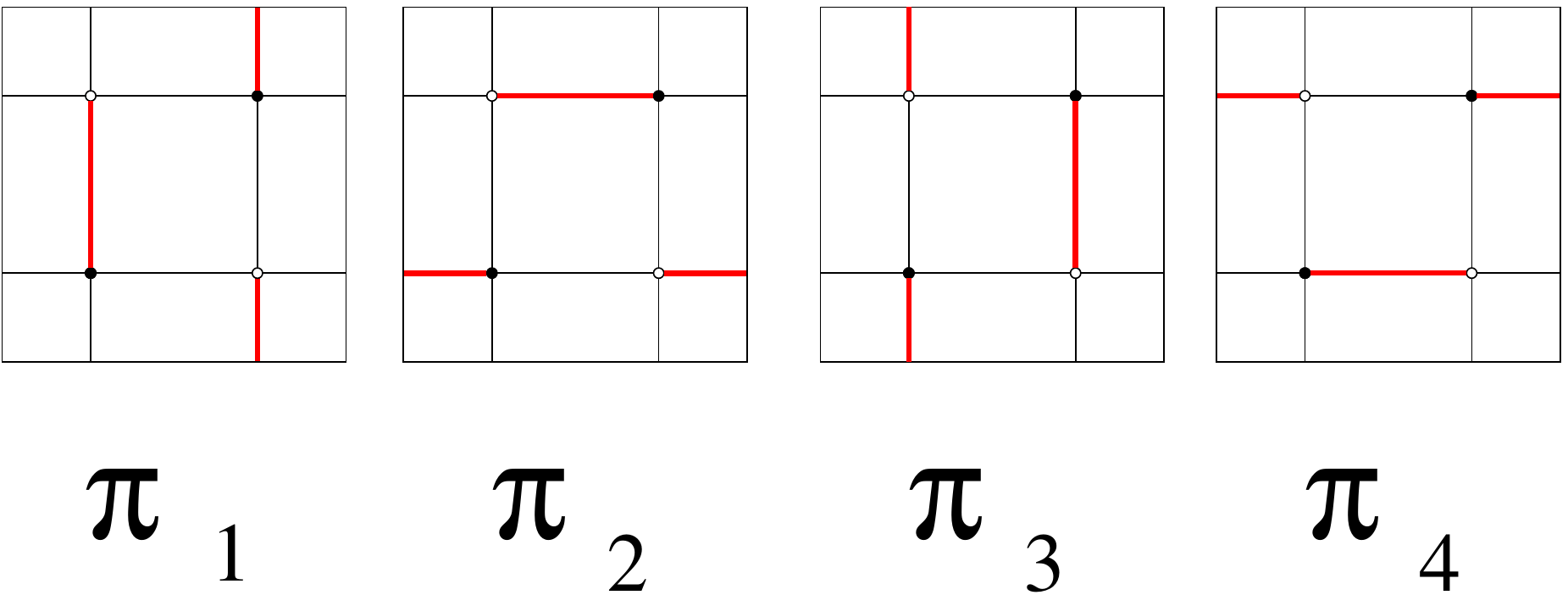}
\caption{External perfect matchings of $\mathbb{F}_0^{(I)}$}
\label{F0fig1}
\end{center}
\end{figure}

\begin{figure}[htpb]
\begin{center}
\includegraphics[width=10cm]{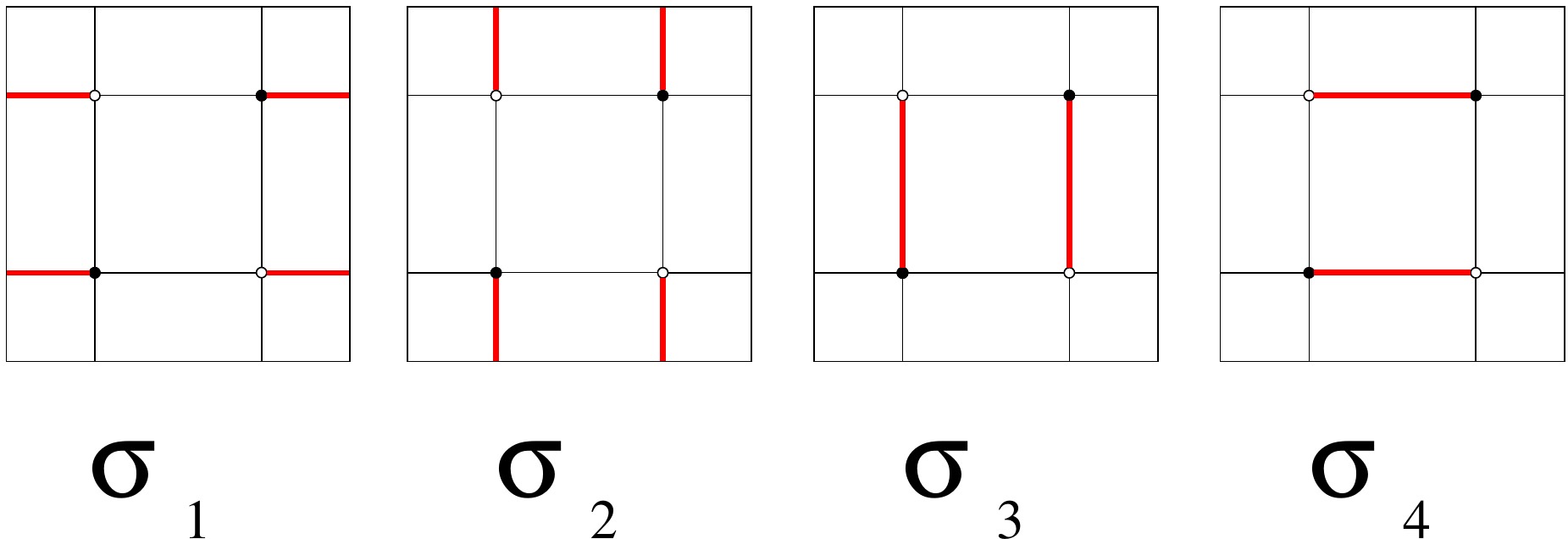}
\caption{Internal  perfect matchings of $\mathbb{F}_0^{(I)}$}
\label{F0fig2}
\end{center}
\end{figure}

The quiver gauge theory describing a stack of $N$ 
 D$3$ branes over the zero Hirzebruch surface is described by
the superpotential
\begin{equation}
W = \epsilon_{ij} \epsilon_{lk} X_{12}^{(i)} X_{23}^{(l)} X_{34}^{(j)} X_{41}^{(k)}
\end{equation}
The quiver, the brane tiling, the toric diagram and the adjacency matrix are specified in the 
Figure \ref{F0fig}.
The four external points of the toric diagram  are associated to the perfect matching in Figure \ref{F0fig1} 
while the four degenerate internal points are given in Figure \ref{F0fig2}.
They respect the relations
\begin{equation}
\pi _2+ \pi _4=\sigma _1+\sigma _4 \quad, \quad \pi _1+\pi _3=\sigma _2+\sigma _3
\end{equation}
The T matrix is then
\begin{equation}
T=
\left(
\begin{array}{cccccccc}
\pi_1&\pi_2&\pi_3&\pi_4&\sigma_1&\sigma_2&\sigma_3&\sigma_4\\
 1 & 0 & 0 & 0 & 0 & 0 & 1 & 0 \\
 0 & 1 & 0 & 0 & 0 & 0 & 0 & 1 \\
 0 & 0 & 1 & 0 & 0 & 0 & 1 & 0 \\
 0 & 0 & 0 & 1 & 0 & 0 & 0 & 1 \\
 0 & 0 & 0 & 0 & 1 & 0 & 0 & -1 \\
 0 & 0 & 0 & 0 & 0 & 1 & -1 & 0
\end{array}
\right)
\end{equation}
We stress that our algorithm is manifestly $SL(g+2,\mathbb{Z})$ invariant
and we are choosing this basis only for having more manageable expressions.
From now one we fix the first external perfect matching as the reference one,
and the $\delta T$ matrix is
\begin{equation}
\delta T =\left(
\begin{array}{cccccccc}
 0 & -1 & -1 & -1 & -1 & -1 & 0 & -1 \\
 0 & 1 & 0 & 0 & 0 & 0 & 0 & 1 \\
 0 & 0 & 1 & 0 & 0 & 0 & 1 & 0 \\
 0 & 0 & 0 & 1 & 0 & 0 & 0 & 1 \\
 0 & 0 & 0 & 0 & 1 & 0 & 0 & -1 \\
 0 & 0 & 0 & 0 & 0 & 1 & -1 & 0
\end{array}
\right)
\end{equation}
The loops $w_A$ are identified with the
differences among the perfect matchings in the matrix $T$,
precisely 
\begin{equation}
w_1= \frac{y_{\delta \sigma_2}}{y_{\delta \sigma_4}},\quad
w_2= \frac{y_{\delta \sigma_1}}{y_{\delta \sigma_2}},\quad
w_3= \frac{y_{\delta \sigma_3}}{y_{\delta \sigma_1}},\quad
w_4= \frac{y_{\delta \sigma_4}}{y_{\delta \sigma_3}}
\end{equation}
The matrix $B$ and $A$ as defined in 
(\ref{base}) and (\ref{funzwA})
are
\begin{equation}
B=
\left(
\begin{array}{cccccc}
 -1 & -1 & -1 & -1 & -1 & 1 \\
 1 & 0 & 0 & -1 & -1 & 1 \\
 0 & 1 & 0 & -1 & -1 & 1 \\
 0 & 0 & 1 & -1 & -1 & 1 \\
 0 & 0 & 0 & 0 & 4 & 1 \\
 0 & 0 & 0 & 4 & 0 & 1
\end{array}
\right)
\quad
A=
\left(
\begin{array}{cccc}
 0 & 0 & 1 & -1 \\
 -1 & 0 & 0 & 1 \\
 0 & 0 & 1 & -1 \\
 -1 & 0 & 0 & 1 \\
 1 & 1 & -1 & -1 \\
 1 & -1 & -1 & 1
\end{array}
\right)
\end{equation}
As explained above the CY condition of the master space
 implies that only $5$ independent $U(1)$ charges
are sufficient to parametrize differences of perfect matchings. i.e. closed cycles on the dimer.
These $5$ variables correspond to the first $5$ columns of the matrix $B$, whereas the
last column is the orthogonal direction.
The first $5$ columns are identified with local coordinates $x_i$
on the Poisson manifold
as
\begin{equation}
\{x_m\} = (c_1,c_2,c_3,q_1,q_2)
\end{equation}
Given the ordering of the column of the $B$ matrix, the first three variables
$c_i$ are associated to Casimir operators, while $q_1$ and $q_2$ are dynamical variables.
By solving the relations (\ref{soluzione}) we find the antisymmetric structure among these charges as
\begin{equation}
\label{structureF0}
\{c_e,c_f\} = 0,\quad  \{c_e,q_{\widetilde n}\}= \{q_{\widetilde m},c_f\}=0, \quad 
\{q_{\widetilde m},q_{\widetilde n}\}= 16 \, \epsilon_{\widetilde m \widetilde n}
\end{equation}
The $y$ variables in terms of the $c_i$ and $q_i$ become
\begin{eqnarray}
&&y_{\delta \pi_1} =1,\quad\quad
y_{\delta \pi_2} =e^{c_1},\quad\quad
y_{\delta \pi_3} =e^{c_2},\quad\quad
y_{\delta \pi_4} =e^{c_3}, \nonumber \\
&&
y_{\delta \sigma_1} =e^{\frac{1}{4} \left(q_2+c_1+c_2+c_3\right)}\quad\quad\,
y_{\delta \sigma_2} =e^{\frac{1}{4} \left(q_1+c_1+c_2+c_3\right)}\nonumber \\
&&
y_{\delta \sigma_3} =e^{\frac{1}{4} \left(-q_1-c_1+3 c_2-c_3\right)}\quad\,
y_{\delta \sigma_4} =e^{\frac{1}{4} \left(-q_2+3 c_1-c_2+3 c_3\right)}
\end{eqnarray}
The only Hamiltonian associated with the single internal point is obtained with the procedure explained in section 
\ref{hamiltoniane}
and it is
\begin{eqnarray}\label{HF01}
H =y_{\delta \sigma_1}+y_{\delta \sigma_2}+y_{\delta \sigma_3}+y_{\delta \sigma_4}
\end{eqnarray}
Finally, by introducing the paths $z_1$ and $z_2$ as 
\begin{equation}
z_1=\frac{y_{ \delta \sigma_{3}}}{y_{\delta \pi_{1}}}
,\quad
z_2=\frac{y_{ \delta \pi_{2}}}{y_{\delta \sigma_5}}
\end{equation}
we reconstruct the algebra of the $w_A$ and $z_1$ and $z_2$ 
defined in \cite{Goncharov:2011hp}
from the algebra of the $c_e$ and $q_{\widetilde m}$. Their 
Poisson brackets,  determined
via the antisymmetric (\ref{structureF0}),
is
\begin{equation}
\left(
\begin{array}{c|c}
\frac{\{w_A,w_B\}}{w_A w_B} & \frac{\{w_A,z_t\}}{w_A z_t} \\
\hline
\frac{\{z_u,w_B\}}{z_u w_B} & \frac{\{z_u,z_t\}}{z_u z_t} 
\end{array}
\right)
=
\left(
\begin{array}{cccccc}
 0 & 2 & 0 & -2 & 1 & -1 \\
 -2 & 0 & 2 & 0 & 1 & 1 \\
 0 & -2 & 0 & 2 & -1 & 1 \\
 2 & 0 & -2 & 0 & -1 & -1 \\
 -1 & -1 & 1 & 1 & 0 & 1 \\
 1 & -1 & -1 & 1 & -1 & 0
\end{array}
\right)
\end{equation}

\subsubsection{$\mathbb{F}_0^{(II)}$}
\begin{figure}[htpb]
\begin{minipage}[b]{0.6\linewidth}
\begin{center}
\includegraphics[width=3cm]{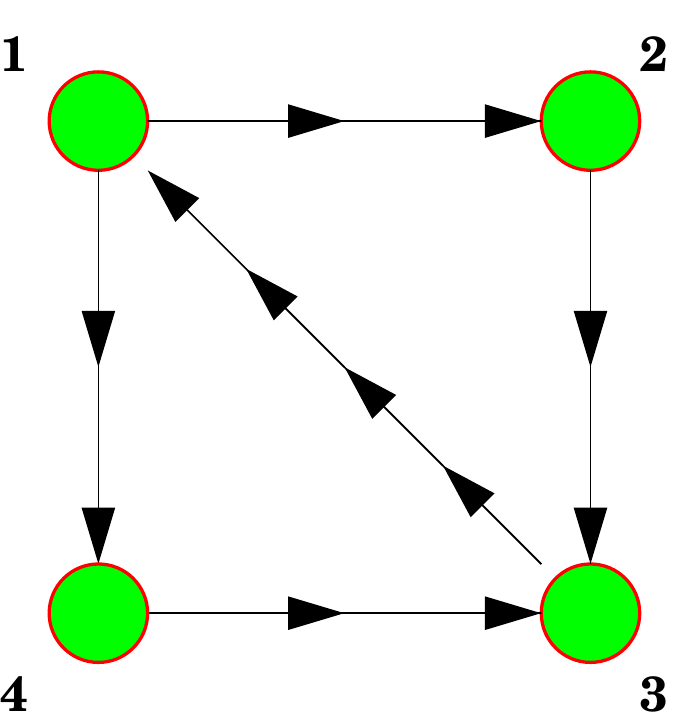}
\end{center}
\end{minipage}
\hspace{-0.5cm}
\begin{minipage}[b]{0.6\linewidth}
\includegraphics[width=3cm]{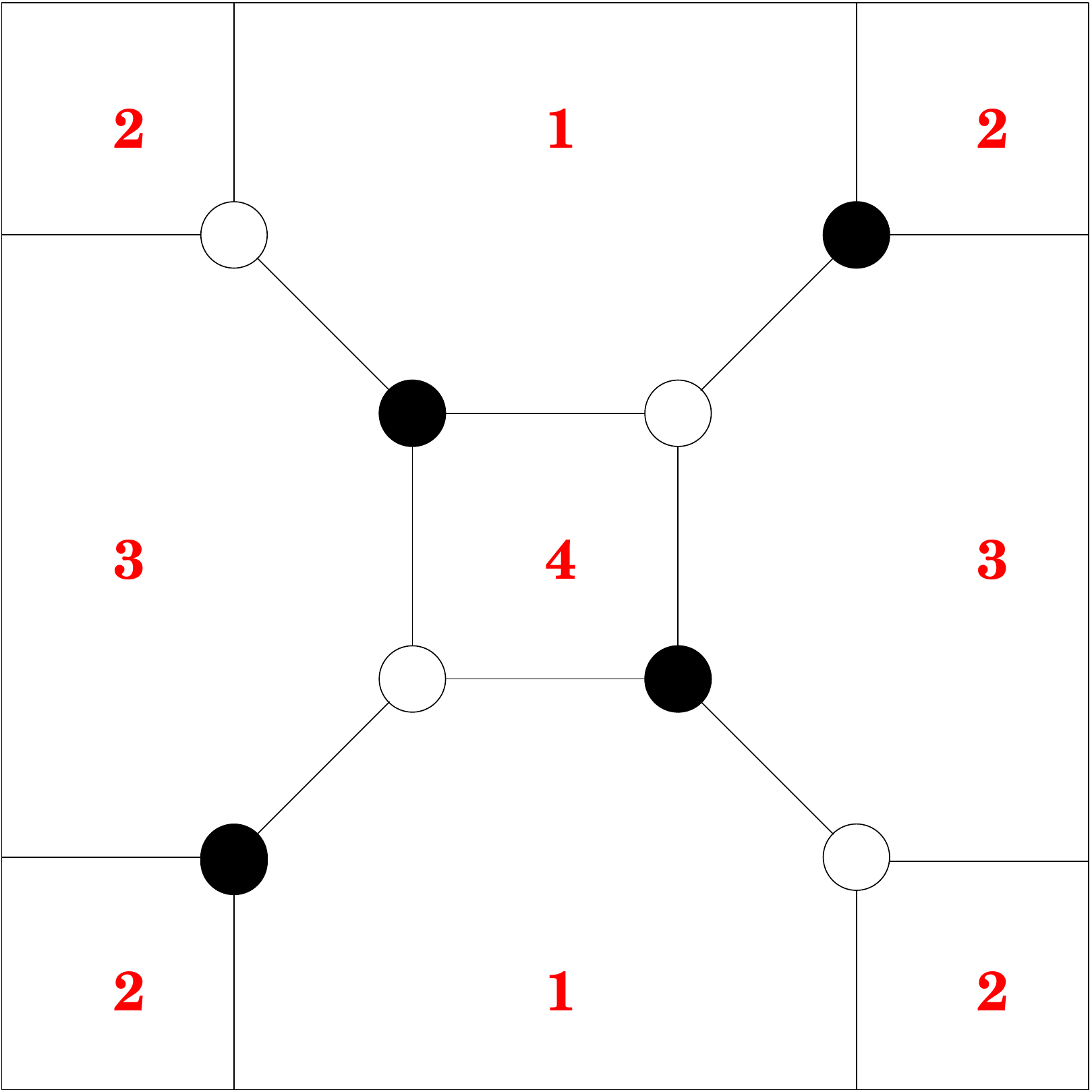}
\end{minipage}
\\
\\
\\
\begin{minipage}[b]{0.6\linewidth}
\begin{center}
\includegraphics[width=3cm]{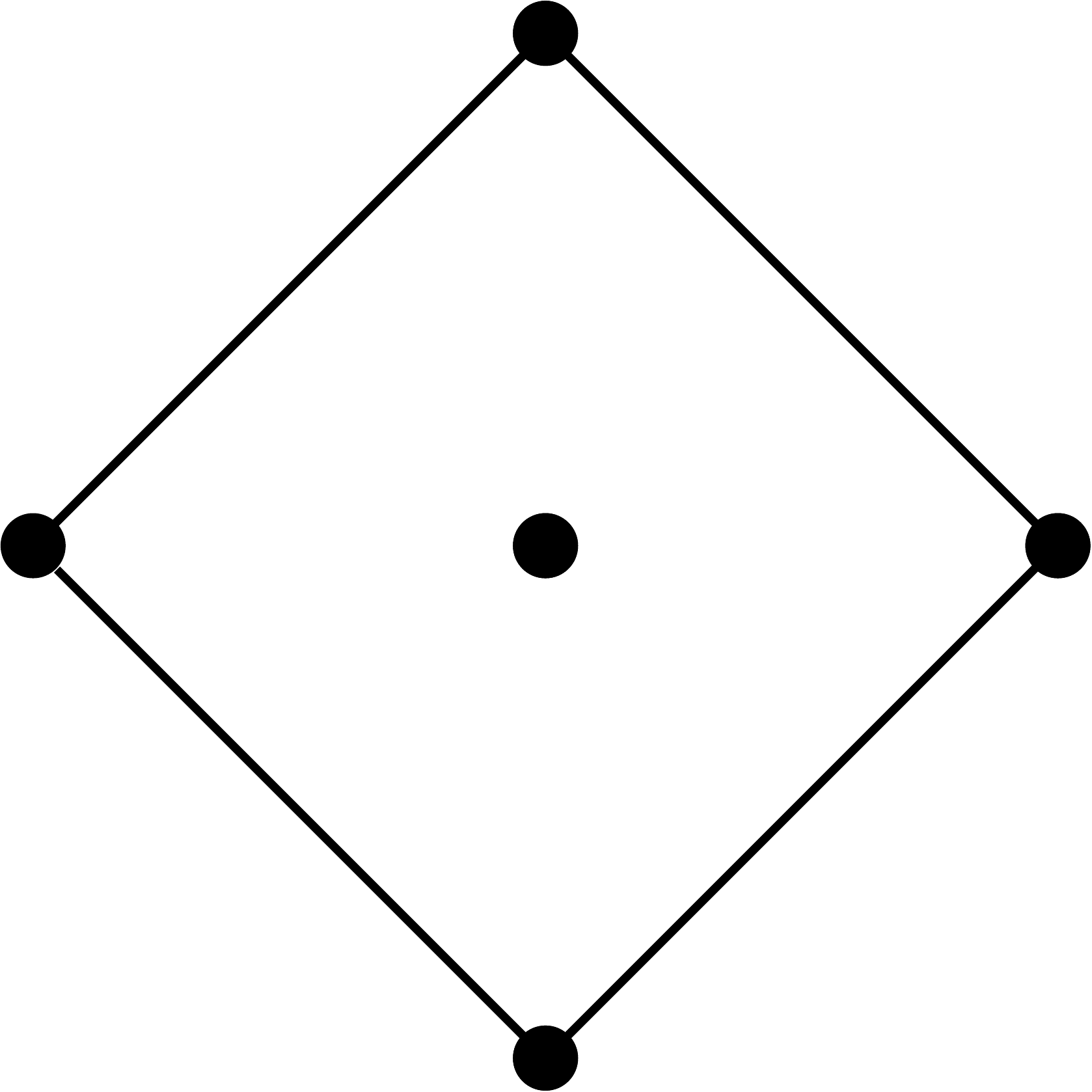}
\end{center}
\end{minipage}
\hspace{-1.6cm}
\begin{minipage}[b]{0.6\linewidth}
\large
$a=
\left(
\begin{array}{cccc}
0&2&-4&2\\
-2&0&2&0\\
4&-2&0&-2\\
-2&0&2&0
\end{array}
\right)
$
\normalsize
\vspace{.5cm}
\end{minipage}
\caption{Quiver, brane tiling, toric diagram and adjacency matrix for $\mathbb{F}_0^{(II)}$}
\label{F02fig}
\end{figure}

\begin{figure}[htpb]
\begin{center}
\includegraphics[width=10cm]{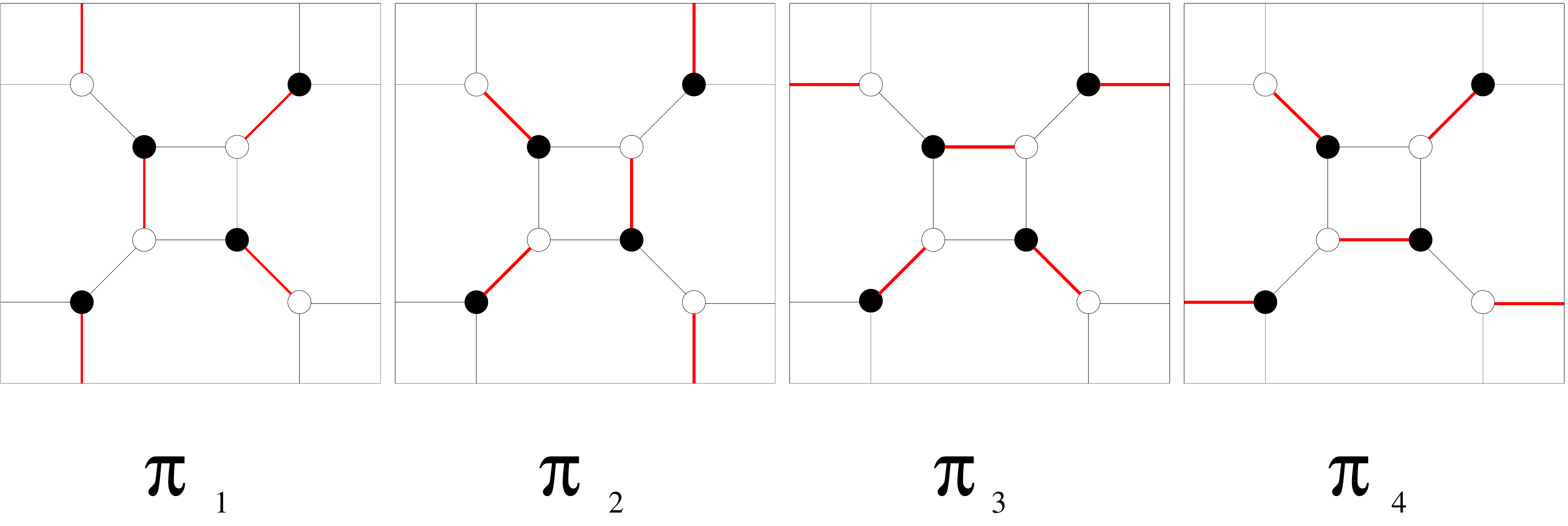}
\end{center}
\caption{External perfect matchings  of $\mathbb{F}_0^{(II)}$}
\label{F02PMEXT}
\end{figure}

\begin{figure}[htpb]
\begin{center}
\includegraphics[width=12cm]{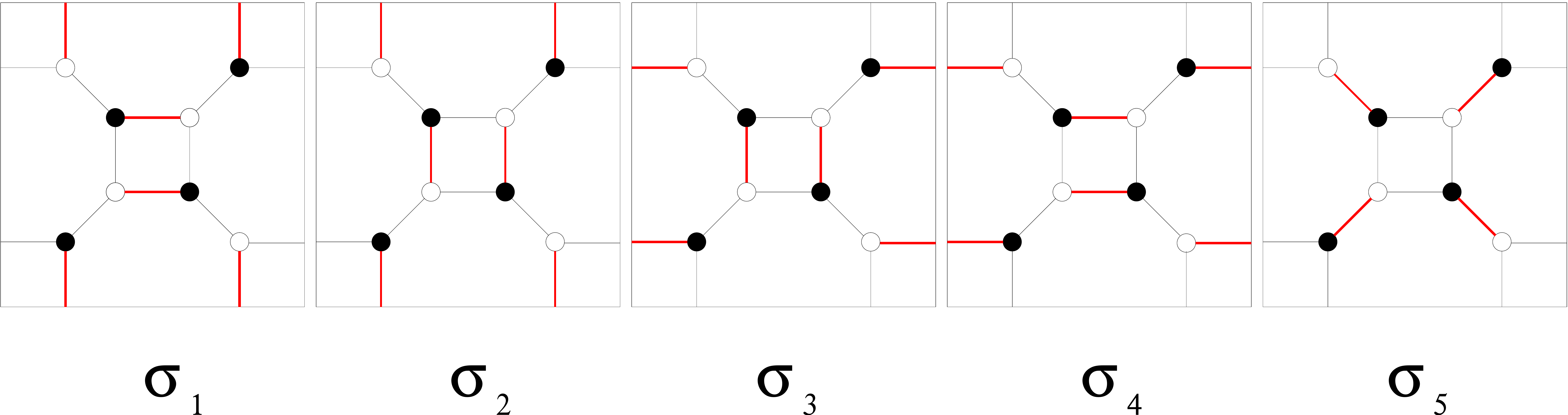}
\end{center}
\caption{Internal perfect matchings  of $\mathbb{F}_0^{(II)}$}
\label{F02PMINT}
\end{figure}

The other phase of the quiver gauge theory describing a stack of 
D$3$ branes over the zero Hirzebruch surface is described by the superpotential
\begin{equation}
W = 
\epsilon_{ij} \epsilon_{lk} X_{12}^{(i)} X_{23}^{(l)} X_{31}^{(jk)}
-
\epsilon_{ij} \epsilon_{lk} X_{14}^{(i)} X_{43}^{(l)} X_{31}^{(kj)}
\end{equation}
The quiver graph, the brane tiling, the toric diagram and the adjacency matrix are
reported in Figure \ref{F02fig}.
The external and internal perfect matchings
 are represented in Figure
\ref{F02PMEXT} and \ref{F02PMINT}
respectively.
They respect the relations
\begin{equation}
\pi _1+\pi _2=\sigma _2+\sigma _5,\quad
\pi _1+\pi _2-\pi _3-\pi _4=\sigma _2-\sigma _4,\quad
\pi _1+\pi _2-\pi _3-\pi _4=2 \sigma _2-\sigma _1-\sigma _3
\nonumber
\end{equation}
while $T$ results
\begin{equation}
T=
\left(
\begin{array}{ccccccccc}
\pi_1&\pi_2&\pi_3&\pi_4&\sigma_1&\sigma_2&\sigma_3&\sigma_4&\sigma_5
\\
 0 & 1 & 0 & 0 & 0 & 0 & -1 & -1 & 1 \\
 1 & 0 & 0 & 0 & 0 & 0 & -1 & -1 & 1 \\
 0 & 0 & 1 & 0 & 0 & 0 & 1 & 1 & 0 \\
 0 & 0 & 0 & 1 & 0 & 0 & 1 & 1 & 0 \\
 0 & 0 & 0 & 0 & 1 & 0 & -1 & 0 & 0 \\
 0 & 0 & 0 & 0 & 0 & 1 & 2 & 1 & -1
\end{array}
\right)
\end{equation}
The loops are identified with the
differences among these perfect matchings as 
\begin{equation}
w_1=\frac{y_{\delta \sigma_1}}{y_{\delta \sigma_5}}, \quad
w_2= \frac{y_{\delta \sigma_3}}{y_{\delta \sigma_2}} , \quad
w_3= \frac{y_{\delta \sigma_5}}{y_{\delta \sigma_3}} , \quad
w_4= \frac{y_{\delta \sigma_2}}{y_{\delta \sigma_1}}
\end{equation}
By fixing the first  perfect matching  as the reference one, we obtain
the 
$\delta T$ matrix as
\begin{equation}
\delta T=
\left(
\begin{array}{ccccccccc}
 0 & 1 & 0 & 0 & 0 & 0 & -1 & -1 & 1 \\
 0 & -1 & -1 & -1 & -1 & -1 & -2 & -2 & 0 \\
 0 & 0 & 1 & 0 & 0 & 0 & 1 & 1 & 0 \\
 0 & 0 & 0 & 1 & 0 & 0 & 1 & 1 & 0 \\
 0 & 0 & 0 & 0 & 1 & 0 & -1 & 0 & 0 \\
 0 & 0 & 0 & 0 & 0 & 1 & 2 & 1 & -1
\end{array}
\right)
\end{equation}
The matrix $B$ and $A$ as defined in 
(\ref{base}) and (\ref{funzwA})
are
\begin{equation}
B=
\left(
\begin{array}{cccccc}
 -1 & -1 & -1 & -1 & -1 & 1 \\
 1 & 0 & 0 & -1 & -1 & 1 \\
 0 & 1 & 0 & -1 & -1 & 1 \\
 0 & 0 & 1 & -1 & -1 & 1 \\
 0 & 0 & 0 & 0 & 4 & 1 \\
 0 & 0 & 0 & 4 & 0 & 1
\end{array}
\right)
\quad
A=\left(
\begin{array}{cccc}
 -1 & -1 & 2 & 0 \\
 -1 & -1 & 2 & 0 \\
 0 & 1 & -1 & 0 \\
 0 & 1 & -1 & 0 \\
 1 & -1 & 1 & -1 \\
 1 & 1 & -3 & 1
\end{array}
\right)
\end{equation}
Once again the CY condition of the master space guarantees that we can express
all the closed cycles on the dimer as combination of the first $5$ columns of the matrix $B$,
whereas the last column is the orthogonal direction.
We associate to the first $5$ columns the following local coordinates $x_m$ 
on the Poisson manifold
\begin{equation}
\{x_m\} = (c_1,c_2,c_3,q_1,q_2)
\end{equation}
where the $c_e$ are associated to Casimir, 
while $q_1$ and $q_2$ are dynamical variables.
By solving (\ref{soluzione}) the antisymmetric structure among these charges is
\begin{eqnarray}
\label{structureF02}
\{c_e,c_f\} = 0,\quad  \{c_e,q_{\widetilde n}\}= \{q_{\widetilde m},c_f\}=0, \quad 
\{q_{\widetilde m},q_{\widetilde n}\}= -16 \, \epsilon_{\widetilde m \widetilde n}
\end{eqnarray}
The $y$ variables in terms of the $c_i$ and $q_i$ become
\begin{eqnarray}
&&
y_{\delta \sigma_1}= e^{\frac{c_1}{4}+\frac{c_2}{4}+\frac{c_3}{4}+\frac{q_2}{4}} \quad \,\,\,,\,\,\,
y_{\delta \sigma_2}=e^{\frac{c_1}{4}+\frac{c_2}{4}+\frac{c_3}{4}+\frac{q_1}{4}},\quad
y_{\delta \sigma_3}=e^{-\frac{3 c_1}{4}+\frac{5 c_2}{4}+\frac{5 c_3}{4}+\frac{q_1}{2}-\frac{q_2}{4}}
\nonumber \\
&&
y_{\delta \sigma_4}=e^{-\frac{3 c_1}{4}+\frac{5 c_2}{4}+\frac{5 c_3}{4}+\frac{q_1}{4}},\, \,\,
y_{\delta \sigma_5}=e^{\frac{3 c_1}{4}-\frac{c_2}{4}-\frac{c_3}{4}-\frac{q_1}{4}}
\end{eqnarray}
The only Hamiltonian corresponding to the single internal point is
\begin{eqnarray} \label{HF02}
H
=
y_{\delta \sigma_1}+y_{\delta \sigma_2}+y_{\delta \sigma_3}+y_{\delta \sigma_4}+y_{\delta \sigma_5}
\end{eqnarray}
Finally, by introducing the paths $z_1$ and $z_2$ as 
\begin{equation}
z_1 =\frac{y_{ \delta \sigma_5}}{y_{\delta \pi_1}}
,\quad
z_2 =\frac{y_{ \delta \sigma_9}}{y_{\delta \pi_3}}
\end{equation}
we reconstruct the algebra of the $w_i$ and $z_1$ and $z_2$
from  (\ref{structureF02}). we have
\begin{equation}
\left(
\begin{array}{c|c}
\frac{\{w_A,w_B\}}{w_A w_B} & \frac{\{w_A,z_t\}}{w_A z_t} \\
\hline
\frac{\{z_u,w_B\}}{z_u w_B} & \frac{\{z_u,z_t\}}{z_u z_t} 
\end{array}
\right)
=
\left(
\begin{array}{cccccc}
 0 & 2 & -4 & 2 & -1 & -1 \\
 -2 & 0 & 2 & 0 & 1 & 1 \\
 4 & -2 & 0 & -2 & -1 & -1 \\
 -2 & 0 & 2 & 0 & 1 & 1 \\
 1 & -1 & 1 & -1 & 0 & 0 \\
 1 & -1 & 1 & -1 & 0 & 0
\end{array}
\right)
\end{equation}

\subsection{Theories with multiple Hamiltonians}

In the sections above we studied models with a single 
internal point, i.e. one Hamiltonian. 
The local coordinate system we have provided reproduces the correct
intersection pairing and allows to extract the canonical variables.
These systems are trivially integrable, 
since there is one Hamiltonian for a phase space of dimension two.
In this section we apply our algorithm to models with 
multiple internal points $I$. 
In such models the integrability is non trivial and is 
guaranteed by the existence of
$I$ Hamiltonians in involution \cite{Goncharov:2011hp} .

We study in detail some $Y^{p0}$ example, which have been shown  in \cite{Eager:2011dp} 
to have the same spectral curve of the Toda chain, in the non relativistic limit.
We introduce the local coordinate system associated to the $U(1)$ charges,
we extract the corresponding Poisson structure,
and we show that the $I$ Hamiltonians are indeed in involutions.
This provide a further consistency check of our algorithm and
of the local coordinatization of the Poisson manifold, based on the master space.
We restrict the examples to the case of the $Y^{p0}$ quiver gauge theories,
but our algorithm can be applied to all the infinite classes $Y^{pq}$\cite{Benvenuti:2004dy}
and $L^{pqr}$ \cite{Benvenuti:2005ja,Butti:2005sw,Franco:2005sm} of toric $\mathcal{N}=1$ SCFT.
With this procedure one can construct explicitly 
the corresponding infinite classes of
integrable systems and express the Hamiltonians 
in terms of local coordinates, and ultimately
of canonical variables.

In the following examples we will not give all the details of the computation
of the the master space.
We leave to the appendix \ref{appC} these straightforward
derivations.

\subsubsection{$Y^{30}$}
\begin{figure}[htpb]
\begin{minipage}[b]{0.6\linewidth}
\begin{center}
\includegraphics[width=6cm]{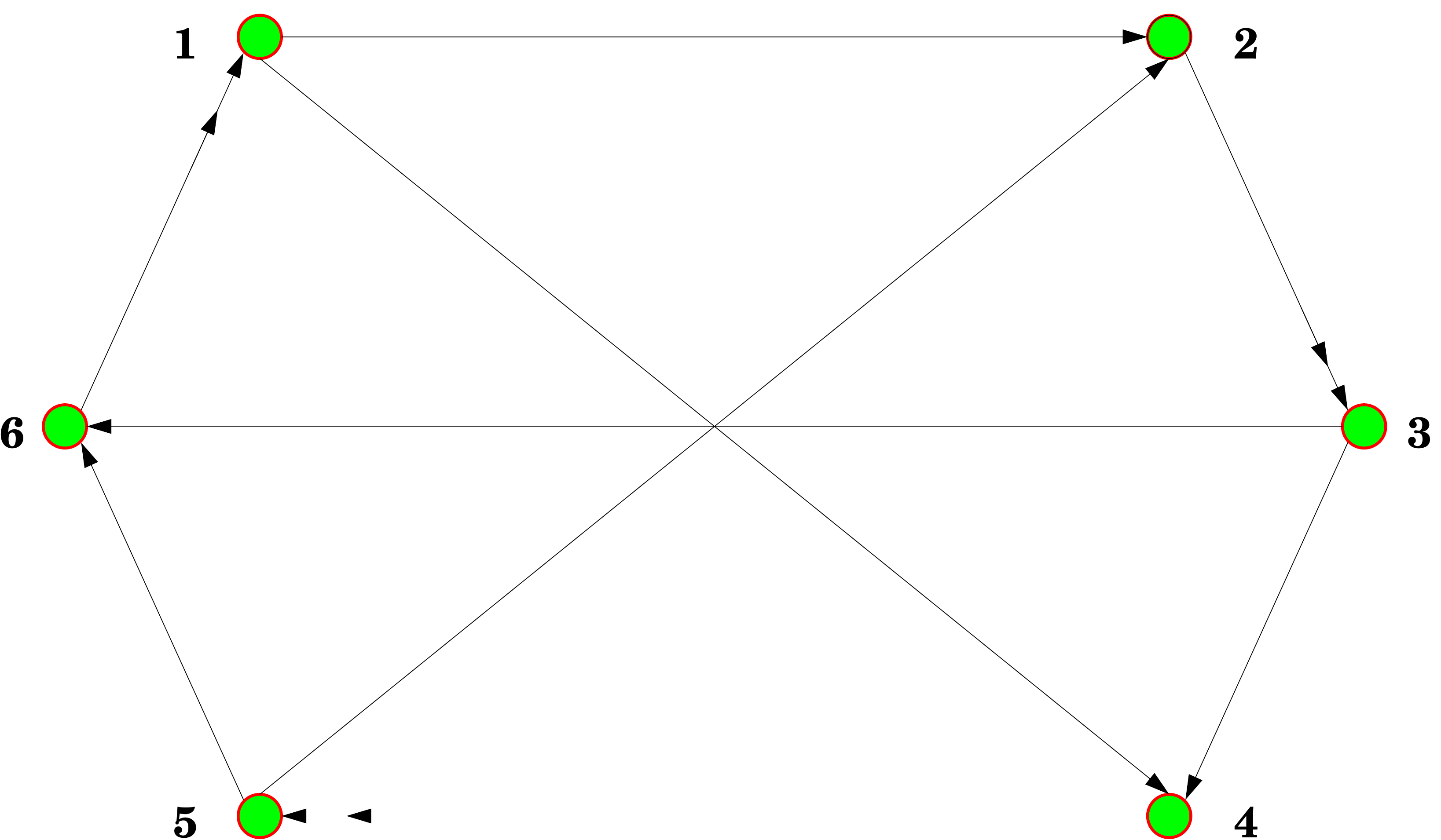}
\end{center}
\end{minipage}
\hspace{-0.5cm}
\begin{minipage}[b]{0.6\linewidth}
\includegraphics[width=6cm]{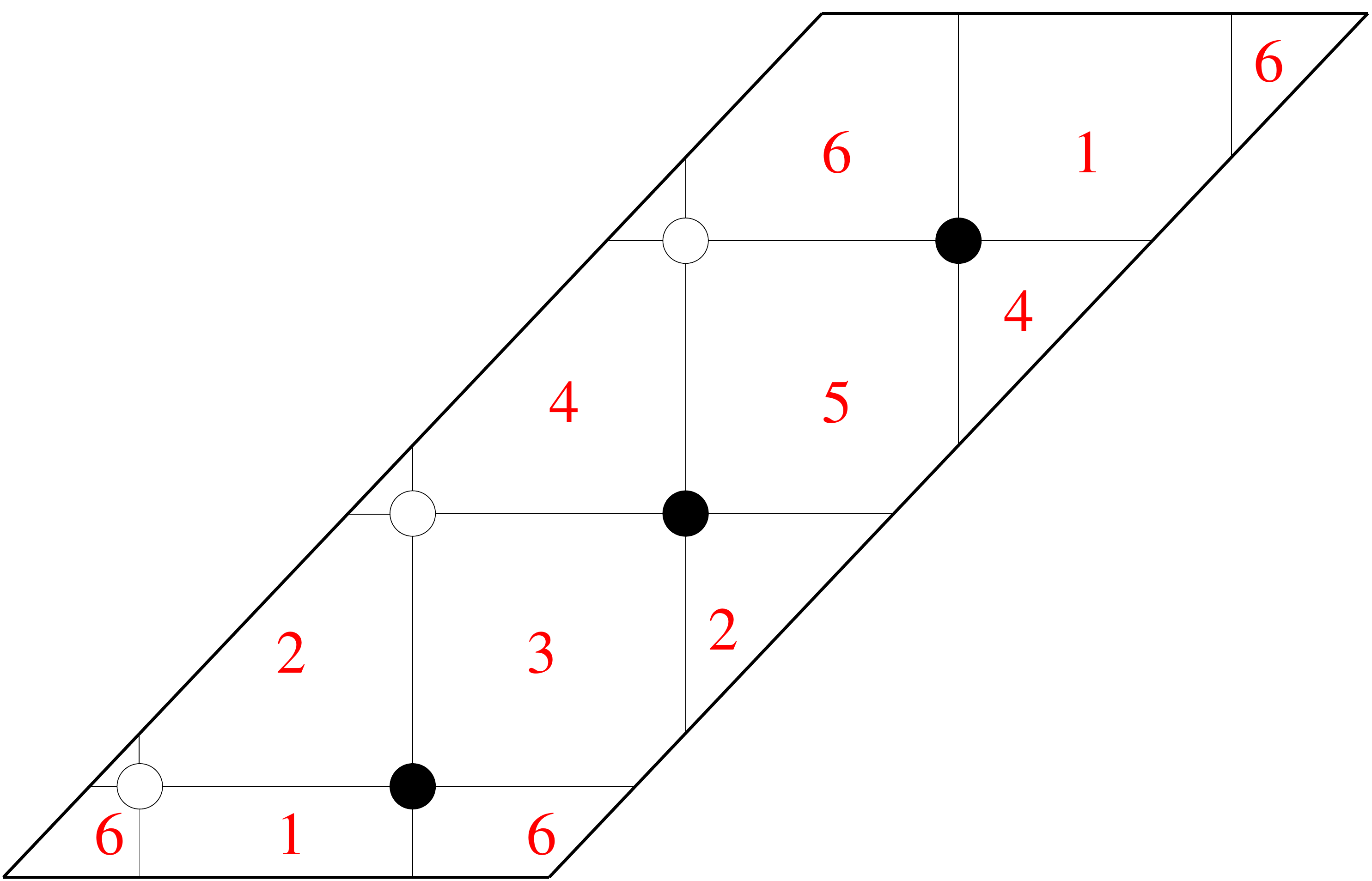}
\end{minipage}
\\
\\
\\
\begin{minipage}[b]{0.6\linewidth}
\begin{center}
\includegraphics[width=5cm]{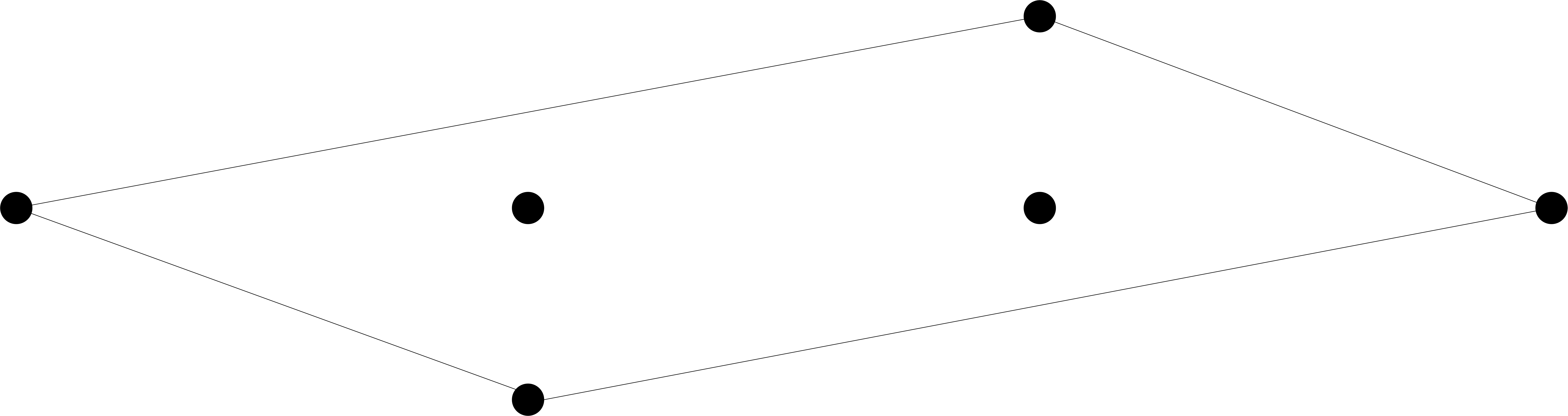}
\end{center}
\end{minipage}
\hspace{-.5cm}
\begin{minipage}[b]{0.6\linewidth}
\small
$a=\left(
\begin{array}{cccccc}
 0 & 1 & 0 & 1 & 0 & -2 \\
 -1 & 0 & 2 & 0 & -1 & 0 \\
 0 & -2 & 0 & 1 & 0 & 1 \\
 -1 & 0 & -1 & 0 & 2 & 0 \\
 0 & 1 & 0 & -2 & 0 & 1 \\
 2 & 0 & -1 & 0 & -1 & 0
\end{array}\right)
$
\normalsize
\vspace{.5cm}
\end{minipage}
\caption{Quiver, brane tiling, toric diagram and adjacency matrix for $Y^{(30)}$}
\label{fig3030}
\end{figure}

Here we study the $\mathcal{N}=1$ SCFT living on D$3$ branes
probing the toric $Y^{30}$ singularity \cite{Benvenuti:2004dy}.
The quiver gauge theory, the tiling, the toric diagram and the adjacency matrix 
are depicted in Figure \ref{fig3030} and the superpotential is
\bea
W&=&
X_{61}^{(1)} X_{12} X_{23}^{(2)} X_{36}
-X_{61}^{(2)} X_{12} X_{23}^{(1)} X_{36}
+X_{61}^{(2)} X_{14} X_{45}^{(1)} X_{56} \nonumber \\
&& 
-X_{61}^{(1)} X_{14} X_{45}^{(2)} X_{56}
+X_{23}^{(1)} X_{34} X_{45}^{(2)} X_{52}
-X_{23}^{(2)} X_{34} X_{45}^{(1)} X_{52}
\eea
In the appendix \ref{appC} we show  both 
the external and the internal perfect matchings.
They are relate by the relations
\begin{eqnarray}
&&
\pi _2+\pi _4=\sigma _1^1+\sigma _4^1,\quad
\pi _4=\sigma _2^1-\sigma _3^1+\sigma _2^2,\quad
\pi _2+\pi _4=\sigma _2^1+\sigma _6^1,\quad
\pi _4=\sigma _2^1+\sigma _4^1-\sigma _5^2\nonumber\\
&&
\pi _1+\pi _2+\pi _3+2 \pi _4=\sigma _2^1+\sigma _4^1+\sigma _5^1+\sigma _1^2+\sigma _2^2,\quad
\pi _4=\sigma _4^1+\sigma _1^2-\sigma _6^2\nonumber \\
&&
\pi _1+\pi _3+\pi _4=\sigma _1^2+\sigma _2^2+\sigma _3^2,\quad
\pi _2+2 \pi _4=\sigma _2^1+\sigma _4^1+\sigma _4^2\nonumber
\end{eqnarray}
The $T$ matrix encoding all the charges information of the $16$ perfect matchings
which generate the master space results
\begin{equation}
T=
\left(
\begin{array}{cccccccccccccccc}
\pi_1&
\pi_2&
\pi_3&
\pi_4&
\sigma^{2}_{2}&
\sigma^{2}_{1}&
\sigma^{1}_{4}&
\sigma^{1}_{2}&
\sigma^{2}_{4}&
\sigma^{2}_{3}&
\sigma^{2}_{6}&
\sigma^{1}_{5}&
\sigma^{2}_{5}&
\sigma^{1}_{6}&
\sigma^{1}_{3}&
\sigma^{1}_{1}
\\
 1 & 0 & 0 & 0 & 0 & 0 & 0 & 0 & 1 & -1 & 1 & 0 & 0 & 1 & -1 & 1 \\
 0 & 1 & 0 & 0 & 0 & 0 & 0 & 0 & 1 & 0 & 1 & 0 & 1 & 1 & 0 & 0 \\
 0 & 0 & 1 & 0 & 0 & 0 & 0 & 0 & 1 & -1 & 1 & 0 & 0 & 1 & -1 & 1 \\
 0 & 0 & 0 & 1 & 0 & 0 & 0 & 0 & 1 & 0 & 0 & 1 & 0 & 1 & 0 & 1 \\
 0 & 0 & 0 & 0 & 1 & 0 & 0 & 0 & 0 & 1 & -1 & 1 & -1 & -1 & 0 & 0 \\
 0 & 0 & 0 & 0 & 0 & 1 & 0 & 0 & -1 & 1 & -1 & 0 & 0 & -1 & 1 & 0 \\
 0 & 0 & 0 & 0 & 0 & 0 & 1 & 0 & -1 & 1 & 0 & 0 & 0 & -1 & 1 & -1 \\
 0 & 0 & 0 & 0 & 0 & 0 & 0 & 1 & -1 & 0 & 0 & -1 & 1 & 0 & 1 & -1
\end{array}
\right)
\end{equation}
The loops  surrounding the faces of the dimer 
are associated to the differences
\begin{eqnarray}
w_1=\frac{y_{\delta \sigma_{4}^{2}}}{y_{\delta \sigma_{6}^{2}}}\quad
w_2=\frac{y_{\delta \sigma_{2}^{2}}}{y_{\delta \sigma_{4}^{2}}}\quad
w_3=\frac{y_{\delta \sigma_{3}^{1}}}{y_{\delta \sigma_{2}^{1}}}\quad
w_4=\frac{y_{\delta \sigma_{1}^{2}}}{y_{\delta \sigma_{3}^{2}}}\quad
w_5=\frac{y_{\delta \sigma_{6}^{1}}}{y_{\delta \sigma_{1}^{1}}}\quad
w_6=\frac{y_{\delta \sigma_{4}^{1}}}{y_{\delta \sigma_{3}^{1}}}
\end{eqnarray}
We consider as the reference perfect matching the first column of the matrix $T$,
we then construct the matrix $\delta T$, and hence the base $B$.
The matrices $B$ and $A$ then results
\begin{equation}
B=
\left(
\begin{array}{cccccccc}
 1 & 0 & 0 & -1 & -1 & -1 & -1 & 1 \\
 0 & 1 & 0 & -1 & -1 & -1 & -1 & 1 \\
 0 & 0 & 1 & -1 & -1 & -1 & -1 & 1 \\
 -1 & -1 & -1 & -1 & -1 & -1 & -1 & 1 \\
 0 & 0 & 0 & 0 & 0 & 0 & 4 & 1 \\
 0 & 0 & 0 & 0 & 0 & 4 & 0 & 1 \\
 0 & 0 & 0 & 0 & 4 & 0 & 0 & 1 \\
 0 & 0 & 0 & 4 & 0 & 0 & 0 & 1
\end{array}
\right)
\quad
\quad
A=
\left(
\begin{array}{cccccc}
 0 & -1 & -1 & 1 & 0 & 1 \\
 0 & -1 & 0 & 0 & 1 & 0 \\
 0 & -1 & -1 & 1 & 0 & 1 \\
 1 & -1 & 0 & 0 & 0 & 0 \\
 1 & 1 & 0 & -1 & -1 & 0 \\
 0 & 1 & 1 & 0 & -1 & -1 \\
 -1 & 1 & 1 & -1 & 0 & 0 \\
 -1 & 1 & 0 & 0 & 1 & -1
\end{array}
\right)
\end{equation}
As usual the last column of the matrix $B$ 
represents the orthogonal direction that we discard.
The first $g+1=7$ columns correspond to local coordinates on
the Poisson manifold, which we label as 
$
\{x_m\} = (c_e,q_{\widetilde m})$, with $e=1,\dots,d-1$ and $\widetilde m=1,\dots,2I$.
Here $d=4$ and $I=2$.

The first $3$ variables $c_e$ are associated to
Casimir operators and they commute with everything.
The other four $q_{\widetilde m}$ variables are dynamical and their
Poisson bracket is found  by solving   (\ref{soluzione}). We
have
\begin{equation} 
\label{algebraY30}
\{q_{\widetilde m} ,q_{\widetilde n} 
\}
=16
\left(
\begin{array}{cccc}
 0 &  \,\,1\,\, & 0 & 0 \\
 -1 & 0 & 0 & 0 \\
 0 & 0 & 0 & -1 \\
 0 & 0 & \,\,1\,\, & 0
\end{array}
\right)
\qquad \widetilde m, \widetilde n=1,\dots,4
\end{equation}
The $y$ variables associated to the external points are
\begin{equation}
 y_{\delta \pi _1}=1 \quad
 y_{\delta \pi _2}=e^{c_1}  \quad
 y_{\delta \pi _3}=e^{c_2}  \quad
 y_{\delta \pi _4}=e^{c_3}
\end{equation}
The $y$ variables associated to the first internal point are

\begin{equation}
\begin{array}{lll}
y_{\delta \sigma _1^1}=e^{c_1+\frac{q_1}{4}-\frac{q_4}{4}},&
y_{\delta \sigma _2^1}=e^{\frac{1}{4} \left(c_1+c_2+c_3-q_2-q_3-q_4\right)},&
y_{\delta \sigma _3^1}=e^{\frac{1}{4} \left(2 c_1+2 c_2-2 c_3-q_3-q_4\right)}\\
y_{\delta \sigma _4^1}=e^{\frac{1}{4} \left(c_1+c_2+c_3-q_2-q_3-q_4\right)},&
y_{\delta \sigma _5^1}=e^{\frac{1}{4} \left(c_1+c_2+c_3+q_3\right)},&
y_{\delta \sigma _6^1}=e^{\frac{1}{4} \left(3 c_1-c_2+3 c_3+q_2+q_3+q_4\right)}
\end{array}
\end{equation}
The $y$ variables associated to the second internal point are 
\begin{equation}
\begin{array}{lll}
y_{\delta \sigma _1^2}=e^{\frac{1}{4} \left(c_1+c_2+c_3+q_1\right)},&
y_{\delta \sigma _2^2}=e^{\frac{1}{4} \left(c_1+c_2+c_3+q_2\right)},&
y_{\delta \sigma _3^2}=e^{\frac{1}{4} \left(-2 c_1+2 c_2+2 c_3-q_1-q_2\right)}\\
y_{\delta \sigma _4^2}=e^{\frac{1}{4} \left(3 c_1-c_2+3 c_3+q_1+q_2+q_3\right)},&
y_{\delta \sigma _5^2}=e^{\frac{1}{4} \left(c_1+c_2+c_3-q_1-q_2-q_3\right)},&
y_{\delta \sigma _6^2}=e^{\frac{1}{4} \left(c_1+c_2+c_3+q_4\right)}
\end{array}
\end{equation}
The Hamiltonians are
\begin{eqnarray}
\label{H1Y30}
H^{a} = \sum_{k_a=1}^{6} y_{\delta \sigma_{k_a}^{a}}
\quad \quad a=1,2
\end{eqnarray}
and they  commute 
given the algebra (\ref{algebraY30}).
The expressions  (\ref{algebraY30}) and (\ref{H1Y30})
determine explicitly the integrable system
associated to the $Y^{30}$ quiver gauge theory.
The local coordinates can be made canonical by transforming the
antisymmetric structure to a canonical block diagonal form.
Finally, by defining the $z_u$ as
\begin{equation}
z_1 =\frac{y_{\delta \sigma_5^2}}{y_{\delta \sigma_6^1}}
,\quad
z_2  =\frac{y_{\delta \pi_1}}{y_{\delta \pi_4}}
\end{equation}
we  obtain the intersection matrix for the base of cycles 
$w_A,z_1,z_2$, by considering them as exponential functions of
the local coordinates $x_m=(c_1,c_2,c_3,q_1,q_2,q_3,q_4)$
 and by using the antisymmetric structure (\ref{algebraY30}). We have
\begin{equation}
\left(
\begin{array}{c|c}
\frac{\{w_A,w_B\}}{w_A w_B} & \frac{\{w_A,z_t\}}{w_A z_t} \\
\hline
\frac{\{z_u,w_B\}}{z_u w_B} & \frac{\{z_u,z_t\}}{z_u z_t} 
\end{array}
\right)
=
\left(
\begin{array}{cccccccc}
 0 & 1 & 0 & 1 & 0 & -2 & 1 & 0 \\
 -1 & 0 & 2 & 0 & -1 & 0 & 1 & 0 \\
 0 & -2 & 0 & 1 & 0 & 1 & -1 & 0 \\
 -1 & 0 & -1 & 0 & 2 & 0 & 0 & 0 \\
 0 & 1 & 0 & -2 & 0 & 1 & 0 & 0 \\
 2 & 0 & -1 & 0 & -1 & 0 & -1 & 0 \\
 -1 & -1 & 1 & 0 & 0 & 1 & 0 & 0 \\
 0 & 0 & 0 & 0 & 0 & 0 & 0 & 0
\end{array}
\right)
\end{equation}
One can check that this matrix 
is the same that can be obtained from the intersection 
index for cycles (\ref{gonceinte}) 
of \cite{Goncharov:2011hp} 
that we review in the appendix \ref{appA}.
This further confirms the validity of our construction of local coordinates 
on the Poisson manifold and their associated Poisson structure.

\subsubsection{$Y^{40}$}

\begin{center}
\begin{figure}[htpb]
\begin{minipage}[b]{0.6\linewidth}
\hspace{1.2cm}
\includegraphics[width=5cm]{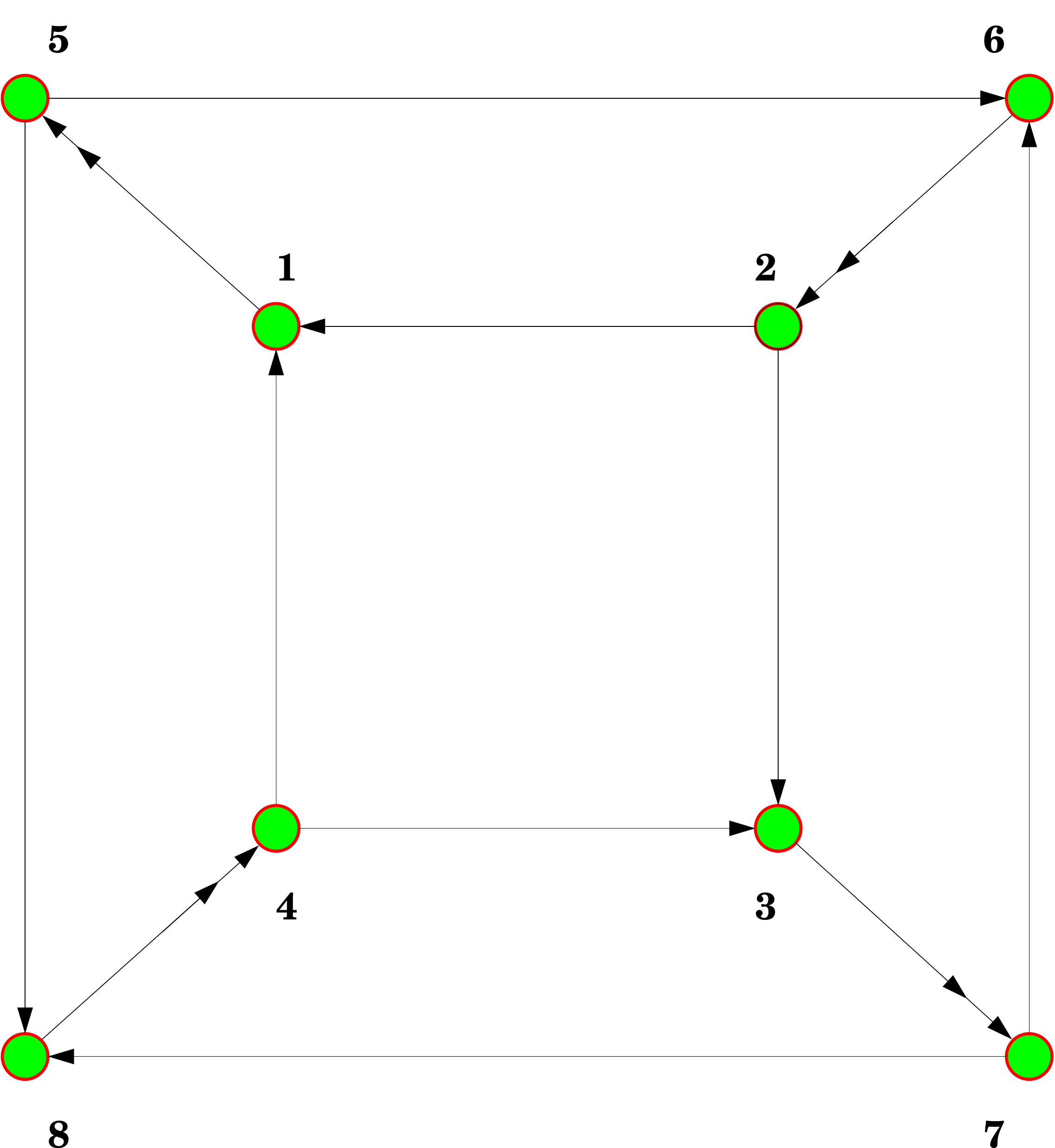}
\vspace{-.2cm}
\end{minipage}
\hspace{-.5cm}
\begin{minipage}[b]{0.6\linewidth}
\includegraphics[width=5cm]{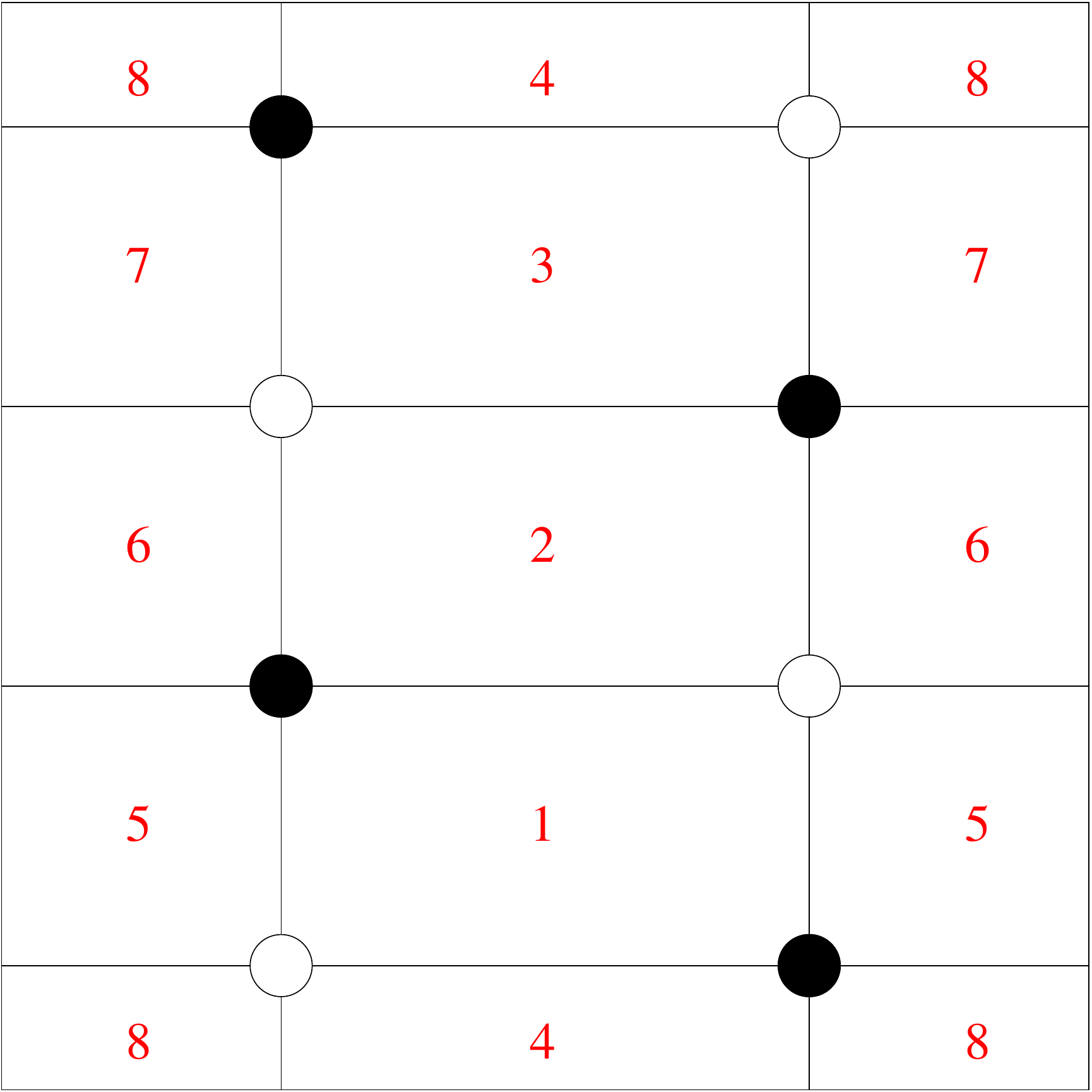}
\end{minipage}
\\
\\
\\
\begin{minipage}[b]{0.6\linewidth}
\hspace{1.1cm}
\includegraphics[width=5cm]{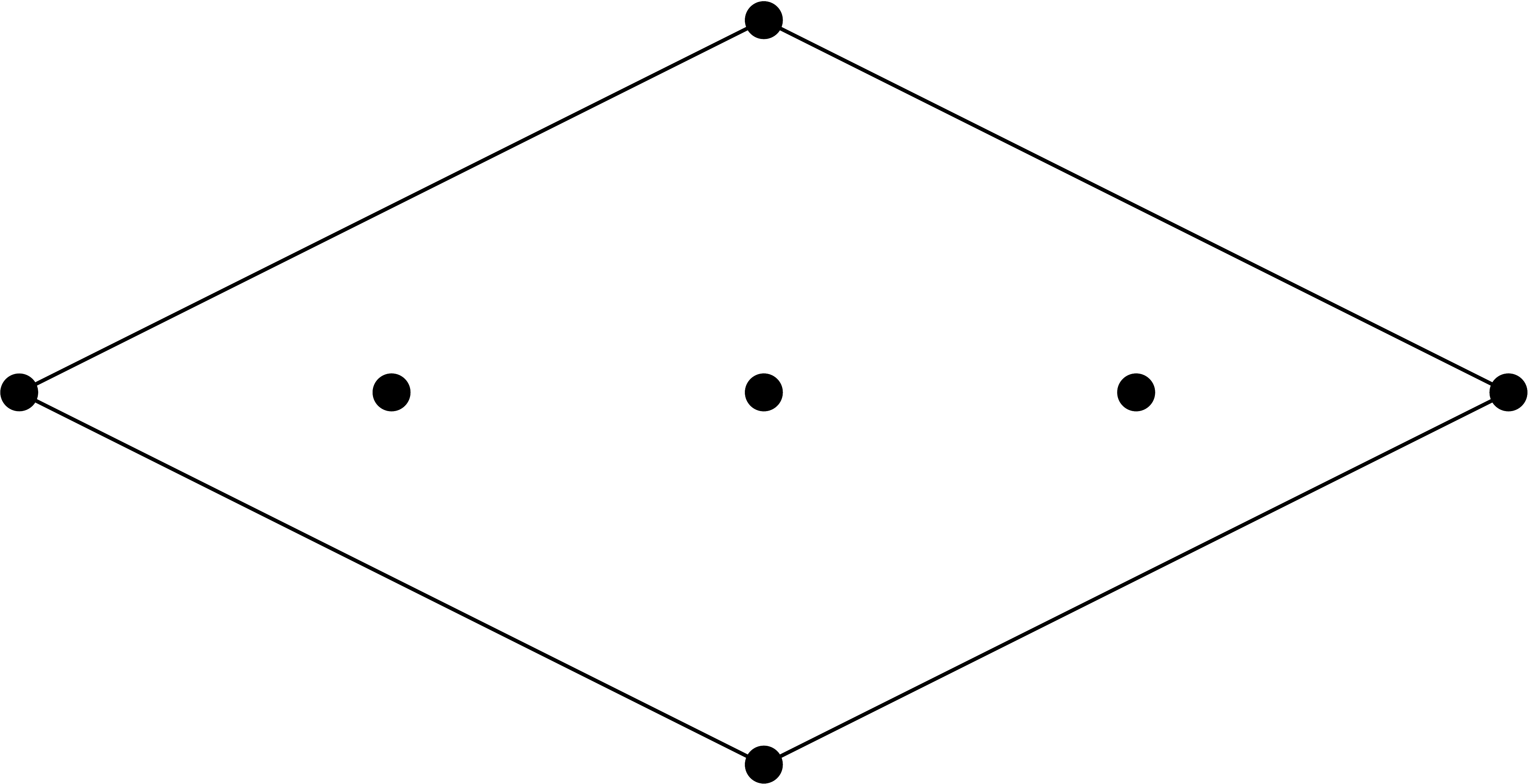}
\vspace{1.8cm}
\end{minipage}
\hspace{-1.6cm}
\begin{minipage}[b]{0.6\linewidth}
\small
$a=
\left(
\begin{array}{cccccccc}
 0 & -1 & 0 & -1 & 2 & 0 & 0 & 0 \\
 1 & 0 & 1 & 0 & 0 & -2 & 0 & 0 \\
 0 & -1 & 0 & -1 & 0 & 0 & 2 & 0 \\
 1 & 0 & 1 & 0 & 0 & 0 & 0 & -2 \\
 -2 & 0 & 0 & 0 & 0 & 1 & 0 & 1 \\
 0 & 2 & 0 & 0 & -1 & 0 & -1 & 0 \\
 0 & 0 & -2 & 0 & 0 & 1 & 0 & 1 \\
 0 & 0 & 0 & 2 & -1 & 0 & -1 & 0
\end{array}
\right)
$
\normalsize
\vspace{.5cm}
\end{minipage}
\caption{Quiver, brane tiling, toric diagram and adjacency matrix for $Y^{(40)}$}
\label{fig4040}
\end{figure}
\end{center}

The last example with multiple internal points is the $\mathcal{N}=1$ SCFT living on D$3$ branes probing the toric $Y^{40}$ 
singularity \cite{Benvenuti:2004dy}.
The quiver gauge theory, the tiling, the toric diagram and the adjacency matrix 
are depicted in Figure \ref{fig4040} and the superpotential is
\bea
W&=&
X_{15}^{(1)} X_{58} X_{84}^{(2)} X_{41} -X_{15}^{(2)} X_{58} X_{84}^{(1)} X_{41}+X_{15}^{(2)}X_{56} X_{62}^{(1)} X_{21}
-X_{15}^{(1)}X_{56} X_{62}^{(2)} X_{21}
\nonumber \\&+&X_{62}^{(2)}X_{23} X_{37}^{(1)} X_{76}-
X_{62}^{(1)}X_{23} X_{37}^{(2)} X_{76}+X_{37}^{(2)}X_{78} X_{84}^{(1)}X_{43}-X_{37}^{(1)}X_{78} X_{84}^{(2)}X_{43}
\nonumber 
\\&&
\eea
In the appendix \ref{appC} we have shown the both 
the external and the internal perfect matchings.
The perfect matchings satisfy the following relations
\footnotesize
\begin{eqnarray}
&&
\pi _2+\pi _3=-\sigma _1^2+\sigma _2^2-\sigma _2^3+\sigma _5^2+\sigma _7^2+\sigma _8^3, \, \,
\pi _3=\sigma _2^3-\sigma _5^2+\sigma _6^3,\, \,
\pi _3=-\sigma _2^2+\sigma _2^3+\sigma _5^3  \nonumber \\
&&
\pi _2+\pi _3=-\sigma _1^2+\sigma _5^2+\sigma _7^2+\sigma _{16}^2 , \, \,
\pi _1+\pi _3+\pi _4=\sigma _1^2+\sigma _1^3+\sigma _7^3
\nonumber \\&&
\pi _2+\pi _3=\sigma _1^2+\sigma _2^2+\sigma _7^2+\sigma _{15}^2 , \, \,
\pi _2+\pi _3=\sigma _1^2+\sigma _2^3+\sigma _7^2+\sigma _8^1 , \, \,
\pi _3=-\sigma _2^2+2 \sigma _2^3-\sigma _5^2+\sigma _{14}^2\nonumber \\
&&
\pi _1+\pi _3+\pi _4=\sigma _1^2+\sigma _1^3+\sigma _2^3-\sigma _5^2+\sigma _{13}^2 \quad
\pi _2+\pi _3=-2 \sigma _1^2+\sigma _2^2+\sigma _5^2+\sigma _7^2+\sigma _{12}^2 \nonumber \\
&&
\pi _3=-\sigma _1^2+\sigma _2^3+\sigma _4^3 , \, \,
\pi _2=\sigma _2^2-2 \sigma _2^3+\sigma _5^2+\sigma _{11}^2 , \, \,
\sigma _1^2-\sigma _2^3+\sigma _3^3-\sigma _7^2=0\nonumber \\
&&
\sigma _1^2-\sigma _2^2-\sigma _7^2+\sigma _{10}^2=0 \quad
\sigma _1^2-\sigma _5^2-\sigma _7^2+\sigma _8^2=0 , \, \,
\pi _1+\pi _4=2 \sigma _1^2+\sigma _1^3-\sigma _2^3-\sigma _7^2+\sigma _9^2 \nonumber \\
&&
\pi _2=\sigma _2^2-\sigma _2^3+\sigma _6^1, \, \,
\pi _2=-\sigma _2^3+\sigma _5^2+\sigma _7^1 \quad
\sigma _1^2-\sigma _2^2+\sigma _2^3-\sigma _5^2+\sigma _5^1-\sigma _7^2=0 \nonumber \\
&&
\pi _1+\pi _4=2 \sigma _1^2+\sigma _1^3+\sigma _4^1-\sigma _5^2-\sigma _7^2 , \, \,
\pi _2=2-\sigma _1^2+\sigma _2^2-\sigma _2^3+\sigma _3^1+\sigma _5^2 , \, \,
\sigma _2^2-\sigma _2^3-\sigma _2^1+\sigma _5^2=0 \nonumber \\
&&
\pi _2=-\sigma _1^2-\sigma _1^3+\sigma _2^2-\sigma _2^3+\sigma _4^2+\sigma _5^2+\sigma _7^2 , \, \,
\pi _2=-\sigma _1^2-\sigma _1^3+\sigma _1^1+\sigma _5^2+\sigma _7^2 \nonumber \\
&&
\sigma _1^3+\sigma _2^2-\sigma _2^3-\sigma _3^2=0 , \, \,
\pi _1+\pi _4=2 \sigma _1^2
\end{eqnarray}
\normalsize
The $T$ matrix encoding all the charges information of the $36$ perfect matchings
which generate the master space results
\tiny
\begin{equation}
T\!=\!
\left(\!\!
\begin{array}{cccccccccccccccccccccccccccccccccccc}
\pi_1 & \pi_2 & \pi_3 &\pi_4 &
\sigma^{2}_{2}&\sigma^{2}_{6}&\sigma^{3}_{1}&\sigma^{2}_{7}&\sigma^{2}_{5}&\sigma^{3}_{2}&\sigma^{2}_{1}&\sigma^{2}_{3}&
\sigma^{1}_{1}&\sigma^{2}_{4}&\sigma^{1}_{2}&\sigma^{1}_{3}&\sigma^{1}_{4}&\sigma^{1}_{5}&\sigma^{1}_{7}&\sigma^{2}_{8}&
\sigma^{1}_{6}&\sigma^{2}_{9}&\sigma^{2}_{10}&\sigma^{3}_{3}&\sigma^{2}_{11}&\sigma^{3}_{4}&\sigma^{2}_{12}&\sigma^{2}_{13}&
\sigma^{2}_{14}&\sigma^{1}_{8}&\sigma^{2}_{15}&\sigma^{3}_{7}&\sigma^{5}_{7}&\sigma^{2}_{16}&\sigma^{6}_{5}&\sigma^{3}_{8}
\\
 1 & 0 & 0 & 0 & 0 & 0 & 0 & 0 & 0 & 0 & 1 & 0 & 0 & 0 & 0 & 0 & 1 & 0 & 0 & 0 & 0 & 1 & 0 & 0 & 0 & 0 & 0 & 1 & 0 & 0 & 0 & 1 & 0 & 0 & 0 & 0 \\
 0 & 1 & 0 & 0 & 0 & 0 & 0 & 0 & 0 & 0 & 0 & 0 & 1 & 1 & 0 & 1 & 0 & 0 & 1 & 0 & 1 & 0 & 0 & 0 & 1 & 0 & 1 & 0 & 0 & 1 & 1 & 0 & 0 & 1 & 0 & 1 \\
 0 & 0 & 1 & 0 & 0 & 0 & 0 & 0 & 0 & 0 & 0 & 0 & 0 & 0 & 0 & 0 & 0 & 0 & 0 & 0 & 0 & 0 & 0 & 0 & 0 & 1 & 1 & 1 & 1 & 1 & 1 & 1 & 1 & 1 & 1 & 1 \\
 0 & 0 & 0 & 1 & 0 & 0 & 0 & 0 & 0 & 0 & 1 & 0 & 0 & 0 & 0 & 0 & 1 & 0 & 0 & 0 & 0 & 1 & 0 & 0 & 0 & 0 & 0 & 1 & 0 & 0 & 0 & 1 & 0 & 0 & 0 & 0 \\
 0 & 0 & 0 & 0 & 1 & 0 & 0 & 0 & 0 & 0 &  \!-1 & 0 & 1 & 1 & 0 & 1 &  \!-2 &  \!-1 & 0 &  \!-1 & 0 &  \!-2 &  \!-1 &  \!-1 & 0 & 1 & 2 &  \!-1 & 0 & 1 & 1 &  \!-1 & 0 & 1 & 0 & 1 \\
 0 & 0 & 0 & 0 & 0 & 1 & 0 & 0 & 0 & 0 & 0 & 1 & 0 &  \!-1 & 1 &  \!-1 & 0 & 1 & 0 & 0 &  \!-1 & 0 & 1 & 0 &  \!-1 & 0 &  \!-1 & 0 & 1 & 0 &  \!-1 & 0 & 1 & 0 & 0 &  \!-1 \\
 0 & 0 & 0 & 0 & 0 & 0 & 1 & 0 & 0 & 0 & 0 & 1 & 1 & 1 & 0 & 0 &  \!-1 & 0 & 0 & 0 & 0 &  \!-1 & 0 & 0 & 0 & 0 & 0 &  \!-1 & 0 & 0 & 0 &  \!-1 & 0 & 0 & 0 & 0 \\
 0 & 0 & 0 & 0 & 0 & 0 & 0 & 1 & 0 & 0 & 0 & 0 &  \!-1 &  \!-1 & 0 & 0 & 1 & 1 & 0 & 1 & 0 & 1 & 1 & 1 & 0 & 0 &  \!-1 & 0 & 0 &  \!-1 &  \!-1 & 0 & 0 &  \!-1 & 0 &  \!-1 \\
 0 & 0 & 0 & 0 & 0 & 0 & 0 & 0 & 1 & 0 & 0 & 0 &  \!-1 &  \!-1 & 1 &  \!-1 & 1 & 1 &  \!-1 & 1 & 0 & 0 & 0 & 0 &  \!-1 & 0 &  \!-1 & 1 & 1 & 0 & 0 & 0 & 0 &  \!-1 & 1 &  \!-1 \\
 0 & 0 & 0 & 0 & 0 & 0 & 0 & 0 & 0 & 1 & 0 & \! \!-1\! & 0 & 1 &  \!-1 & 1 & 0 &  \!-1 & 1 & 0 & 1 & 1 & 0 & 1 & 2 &  \!-1 & 0 &  \!-1 &  \!-2 &  \!-1 & 0 & 0 &  \!-1 & 0 &  \!-1 & 1
\end{array}\!
\right)
\end{equation}
\normalsize
The loops  around the faces of the tiling can be written
in term of the $y$ variables as
\begin{eqnarray}
&&
w_1=\frac{y_{\delta \sigma_{4}^{2}}} {y_{\delta \sigma_{1}^{2}}},\,
w_2=\frac{y_{\delta \sigma_{3}^{1}}} {y_{\delta \sigma_{6}^{1}}},\,
w_3=\frac{y_{\delta \sigma_{14}^{2}}} {y_{\delta \sigma_{3}^{2}}},\,
w_4=\frac{y_{\delta \sigma_{2}^{1}}} {y_{\delta \sigma_{8}^{1}}},\,
w_5=\frac{y_{\delta \sigma_{4}^{1}}} {y_{\delta \sigma_{5}^{1}}},\,
w_6=\frac{y_{\delta \sigma_{6}^{2}}} {y_{\delta \sigma_{2}^{2}}},\,
w_7=\frac{y_{\delta \sigma_{2}^{3}}} {y_{\delta \sigma_{10}^{2}}},\,
w_8=\frac{y_{\delta \sigma_{5}^{1}}} {y_{\delta \sigma_{2}^{1}}}
\nonumber
\end{eqnarray}
We consider as the reference perfect matching the first column of the matrix $T$,
we then construct the matrix $\delta T$, and hence the base $B$.
The matrices $B$ and $A$ then results
\small
\begin{equation}
B=
\left(
\begin{array}{cccccccccc}
 -1 & -1 & -1 & -1 & -1 & -1 & -1 & -1 & -1 & 1 \\
 1 & 0 & 0 & -1 & -1 & -1 & -1 & -1 & -1 & 1 \\
 0 & 1 & 0 & -1 & -1 & -1 & -1 & -1 & -1 & 1 \\
 0 & 0 & 1 & -1 & -1 & -1 & -1 & -1 & -1 & 1 \\
 0 & 0 & 0 & 0 & 0 & 0 & 0 & 0 & 4 & 1 \\
 0 & 0 & 0 & 0 & 0 & 0 & 0 & 4 & 0 & 1 \\
 0 & 0 & 0 & 0 & 0 & 0 & 4 & 0 & 0 & 1 \\
 0 & 0 & 0 & 0 & 0 & 4 & 0 & 0 & 0 & 1 \\
 0 & 0 & 0 & 0 & 4 & 0 & 0 & 0 & 0 & 1 \\
 0 & 0 & 0 & 4 & 0 & 0 & 0 & 0 & 0 & 1
\end{array}
\right)
\quad
\quad
A=
\left(
\begin{array}{cccccccc}
 -1 & 0 & 0 & 0 & 1 & 0 & 0 & 0 \\
 1 & 0 & 0 & -1 & 0 & 0 & 0 & 0 \\
 0 & 0 & 1 & -1 & 0 & 0 & 0 & 0 \\
 -1 & 0 & 0 & 0 & 1 & 0 & 0 & 0 \\
 2 & 1 & 0 & -1 & -1 & -1 & 1 & -1 \\
 -1 & 0 & 0 & 1 & -1 & 1 & 0 & 0 \\
 1 & 0 & -1 & 0 & -1 & 0 & 1 & 0 \\
 -1 & 0 & 0 & 1 & 0 & 0 & -1 & 1 \\
 -1 & -1 & 1 & 1 & 0 & 0 & 0 & 0 \\
 1 & 0 & -1 & 0 & 1 & 0 & -1 & 0
\end{array}
\right)
\end{equation}
\normalsize
As usual we can define the $c$ and $q$ variables, and we have 
$
\{x_m\} = (c_e,q_{\widetilde m})$, with $e=1,\dots,d-1$ and $\widetilde m=1,\dots,2I$.
Here $d=4$ and $I=3$.
The $c_e$ commute with everything and they are related to the Casimir
operators while
the $q_{\widetilde m}$ algebra becomes
\begin{equation} \label{algebra40}
\{q_{\widetilde m},q_{\widetilde n}\}=
16
\left(
\begin{array}{cccccc}
 \,0\, & \,0\, & -1 & -1 & \,0\, & \,0\, \\
 \,0\, & \,0\, & -2 & \,0\, & \,0\, & -1 \\
 \,1\, & \,2\, & \,0\, & \,0\, & \,0\, & \,1\, \\
 \,1\, & \,0\, & \,0\, & \,0\, & 1 & \,0\, \\
 \,0\, & \,0\, & \,0\, & -1 & \,0\, & \,1\, \\
 \,0\, & \,1\, & -1 & \,0\, & -1 & \,0\,
\end{array}
\right)
\quad \quad 
\widetilde m,\widetilde n=1,\dots,6
\end{equation}
The $y$ variables can be expressed in terms of the $c_e$ and $q_{\widetilde m}$ 
by looking at the matrix $B^{-1} \delta T$.
The $y$ associated to the external perfect matchings are
\begin{equation}
 y_{\delta \pi _1}=1 \quad
 y_{\delta \pi _2}=e^{c_1}  \quad
 y_{\delta \pi _3}=e^{c_2}  \quad
 y_{\delta \pi _4}=e^{c_3}
\end{equation}
The $y$ associated to the internal perfect matchings are
the exponential of the sub-sector of  $B^{-1} \delta T$
describing the  internal perfect matchings. This is
\tiny
\begin{equation}
\left(\!\!
\begin{array}{c|cccccccccccccccccccccccccccccccc}
&
\delta \sigma^{2}_{2}\!&\!\delta \sigma^{2}_{6}\!&\!\delta \sigma^{3}_{1}\!&\!\delta \sigma^{2}_{7}\!&\!\delta \sigma^{2}_{5}\!&\!\delta \sigma^{3}_{2}\!&\!\delta \sigma^{2}_{1}\!&\!\delta \sigma^{2}_{3}\!&\!
\delta \sigma^{1}_{1}\!&\!\delta \sigma^{2}_{4}\!&\!\delta \sigma^{1}_{2}\!&\!\delta \sigma^{1}_{3}\!&\!\delta \sigma^{1}_{4}\!&\!\delta \sigma^{1}_{5}\!&\!\delta \sigma^{1}_{7}\!&\!\delta \sigma^{2}_{8}\!&\!
\delta \sigma^{1}_{6}\!&\!\delta \sigma^{2}_{9}\!&\!\delta \sigma^{2}_{10}\!&\!\delta \sigma^{3}_{3}\!&\!\delta \sigma^{2}_{11}\!&\!\delta \sigma^{3}_{4}\!&\!\delta \sigma^{2}_{12}\!&\!\delta \sigma^{2}_{13}\!&\!
\delta \sigma^{2}_{14}\!&\!\delta \sigma^{1}_{8}\!&\!\delta \sigma^{2}_{15}\!&\!\delta \sigma^{3}_{7}\!&\!\delta \sigma^{5}_{7}\!&\!\delta \sigma^{2}_{16}\!&\!\delta \sigma^{6}_{5}\!&\!\delta \sigma^{3}_{8}
\\ \hline
 c_1 & \frac{1}{4} & \frac{1}{4} & \frac{1}{4} & \frac{1}{4} & \frac{1}{4} & \frac{1}{4} & \!\!-\frac{1}{4} & \frac{1}{4} & 1 & 1 & \frac{1}{4} & 1 & \!\!-\frac{1}{4} & \frac{1}{4} & 1 & \frac{1}{4} & 1 & \!\!-\frac{1}{4} & \frac{1}{4} & \frac{1}{4} & 1 & 0 & \frac{3}{4} & \!\!-\frac{1}{2} & 0 & \frac{3}{4} & \frac{3}{4} & \!\!-\frac{1}{2} & 0 & \frac{3}{4} & 0 & \frac{3}{4} \\
 c_2 & \frac{1}{4} & \frac{1}{4} & \frac{1}{4} & \frac{1}{4} & \frac{1}{4} & \frac{1}{4} & \!\!-\frac{1}{4} & \frac{1}{4} & 0 & 0 & \frac{1}{4} & 0 & \!\!-\frac{1}{4} & \frac{1}{4} & 0 & \frac{1}{4} & 0 & \!\!-\frac{1}{4} & \frac{1}{4} & \frac{1}{4} & 0 & 1 & \frac{3}{4} & \frac{1}{2} & 1 & \frac{3}{4} & \frac{3}{4} & \frac{1}{2} & 1 & \frac{3}{4} & 1 & \frac{3}{4} \\
 c_3 & \frac{1}{4} & \frac{1}{4} & \frac{1}{4} & \frac{1}{4} & \frac{1}{4} & \frac{1}{4} & \frac{3}{4} & \frac{1}{4} & 0 & 0 & \frac{1}{4} & 0 & \frac{3}{4} & \frac{1}{4} & 0 & \frac{1}{4} & 0 & \frac{3}{4} & \frac{1}{4} & \frac{1}{4} & 0 & 0 & \!\!-\frac{1}{4} & \frac{1}{2} & 0 & \!\!-\frac{1}{4} & \!\!-\frac{1}{4} & \frac{1}{2} & 0 & \!\!-\frac{1}{4} & 0 & \!\!-\frac{1}{4} \\
 q_1 & 0 & 0 & 0 & 0 & 0 & \frac{1}{4} & 0 & \!\!-\frac{1}{4} & 0 & \frac{1}{4} & \!\!-\frac{1}{4} & \frac{1}{4} & 0 & \!\!-\frac{1}{4} & \frac{1}{4} & 0 & \frac{1}{4} & \frac{1}{4} & 0 & \frac{1}{4} & \frac{1}{2} & \!\!-\frac{1}{4} & 0 & \!\!-\frac{1}{4} & \!\!-\frac{1}{2} & \!\!-\frac{1}{4} & 0 & 0 & \!\!-\frac{1}{4} & 0 & \!\!-\frac{1}{4} & \frac{1}{4} \\
 q_2 & 0 & 0 & 0 & 0 & \frac{1}{4} & 0 & 0 & 0 & \!\!-\frac{1}{4} & \!\!-\frac{1}{4} & \frac{1}{4} & \!\!-\frac{1}{4} & \frac{1}{4} & \frac{1}{4} & \!\!-\frac{1}{4} & \frac{1}{4} & 0 & 0 & 0 & 0 & \!\!-\frac{1}{4} & 0 & \!\!-\frac{1}{4} & \frac{1}{4} & \frac{1}{4} & 0 & 0 & 0 & 0 & \!\!-\frac{1}{4} & \frac{1}{4} & \!\!-\frac{1}{4} \\
 q_3 & 0 & 0 & 0 & \frac{1}{4} & 0 & 0 & 0 & 0 & \!\!-\frac{1}{4} & \!\!-\frac{1}{4} & 0 & 0 & \frac{1}{4} & \frac{1}{4} & 0 & \frac{1}{4} & 0 & \frac{1}{4} & \frac{1}{4} & \frac{1}{4} & 0 & 0 & \!\!-\frac{1}{4} & 0 & 0 & \!\!-\frac{1}{4} & \!\!-\frac{1}{4} & 0 & 0 & \!\!-\frac{1}{4} & 0 & \!\!-\frac{1}{4} \\
 q_4 & 0 & 0 & \frac{1}{4} & 0 & 0 & 0 & 0 & \frac{1}{4} & \frac{1}{4} & \frac{1}{4} & 0 & 0 & \!\!-\frac{1}{4} & 0 & 0 & 0 & 0 & \!\!-\frac{1}{4} & 0 & 0 & 0 & 0 & 0 & \!\!-\frac{1}{4} & 0 & 0 & 0 & \!\!-\frac{1}{4} & 0 & 0 & 0 & 0 \\
 q_5 & 0 & \frac{1}{4} & 0 & 0 & 0 & 0 & 0 & \frac{1}{4} & 0 & \!\!-\frac{1}{4} & \frac{1}{4} & \!\!-\frac{1}{4} & 0 & \frac{1}{4} & 0 & 0 & \!\!-\frac{1}{4} & 0 & \frac{1}{4} & 0 & \!\!-\frac{1}{4} & 0 & \!\!-\frac{1}{4} & 0 & \frac{1}{4} & 0 & \!\!-\frac{1}{4} & 0 & \frac{1}{4} & 0 & 0 & \!\!-\frac{1}{4} \\
 q_6 & \frac{1}{4} & 0 & 0 & 0 & 0 & 0 & \!\!-\frac{1}{4} & 0 & \frac{1}{4} & \frac{1}{4} & 0 & \frac{1}{4} & \!\!-\frac{1}{2} & \!\!-\frac{1}{4} & 0 & \!\!-\frac{1}{4} & 0 & \!\!-\frac{1}{2} & \!\!-\frac{1}{4} & \!\!-\frac{1}{4} & 0 & \frac{1}{4} & \frac{1}{2} & \!\!-\frac{1}{4} & 0 & \frac{1}{4} & \frac{1}{4} & \!\!-\frac{1}{4} & 0 & \frac{1}{4} & 0 & \frac{1}{4}
\end{array}
\right)
\end{equation}
\normalsize
The three Hamiltonians are
\begin{eqnarray}
H^1=\sum_{k_1=1}^{8} y_{\delta \sigma_{k_1}^1}\quad ,\quad
H^2=\sum_{k_2=1}^{16} y_{\delta \sigma_{k_2}^2}\quad ,\quad
H^3=\sum_{k_3=1}^{8} y_{\delta \sigma_{k_3}^3}
\end{eqnarray}
and given the algebra (\ref{algebra40}) they commute one each other.
 Finally, by defining the $z$ as
\begin{equation}
z_1 =\frac{y_{\delta \pi_{3}}}{y_{\delta \sigma_{5}^{3}}}
,\quad
z_2 =\frac{y_{\delta \pi_{4}}}{y_{\delta \sigma_{1}^{2}}}
\end{equation}
we can obtain the intersection matrix for the base of cycles 
$w_A,z_1,z_2$. By considering them as exponential functions of
the local coordinates $x_m=(c_1,c_2,c_3,q_1,q_2,q_3,q_4,q_5,q_6)$ and by
using the antisymmetric structure (\ref{algebra40}) we have
\begin{equation}
\left(
\begin{array}{c|c}
\frac{\{w_A,w_B\}}{w_A w_B} & \frac{\{w_A,z_t\}}{w_A z_t} \\
\hline
\frac{\{z_u,w_B\}}{z_u w_B} & \frac{\{z_t,z_u\}}{z_t z_u} 
\end{array}
\right)
=
\left(
\begin{array}{cccccccccc}
 0 & -1 & 0 & -1 & 2 & 0 & 0 & 0 & 0 & 0 \\
 1 & 0 & 1 & 0 & 0 & -2 & 0 & 0 & 0 & 0 \\
 0 & -1 & 0 & -1 & 0 & 0 & 2 & 0 & 2 & 0 \\
 1 & 0 & 1 & 0 & 0 & 0 & 0 & -2 & 0 & -2 \\
 -2 & 0 & 0 & 0 & 0 & 1 & 0 & 1 & 0 & 1 \\
 0 & 2 & 0 & 0 & -1 & 0 & -1 & 0 & -1 & 0 \\
 0 & 0 & -2 & 0 & 0 & 1 & 0 & 1 & 0 & 1 \\
 0 & 0 & 0 & 2 & -1 & 0 & -1 & 0 & -1 & 0 \\
 0 & 0 & -2 & 0 & 0 & 1 & 0 & 1 & 0 & 1 \\
 0 & 0 & 0 & 2 & -1 & 0 & -1 & 0 & -1 & 0
\end{array}
\right)
\end{equation}
Even in this case  this matrix 
coincide with the one  obtained from the intersection 
index for cycles (\ref{gonceinte}) 
of \cite{Goncharov:2011hp} 
that we review in the appendix \ref{appA}.

\section{Seiberg duality as a canonical transformation}
\label{sec6}

In this section we discuss the interpretation of Seiberg 
duality as a canonical transformation that glues the
different local patches of the cluster integrable model.
The dimer integrable system depends on the multiplicity of the internal
point of the toric diagram, in the structure of the Hamiltonians,
and the Poisson structure is related to the anomalous global symmetries of the theory.
Hence for chiral theories the dimer integrable systems is modified 
under Seiberg duality. However, by using the local parametrization
introduced in the previous sections, one can verify that
Seiberg duality acts on the coordinates as a canonical transformation
\begin{equation}
(c_e, q_{\widetilde m})  =( c_e (c_f^{S.d.}), q_{\widetilde m}(c_f^{S.d.},q_{\widetilde n}^{S.d.})) 
\end{equation}
Moreover, as we discussed in section \ref{sec4}   the toric data of the master
spaces  give canonical transformation that are linear maps among the
Casimir operators and non linear maps for the $q$ variables.
The first are realized by mapping
the external perfect matching in the usual toric way and 
the second relations, among the internal 
perfect matchings, are obtained by equating
the Hamiltonian functions and their flows.

 Here we show the equivalence between this interpretation of the duality as a 
 symplectic morphism on the master space and the description of
 \cite{Goncharov:2011hp} in terms of mutation of the seed defining the cluster algebra.

\subsection{Seiberg duality on  $\mathbb{F}_0$}

The transformation on the $c_e$ coordinates can be understood from the T matrix and equation (\ref{casimirmap}). 
We have
\begin{equation} \label{t1}
c_1= c_2^{S.d.}\quad \quad
c_2 = c_1^{S.d.}\quad \quad
c_3= c_3^{S.d.} 
\end{equation}
The transformation among the $q$ variables is obtained by solving the equations
(\ref{qpmap}) and (\ref{qpmap2}).
The first equation is an algebraic equation and it reads
\begin{equation}
\label{eqal}
y_{\delta \sigma_1}+y_{\delta \sigma_2}+y_{\delta \sigma_3}+y_{\delta \sigma_4}=
 y_{\delta \sigma_1^{S.d.}}+ y_{\delta \sigma_2^{S.d.}}+ y_{\delta \sigma_3^{S.d.}}+ y_{\delta \sigma_4^{S.d.}}+ y_{\delta \sigma_5^{S.d.}}\end{equation}
The other two differential equations are
\begin{eqnarray}
\label{eqdiff} 
&&\!\!
y_{\delta \sigma _7}\!-\!y_{\delta \sigma _6}=\left(y_{\delta \sigma _6^{S.d.}}+2 y_{\delta \sigma _7^{S.d.}}+y_{\delta \sigma _8^{S.d.}}-y_{\delta \sigma _9^{S.d.}}\right) \partial _{q_2^{S.d.}}q_2+\left(y_{\delta \sigma _7^{S.d.}}-y_{\delta \sigma _5^{S.d.}}\right) \partial _{q_1^{S.d.}}q_2
 \nonumber
\\
&&\!\!
y_{\delta \sigma _5}\!-\!y_{\delta \sigma _8}=\left(y_{\delta \sigma _6^{S.d.}}+2 y_{\delta \sigma _7^{S.d.}}+y_{\delta \sigma _8^{S.d.}}-y_{\delta \sigma _9^{S.d.}}\right) \partial _{q_2^{S.d.}}q_1+\left(y_{\delta \sigma _7^{S.d.}}-y_{\delta \sigma _5^{S.d.}}\right) \partial _{q_1^{S.d.}}q_1 \nonumber \\
  \end{eqnarray}
 Once expressed in terms of the $q_{\widetilde m}$ variables these equations look 
complicate partial differential equations. 
 We will show in the next section that the functions
 $q_1(c_e^{S.d.},q_{\widetilde m}^{S.d.})$ and  $q_2(c_e^{S.d.},q_{\widetilde m}^{S.d.})$
 which solve these equations coincide with the functions obtained by using the procedure of
  \cite{Goncharov:2011hp}.

\subsection{Duality as seed mutation}

\begin{figure}
\includegraphics[width=14cm]{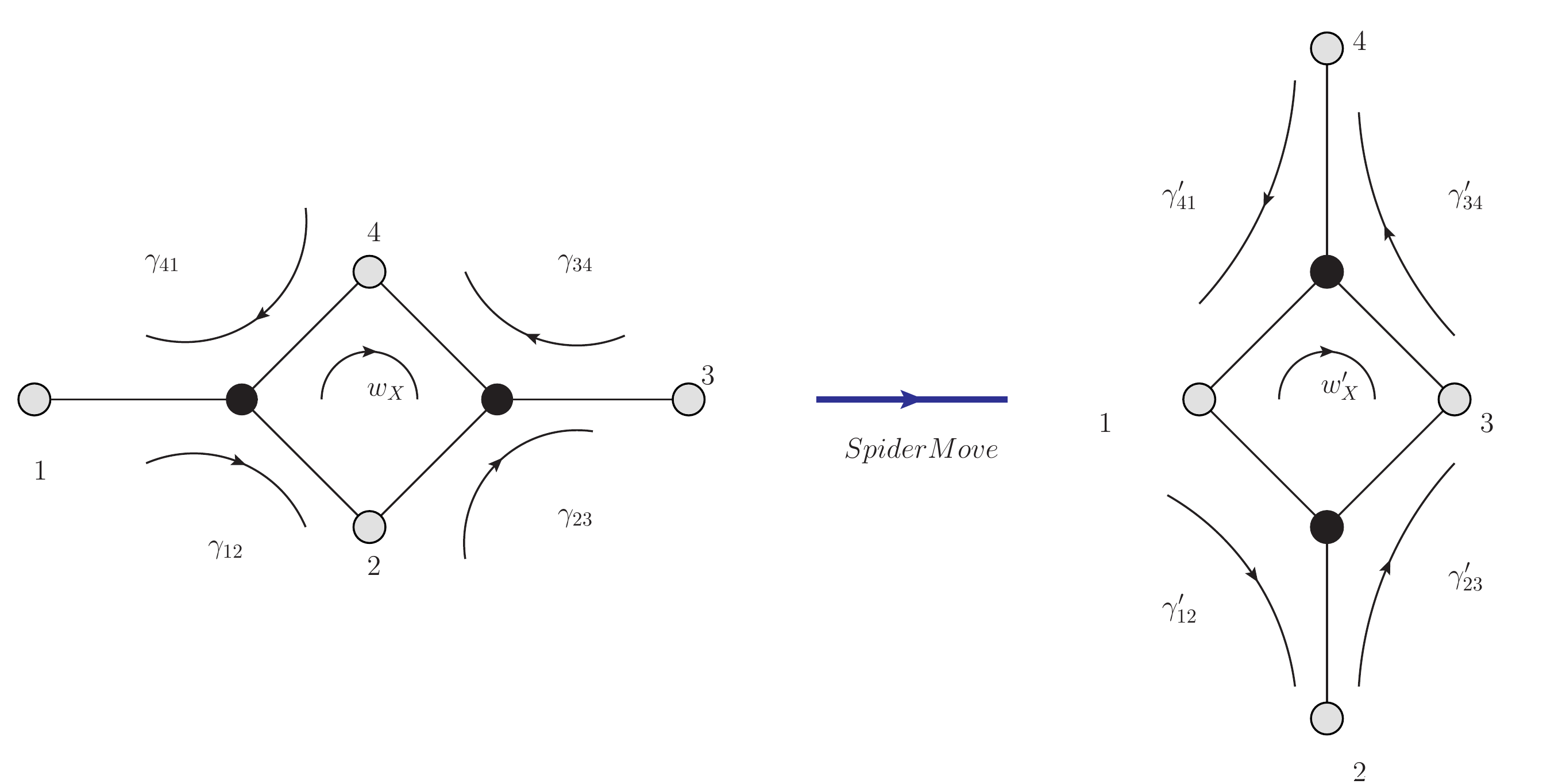}
\caption{Spider move}
\label{spider}
\end{figure}

 In this section we are describing the mapping among the dual phases 
 as discussed in \cite{Goncharov:2011hp}.
For a more detailed derivation we refer the reader to the appendix \ref{appB}.
The Seiberg duality on the dimer is referred to as a 
cluster Poisson transformation 
obtained by a mutation of the seed.
This transformation is based on an operation on the bipartite
graphs called spider move.
This is a local operation on the bipartite
graphs  which acts as in Figure \ref{spider}.
The original dimer and the resulting dimer correspond to two different (Seiberg) phases of
the same gauge theories. The adjacency
matrix is different and the number of perfect matching is different. The dimension of the corresponding
Poisson manifold is however the same ($g+1$), and also the number of Casimir
operators ($d-1$).
There is a precise mapping between the two phases that makes the integrable system equivalent.
This is a map between functions on the Poisson manifolds. 
Functions which correspond to cycles which do  not intersect the cycle $w_X$ do not change
under the spider move and should be the same in the two phases.
Functions corresponding to cycles which intersect non trivially with the cycle $w_X$
are modified as follows.
We can decompose every cycle which intersects with $w_X$ in two pieces:
one part which is external to the spider move and one part which is involved in the spider move.
The former is not modified by the spider move and it is the same in the two phases.
The latter part can be written as combination of the paths $\gamma, w_X$ in the Figure.
In order for the two phases to describe two equivalent integrable systems, 
the paths in the two phases $\gamma,w_X$ and $\gamma',w'_X$ should be mapped as
\begin{equation}
\label{rules}
 \gamma_{41} \to  \frac{\gamma_{41}'}{(1+\frac{1}{w_X'})} ,\,
 \gamma_{12} \to \gamma_{12}' (1+w_X'),\,
 \gamma_{23} \to  \frac{\gamma_{23}'}{(1+\frac{1}{w_X'})} ,\,
 \gamma_{34} \to \gamma_{34}' (1+w_X') ,\,
 w_X \to \frac{1}{w_X'} 
\end{equation}
This mapping is translated in a map between functions in the two different phases.
This mapping guarantees that the two integrable systems are equivalent.

\subsubsection{The dual phases of $\mathbb{F}_0$}

\begin{figure}[htpb]
\begin{center}
\includegraphics[width=13cm]{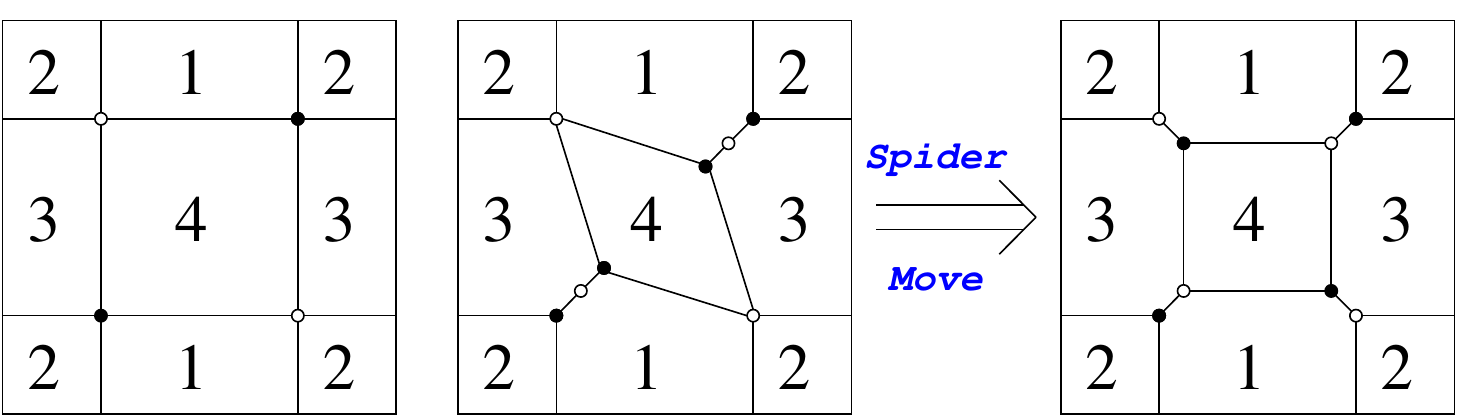}
\end{center}
\caption{Seiberg duality for $\mathbb{F}_0$:
insertion of massive bifundamentals in $\mathbb{F}_0^{(I)}$ and spider move}
\label{fignuova}
\end{figure}

In this section we show the equivalence between
the cluster Poisson transformation  of \cite{Goncharov:2011hp} and 
our derivation of Seiberg duality as a canonical transformation.
The  cluster Poisson transformation on the dimer 
is described by the  spider move, combined with the 
integration of some massive field (if necessary).
Indeed  the two bipartite 
graphs describing the two phases of $\mathbb{F}_0$ 
cannot be immediately mapped by a spider move.
However there is a simple operation 
in field theory which lead to recover the set up
of spider move we have just explained.
This operation 
consist of integrating in massive fields, 
which does not change the moduli space of the theory 
(and also the corresponding integrable system \cite{Goncharov:2011hp}).
We integrate in some massive field on dimer of the first phase of 
$\mathbb{F}_0$  as in Figure \ref{fignuova}.
In this description of $\mathbb{F}_0^{(I)}$ we can easily identify the face
$w_4$ as face on which we can act with a spider move, i.e. that we can dualize.
Indeed, by
acting with a spider move on the fourth face one obtains the tiling of the dual phase,
see Figure \ref{fignuova}.
This Figure shows that this combination of operation on the bipartite graphs 
gives the \emph{urban renewal} \cite{2001math.....11034P} transformation, already observed to
describe Seiberg duality on the bipartite graphs \cite{Franco:2005rj}.

We are now interested in finding the transformation which map the cycles in the two
dual phases.
We consider the basis of cycles $w_A,z_1, z_2$ and $w'_A,z'_1,z'_2$ 
as in section \ref{sec5} for the two phases respectively.
By following the discussion in \cite{Goncharov:2011hp}
that we just reviewed,
the cycles 
can be factorized in two terms. 
One is related to the paths involved in the spider move and the second part is 
invariant under this transformation. 
In the phase $I$ this parametrization gives
\beq
w_1=w_1^* \gamma_{34} \gamma_{12}, \quad
w_2=w_2^*, \quad
\quad
w_3=w_3^* \gamma_{41} \gamma_{23} ,\quad
w_4=w_X \quad
z_1=z_1^* \gamma_{23}, \quad
z_2=z_2^* \gamma_{34}^{-1}
\eeq
whereas in phase $II$ gives 
\beq
w_1'=w_1^* \gamma'_{34} \gamma'_{12}, \quad
w_2'=w_2^*, \quad
w_3'=w_3^* \gamma'_{41} \gamma'_{23}, \quad
w_4'=w'_X, \quad
z_1'=z_1^* \gamma'_{23}, \quad
z_2'=z_2^* \frac{1}{\gamma'_{34}}
\eeq
where we have denoted with $w^*$ the part of the cycle which is not modified by the spider move.
The transformation we have explained in the previous section on the $\gamma,w_X$ then implies that
the mapping among these cycles is
\begin{eqnarray} 
\label{SBSD2}
w_1 &=& w_1'(1+w_4')^2, \quad 
w_2 = w_2',\quad
w_3 = \frac{w_3'}{(1+{w_4'}^{-1})^{2}},\quad
w_4 = {w_4'}^{-1} \nonumber
\\
z_1 ~&=& \frac{z_1' }{(1+{w_4'}^{-1})} , \quad
z_2 = \frac{z_1' }{(1+{w_4'}^{-1})}
\end{eqnarray}
In these expressions the mapping between cycles should be understood as a map between the
corresponding functions, using the local coordinate systems that
we introduced in the previous sections.
For instance
\beq
w_1=e^{-\frac{c_1}{2}+\frac{c_2}{2}-\frac{c_3}{2}+\frac{q_1}{4}+\frac{q_2}{4}}
= w_1'(1+w_4')^2=
e^{-\frac{c_1^{S.d.}}{2}+\frac{c_2^{S.d.}}{2}+\frac{c_3^{S.d.}}{2}+\frac{q_1^{S.d.}}{4}+\frac{q_2^{S.d.}}{4}} \left(
1+ e^{\frac{q_1^{S.d.}}{4}-\frac{q_2^{S.d.}}{4}}
\right)^2
\eeq
and so on for the other $w_A$ and $z_1,z_2$.
 A solution to these equations is given by
\bea \label{t2}
&&
c_1=c_2^{S.d.}, \quad
c_2=c_1^{S.d.}, \quad 
 c_3=c_3^{S.d.},\quad
q_1= 4 \log \left( e^{\frac{q_1^{S.d.}}{4}} + e^{\frac{q_2^{S.d.}}{4}} \right)\\
&&
q_2 = 4 \log \left( e^{\frac{q_1^{S.d.}}{4}} + e^{\frac{q_2^{S.d.}}{4}} \right)+q_1^{S.d.}-q_2^{S.d.}
-4\left( c_1^{S.d.}- c_2^{S.d.}- c_3^{S.d.}\right)
\nonumber 
\eea
The solutions for the Casimir are the same as the one found in (\ref{t1})
while the solutions for $q_1$ and $q_2$ solve the equations
(\ref{eqal}) and (\ref{eqdiff}) as can be easily checked.
This corroborates the claim that Seiberg duality acts as a 
canonical transformation on the integrable system described through
 the master space and it is identified with the cluster Poisson transformation
 that glues the different patches of the integrable system.

\section{Conclusions and future directions}
\label{sec7}

In this paper we investigated the relation between  
the cluster integrable system of \cite{Goncharov:2011hp} 
and the coherent component of the master space of toric SCFT.
More precisely the irreducible component of the 
master space $\firr~$ is the variety describing 
a seed of the cluster integrable dimer model.
This local description is then globally
extended by acting with cluster Poisson transformations 
that in our language are associated to non toric maps 
among Seiberg dual $\firr~$. 
Many extensions of our work can be explored.

One may wonder if a similar description exists in three dimensional 
field theories. Indeed  the AdS/CFT has been extended to 
three dimensional
supersymmetric  quiver gauge theories in \cite{Aharony:2008ug}.
These theories are CS gauge theories and they are toric 
if the four dimensional CY has a $U(1)^4$ isometry.
In \cite{Hanany:2008cd,Hanany:2008fj,Martelli:2008si,Franco:2008um} 
many quiver gauge theories have been 
conjectured to describe this class
of singularity.
Extending the results of \cite{Goncharov:2011hp}
in three dimensions have two non trivial 
problems.
First the absence of global $U(1)$ anomalies in field 
theory hides the role of the master space and its relation
with the antisymmetric structure obtained from the 
skew-symmetric adjacency matrix.
Then it is not immediate to understand if the 
the candidate theories should be associated
to models with internal points or models with 
points on the faces. In the first case there exists 
models in which the antisymmetric structure 
 is not vanishing (chiral models) but without 
 internal points. In the second case there exist
 vector like models where the toric diagram has 
 points on the external faces. 
\footnote{Another problem is that it is not yet clear if  
chiral like models really describe the CY$_4$ 
geometry \cite{Jafferis:2011zi,Gulotta:2011aa,Amariti:2011uw}}

Then one can also  study  the 
integrability of these models from a geometrical
perspective. Indeed  the relation among the perfect matchings,
given by the symplectic quotient, should led to a 
geometrical proof of the integrability of the model.
This may be associated to the definition of the system as a 
polynomial Poisson algebra. In the case of a single internal point 
 the Poisson structure
between two generic function is 
\begin{equation} \label{bella}
\{f,g \}=\frac{df \wedge dg\wedge dQ_1 \wedge \dots \wedge dQ_{n-2}}{dx_1 \wedge dx_2 \wedge \dots \wedge d x_n}
\end{equation} 
where the $n$ coordinates $x_i$ are constrained by the polynomial relations
\begin{equation} 
Q_i (x_1, \dots, x_n)=0 \qquad i=1, n-2
\end{equation} 
In the case of the integrable  dimer  models these 
polynomial constraints are the relations among the perfect matchings.
It would be interesting to study the generalization of (\ref{bella})
 to the case of multiple internal points,
in which  a multilinear antisymmetric structure
is involved. We hope to come back to this
topic in future works.

Another further development of the work is the analysis of Seiberg dualities
as canonical transformations in the cases with multiple internal points. In that cases
the equation of the conservation of the Hamiltonian flow become more complicate, 
because they involve systems of coupled differential equations.
It would be interesting to study a general method for solving these equations and if they 
can be written as an algebraic set of equation as in 
\cite{Goncharov:2011hp}.
Another interesting development of our equations on Seiberg duality 
is related to the models with a cascading behavior. Indeed in a cascading gauge theory
a finite set of Seiberg dualities lead to the original quiver.  
In this case the duality not only 
preserves the Hamiltonian flow but also the functional form of the Hamiltonians.
These transformation have been observed in \cite{Eager:2011dp}
to be related to auto Backlund-Darboux transformations \cite{auto}
of the integrable system. It would be interesting to find the connections 
among these transformation and our equation (\ref{qpmap2}).

A last topic that we did not address in the paper concerns the role
of theories with points on the perimeter.
This question has been investigated in \cite{Eager:2011dp} where it was shown that
models with points on the perimeter can be obtained by partially resolving the singularity.
It was observed that in this way new integrable systems are obtained from known ones.
It would be interesting to study this mechanism on the master space.

\section*{Acknowledgements}

We are happy to thank Sebastian Franco,  Yang-Hui He, Kenneth Intriligator, Dan Thompson
and Alberto Zaffaroni   for nice discussions.
A.A. is supported by UCSD grant DOE-FG03-97ER40546; the work of D.F. is partially
supported by IISN - Belgium (convention 4.4514.08), by the Belgian Federal Science 
Policy Oce through the Interuniversity Attraction Pole P6/11 and by the Communaute
Francaise de Belgique through the ARC program; 
A.M. is a Postdoctoral Researcher of FWO-Vlaanderen. A. M. is also supported in part by
 the FWO-Vlaanderen through the project G.0114.10N, and in part by
the Belgian Federal Science Policy Oce through the Interuniversity Attraction Pole
IAP VI/11.

\appendix

\section{Intersection pairing}
\label{appA}

\begin{figure}
\begin{center}
\includegraphics[width=12cm]{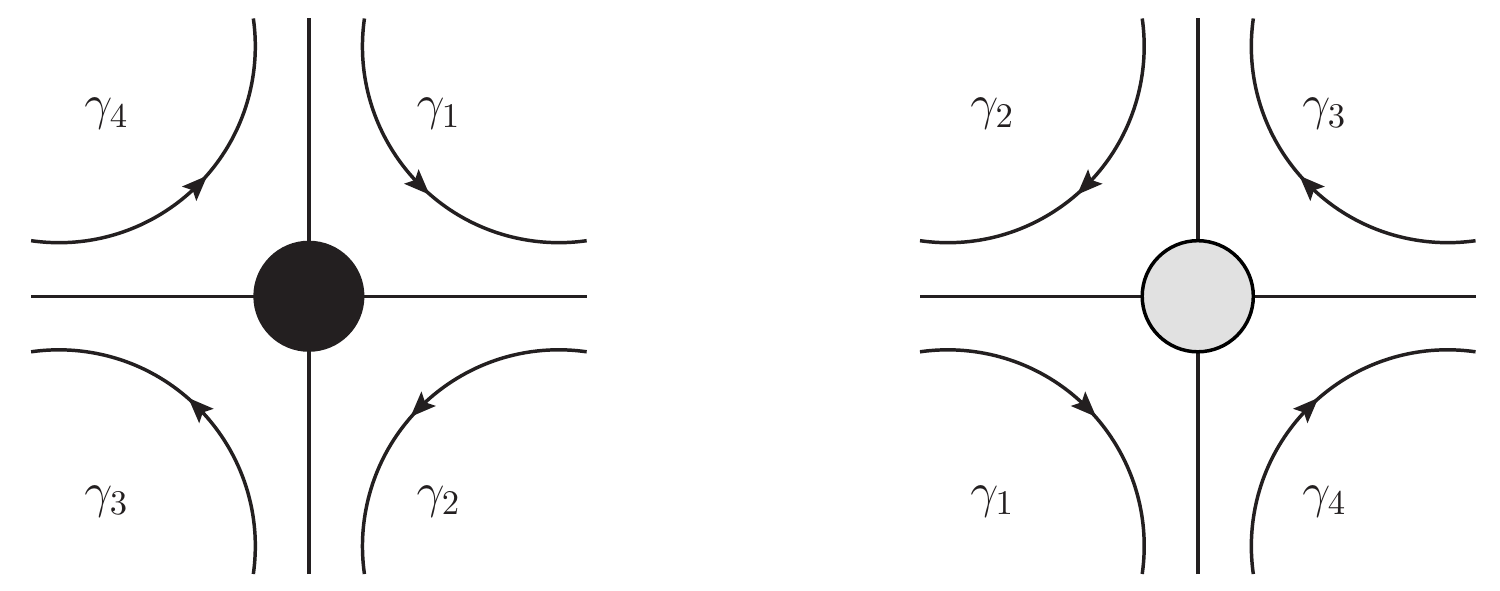}
\caption{
Base of cycles for black and white vertices.}
\label{gamma}
\end{center}
\end{figure}

\begin{figure}
\includegraphics[width=14cm]{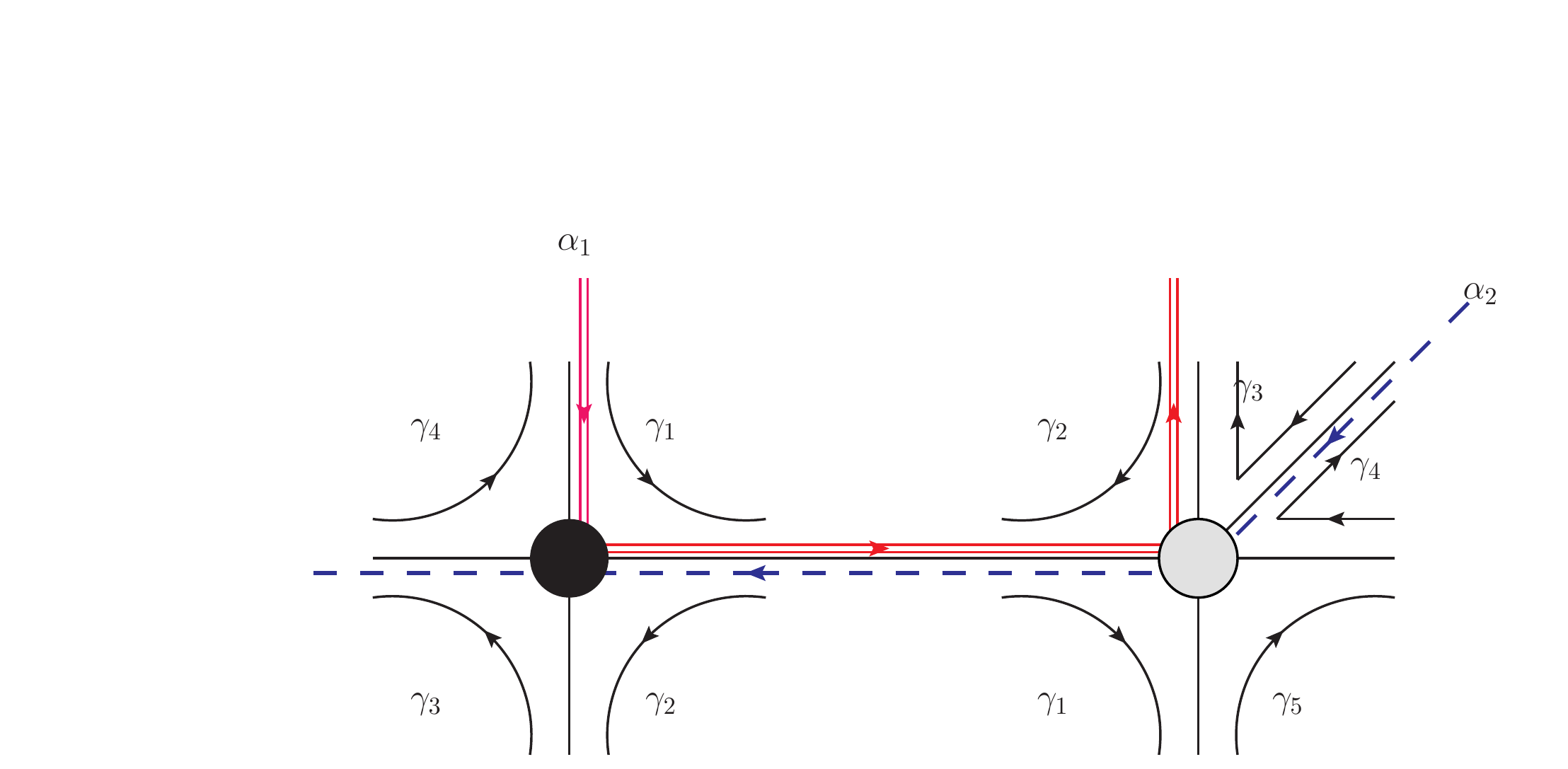}
\caption{
$\epsilon_{\alpha_1,\alpha_2}=1$}
\label{example}
\end{figure}

In this appendix we review the construction of \cite{Goncharov:2011hp}
for an antisymmetric pairing between cycles on a dimer model.
As explained in the main text,
the basic objects which can be interpreted as functions on the Poisson manifolds
are oriented closed cycles on the dimer.
The dimer is equipped with a precise orientation, given by the bipartite structure.
We take the convention in which the black nodes are oriented clockwise (and with a $+$ sign) 
and the white nodes
are oriented anti clockwise (with a $-$ sign).
Given two oriented cycles $\alpha_i$, $\alpha_j$ on a dimer, 
their poisson bracket is defined as
\beq
\{ \alpha_i, \alpha_j \}=\epsilon_{\alpha_i,\alpha_j} \alpha_i \alpha_j
\eeq
where $\epsilon_{\alpha_i,\alpha_j}$ is an antisymmetric pairing, that we now define.
In order to introduce the antisymmetric pairing we have to define 
an antisymmetric index that characterize the intersection of two cycles at every vertex.
Indeed the two cycles $\alpha_i$, $\alpha_j$ intersect on a finite number of vertices
on the dimer. 
Labeling with $v$ the vertices of the dimer, 
we define
\beq
\label{inte}
\epsilon_{\alpha_i,\alpha_j}= \sum_{v \in \alpha_i \& v \in \alpha_j} [v] \delta_v (\alpha_i \wedge \alpha_j)
\eeq
where $[v]=\pm 1$ if $v$ is a black or white vertex respectively.
$\delta_v (\alpha_i \wedge \alpha_j)$ is the antisymmetric index associated to the vertex $v$
and depends on the orientation and on the shape of $\alpha_i$, $\alpha_j$
around the vertex.
At every vertex we shall provide a base of local paths
which determine
the antisymmetric index.
Indeed,
the clockwise or anti-clockwise orientation of the vertex induces an orientation for the possible
paths passing through the vertex. 
We label the paths induced by this orientation with $\gamma^v_m$.
The index $m$ runs from $1$ to the valence of the vertex $v$,
named $val[v]$.
We give the parametrization in terms of these paths in
 Figure \ref{gamma}. 
 The orientation and also the enumeration of the path $\gamma_v^m$
 are relevant to define the intersection pairing.
We enumerate them always in the same order\footnote{Alternatively one can define the enumeration
with clockwise or anti-clockwise orientation, and the sign $[v]$ is always set to 1.}, 
that is clockwise.
A cycle $\alpha_i$ 
passing through the vertex $v$ can be always decomposed in sum or differences of 
the $\gamma^v_m$.
We define the antisymmetric index $\delta_v$ on the $\gamma^v_m$ basis as 
follows
\bea
&&
\delta_v(\gamma_m \wedge \gamma_p)= -\delta_v(\gamma_p \wedge \gamma_m) \nonumber \\ 
&&
\label{defi}
\delta_v(\gamma_m \wedge \gamma_{m+1})= \frac{1}{2} = \delta_v (\gamma_{m-1} \wedge \gamma_{m}) \\
&&
\delta_v(\gamma_m \wedge \gamma_{n \neq m+1 \& n \neq m-1})=\delta_v(\gamma_m \wedge \gamma_m)= 0 \nonumber
\eea
where a periodic enumeration of the basis, i.e. $(m+val[v] = m)$, is implicit.
Now we can find the index 
$\delta_v (\alpha_i \wedge \alpha_j)$
for two arbitrary cycles passing through the
vertex $v$.
We decompose the cycles $\alpha_i$ and $\alpha_j$ around $v$ on the base $\gamma^v_m$
\beq
\alpha_i=a_i^m \gamma^v_m \qquad \alpha_j=a_j^m \gamma^v_m
\eeq
where $a_i^m$ are $0,\pm 1$ depending on the edges and on the orientation of $\alpha_i$
with respect to $\gamma_v^m$.
Then the index is
\beq
\delta_v(\alpha_i \wedge \alpha_j) =  a_i^m a_j^n \delta_v(\gamma_m \wedge \gamma_n)
\eeq
Using these rules we can obtain the index $\delta_v$ at each vertex.
Then summing on the common
vertices as in (\ref{inte}),
we can obtain the antisymmetric intersection pairing $\epsilon_{\alpha_i,\alpha_j}$.
We provide an example in Figure \ref{example}.
In the Figure the cycle $\alpha_1$ is the red one (double line), and the cycle $\alpha_2$ 
is the blu one (dashed line). 
They can be expanded on the basis of the $\gamma_m$ around each vertex.
The intersection is
\bea
\epsilon_{\alpha_1,\alpha_2}&=&
+\delta_{black}(\alpha_1 \wedge \alpha_2)- \delta_{white}(\alpha_1 \wedge \alpha_2)= \\
&&
+\delta_{black}(\gamma_1 \wedge (\gamma_2+\gamma_3))-\delta_{white}(-\gamma_2 \wedge (-\gamma_1-\gamma_4-\gamma_5))=1 \non
\eea
One can check that these rules reproduce the intersection numbers 
in
\cite{Goncharov:2011hp,Franco:2011sz,Eager:2011dp}
.

\section{Spider move transformations}
\label{appB}

\begin{figure}[htpb]
\begin{center}
\includegraphics[width=14cm]{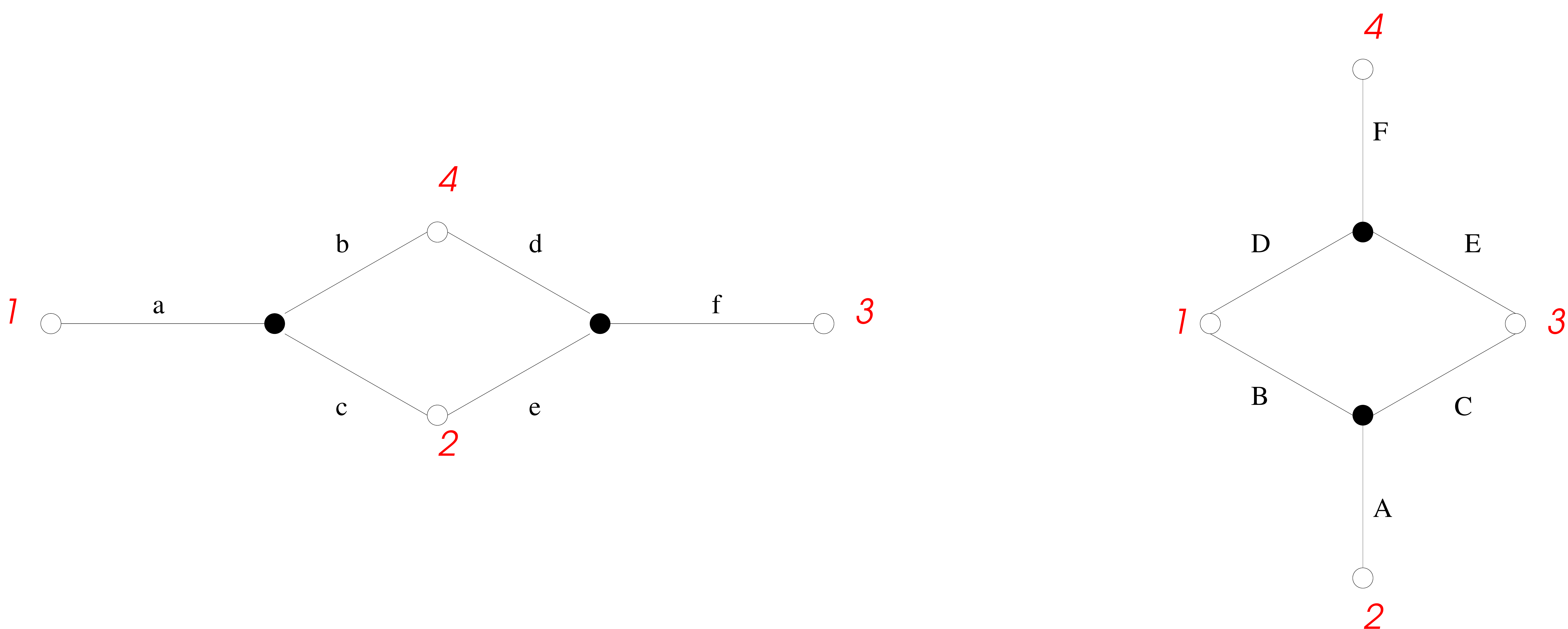}
\end{center}
\caption{Details  of the Spider Move }
\label{fignuova2}
\end{figure}

This is essentially a review section of the result of \cite{Goncharov:2011hp}
on Seiberg duality 
on the integrable dimer model.
First we introduce a parametrization of the loops in terms of ratio 
of edges (of fields), see also \cite{Franco:2011sz,Eager:2011dp}.
A loop is given by the difference of the $I$-th 
perfect matching and the $J$-th and we can parameterize it as in \ref{sec5}.
Then we define a new  $G+2 \times n $ matrix $F$, where $n$ is the number of fields,
such that $A= F d^T$
\begin{equation} \label{equo}
w_A = e^{ x_m B_{m,s}^{-1} F_{s,i} d_{i,A}^T}
\end{equation}
where the sums are understood.
Then we define $t_i = x_m B_{m,s}^{-1} F_{s,i} $ and (\ref{equo}) becomes
\begin{equation} 
w_A =\prod_{i=1}^{n} e^{t_i d_{i,A}^T}
\end{equation}
 Then we  associate to every field
a factor $\phi_{i} = e^{t_i}$. Then the $A$ index represents the loop. By dividing $n$ in the values of its entries 
$n=\{n_+,n_-,0\}$
\begin{equation}  \label{fortmu}
w_A =\frac{ {\prod}_{i \in n_+}  \phi_{i}}{ {\prod}_{j \in n_-}  \phi_{j}}
\end{equation}
where $n_+$ and $n_-$ refers to the $\pm1$ entries of the $A$-th row of the incidence matrix $d$.
Now, the \emph{spider move} is a local transformation on the tiling, and it corresponds to a Seiberg duality
on the dual field theory. It is represented by the transformation  
depicted in Figure \ref{fignuova}.
The white node labelled $1,2,3,4$ are the ones connected with the rest of the tiling, which is
invariant under the spider move.
Hence the entire characterization of this transformation can be encoded in the modifications
of the structure of the edges connecting the nodes $1,2,3,4$.
We label the edges involved in the spider move as in the Figure \ref{fignuova2}.
In order for the two phases to describe the same integrable system, we should match the 
perfect matchings in the two descriptions.
A perfect matching can be decomposed in an external part, which is not modified by the spider move, 
and an internal part, which is made by edges connecting the various vertices.
We can parametrize the internal perfect matchings in the following way.
Given two external vertices $i,j$, we denote with $p_{ij}$ the internal
perfect matchings which do not touch the vertices $i,j$.
This parametrization can be done for the two phases of the dimer, leading to
\begin{eqnarray}
&&
p_{12}= b f, \quad \quad
p_{13}=b e + c d,\quad \quad
p_{14}=c f 
\nonumber \\
&&
p_{23}= a d, \quad \quad
p_{24}=a f ,\quad \quad\quad\,\,\,\,\,
p_{34}=a e
\end{eqnarray}
before the transformation
\begin{eqnarray}
&&
p_{12}'= C F,\quad \quad
p_{13}'=A F,\quad \quad \quad \quad \quad
p_{14}'=A E
\nonumber \\
&&
p_{23}'= B F ,\quad \quad
p_{24}'=B E + C D ,\quad \quad
p_{34}'=A D
\end{eqnarray}
after the the transformation.
In order for the two integrable systems to be equivalent the internal perfect matchings 
of the two phases should be proportional one to each other \cite{Goncharov:2011hp}
\beq
\label{condgonce}
p_{ij} \sim p_{ij}'
\eeq
Ultimately
we are interested in finding the mapping between cycles of the two phases.
Cycles which do not intersect the dualized face $w_X$ are not modified by the spider move,
and hence are identified in the two dual phases. 
Cycles which intersect the face $w_X$ are instead involved in the spider move.
Such cycles can be decomposed in a basis of local path
given in Figure:
$\gamma_{12},\gamma_{23},\gamma_{34},\gamma_{41},w_X$ for phase (I) and 
$\gamma'_{12},\gamma'_{23},\gamma'_{34},\gamma'_{41},w'_X$ for phase (II).
These paths can be understood as ratio of edges
\bea
&&
\gamma_{12}=\frac{a}{c}, \quad \,\gamma_{23}=\frac{e}{f},
\quad \, \,\gamma_{34}=\frac{f}{d}, \quad \, \,  \gamma_{41}=\frac{b}{a},
\quad w_X=\frac{1}{\gamma_{12} \gamma_{23} \gamma_{34} \gamma_{41}}\\
&&
\gamma'_{12}=\frac{B}{A}, \quad \gamma'_{23}=\frac{A}{C},
\quad \gamma'_{34}=\frac{E}{F},
\quad \gamma'_{41}=\frac{F}{D},
\,\,\,\,\,
w'_X=\frac{1}{\gamma'_{12} \gamma'_{23} \gamma'_{34} \gamma'_{41}}
\eea
The requirement (\ref{condgonce}) can then be translated in a map between cycles in the two dual phases.
For instance
\beq
\gamma_{12}=\frac{p_{24}}{p_{14}}=\frac{p'_{24}}{p'_{14}}=
\frac{B E +  CD }{A E }=
\gamma'_{12} \left(1+w'_X\right)
\eeq
and so on for the other ratios of $p_{ij}$ and $p'_{ij}$ leading to
(\ref{rules}).

\section{Perfect Matchings of $Y^{30}$ and $Y^{40}$}
\label{appC}
\subsection{$Y^{30}$}
\begin{figure}
\begin{center}
\includegraphics[width=15cm]{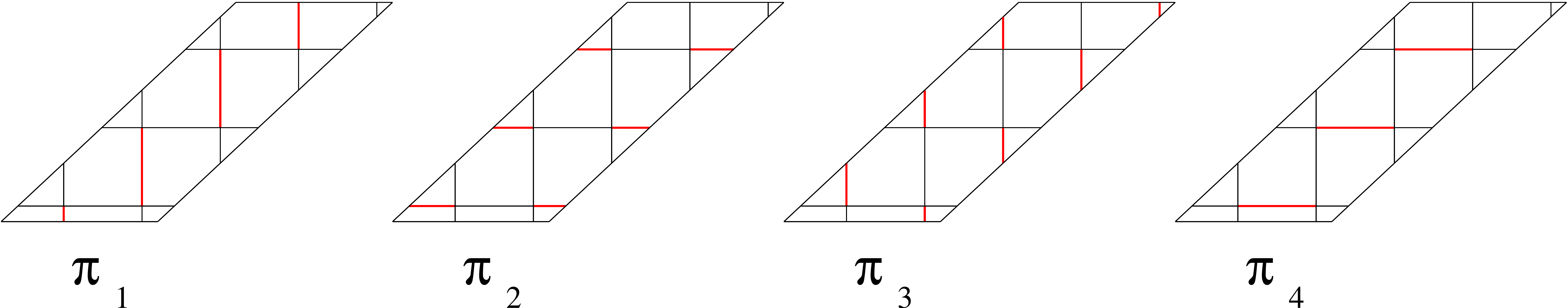}
\caption{External perfect matching for $Y^{30}$}
\label{ext30} 
\end{center}
\end{figure}

\begin{figure}
\begin{center}
\includegraphics[width=12cm]{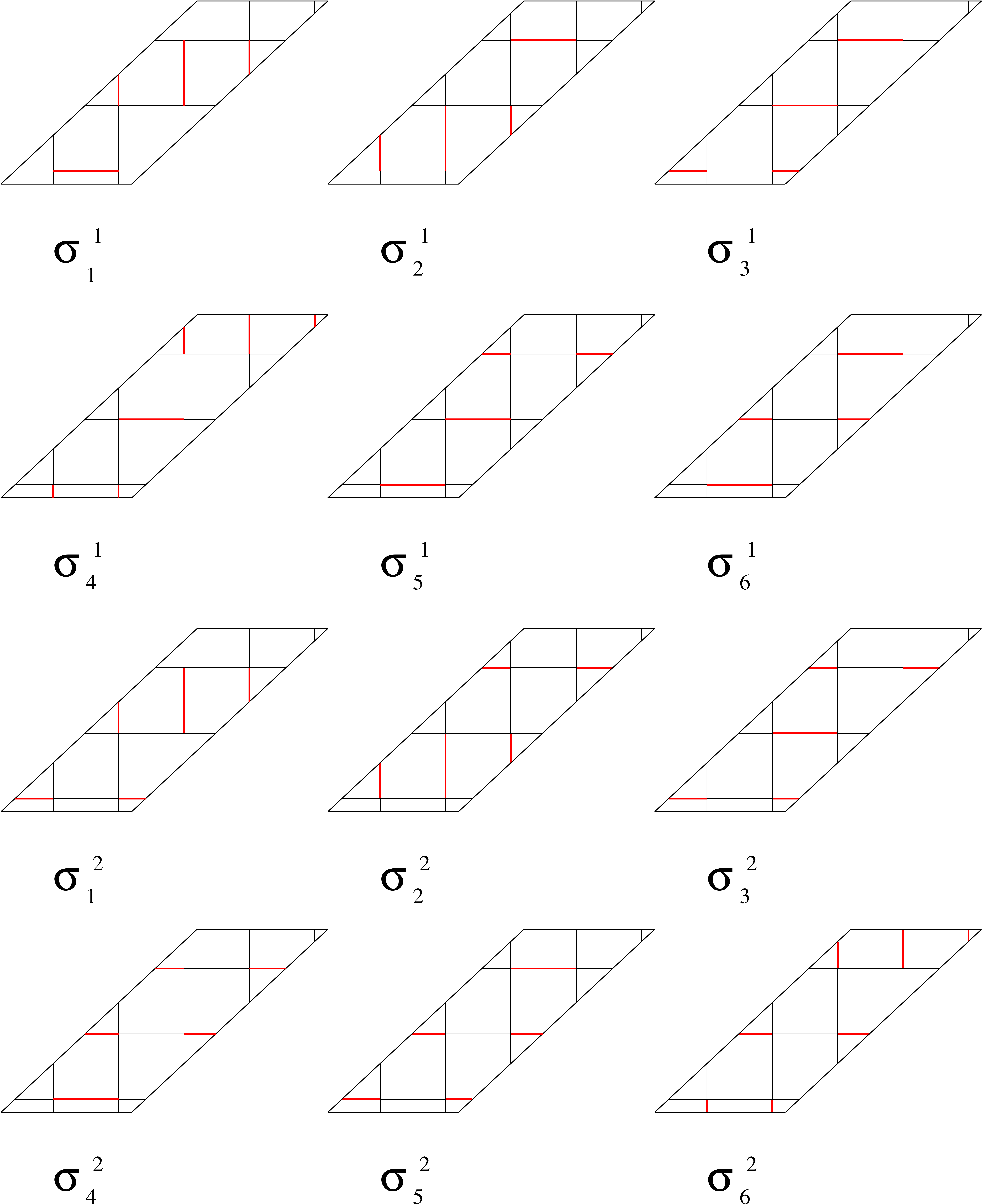}
\caption{Internal perfect matching for $Y^{30}$}
\label{int30} 
\end{center}
\end{figure}

The quiver gauge theory, the tiling and the toric diagram for $Y^{30}$
are reported in the Figure \ref{fig3030}.
The superpotential is
\bea
W&=&
X_{61}^{(1)} X_{12} X_{23}^{(2)} X_{36}
-X_{61}^{(2)} X_{12} X_{23}^{(1)} X_{36}
+X_{61}^{(2)} X_{14} X_{45}^{(1)} X_{56} \nonumber \\
&& 
-X_{61}^{(1)} X_{14} X_{45}^{(2)} X_{56}
+X_{23}^{(1)} X_{34} X_{45}^{(2)} X_{52}
-X_{23}^{(2)} X_{34} X_{45}^{(1)} X_{52}
\eea
The external perfect matchings are given in Figure \ref{ext30}
while the internal  perfect matchings are given in Figure \ref{int30}.
\subsection{$Y^{40}$}
\begin{figure}
\begin{center}
\includegraphics[width=8cm]{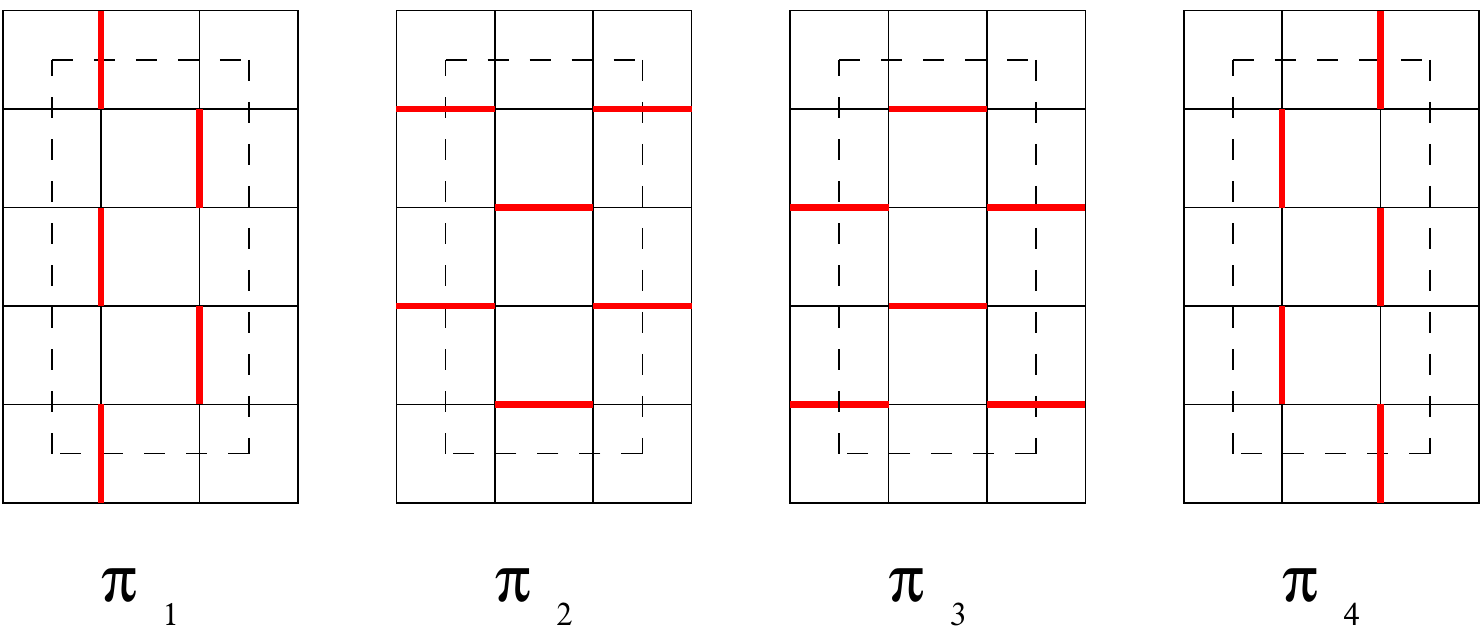}
\caption{External perfect matching for $Y^{40}$}
\label{ext40} 
\end{center}
\end{figure}

\begin{figure}
\begin{center}
\includegraphics[width=14cm]{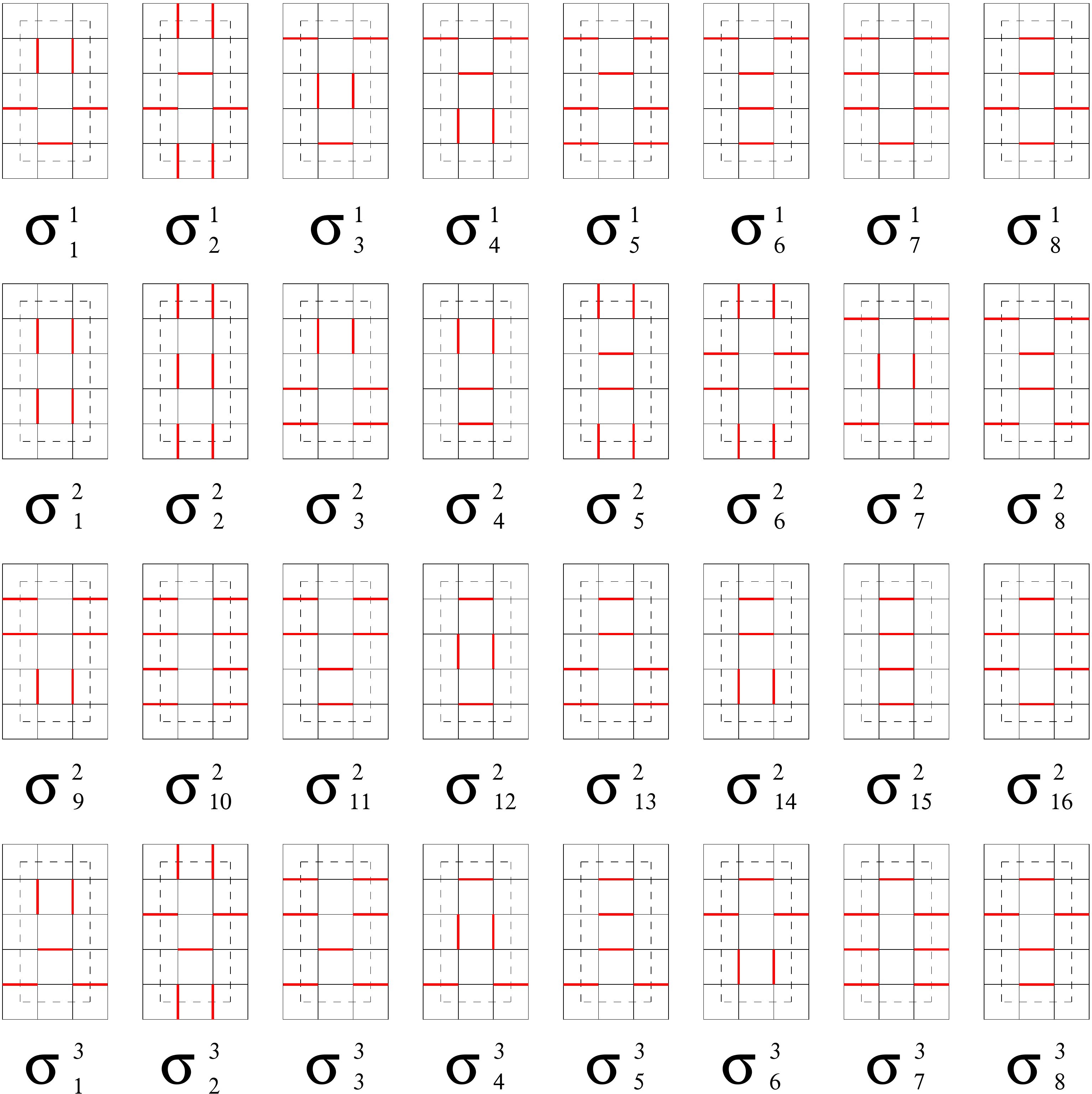}
\caption{Internal perfect matching for $Y^{40}$}
\label{int40} 
\end{center}
\end{figure}

The quiver gauge theory, the tiling and the toric diagram for $Y^{40}$
are reported in the Figure \ref{fig4040}.
The superpotential is
\bea
W&=&
X_{15}^{(1)} X_{58} X_{84}^{(2)} X_{41} -X_{15}^{(2)} X_{58} X_{84}^{(1)} X_{41}+X_{15}^{(2)}X_{56} X_{62}^{(1)} X_{21}
-X_{15}^{(1)}X_{56} X_{62}^{(2)} X_{21}
\nonumber \\&+&X_{62}^{(2)}X_{23} X_{37}^{(1)} X_{76}-
X_{62}^{(1)}X_{23} X_{37}^{(2)} X_{76}+X_{37}^{(2)}X_{78} X_{84}^{(1)}X_{43}-X_{37}^{(1)}X_{78} X_{84}^{(2)}X_{43}
\nonumber 
\\&&
\eea
The external perfect matchings are given in Figure \ref{ext40}
while the internal  perfect matchings are given in Figure \ref{int40}.
\newpage

\clearpage
\bibliographystyle{JHEP}
\bibliography{BibFile}

\end{document}